\documentclass[fleqn,usenatbib]{mnras}

\usepackage{newtxtext,newtxmath}

\usepackage[T1]{fontenc}
\usepackage{ae,aecompl}

\usepackage{graphicx}	
\usepackage{amsmath}	
\usepackage{enumerate}   
\usepackage{color}
\usepackage{ulem}
\usepackage{multirow}   
\usepackage{booktabs}   
\usepackage{siunitx}    
\usepackage{xcolor}     
\usepackage{xspace}
\usepackage{tabularx}
\usepackage{lipsum} 

\usepackage{amssymb}	
\usepackage{threeparttablex} 
\usepackage{longtable} 

\defcitealias{keszthelyi2019}{\nobreak{Paper~I}} 
\newcommand{\PaperI}{\citetalias{keszthelyi2019}\xspace} 

\defcitealias{keszthelyi2020}{\nobreak{Paper~II}} 
\newcommand{\PaperII}{\citetalias{keszthelyi2020}\xspace} 

\defcitealias{keszthelyi2021}{\nobreak{Paper~III}} 
\newcommand{\PaperIII}{\citetalias{keszthelyi2021}\xspace} 

\defcitealias{keszthelyi2022}{\nobreak{Paper~IV}} 
\newcommand{\PaperIV}{\citetalias{keszthelyi2022}\xspace}

\title[The effects of surface fossil magnetic fields on massive star evolution: 
V.]{The effects of surface fossil magnetic fields on massive star evolution: 
V. Models at low metallicity}
\author[Z. Keszthelyi et al.]{
Z. Keszthelyi$^{1}$\thanks{E-mail: 
 \href{mailto:zsolt.keszthelyi@nao.ac.jp}{zsolt.keszthelyi@nao.ac.jp}},
J. Puls$^{2}$,
G. Chiaki$^{1,3}$,
H. Nagakura$^{1}$,
A. ud-Doula$^{4}$,
\newauthor
T. Takiwaki$^{1}$, 
N. Tominaga$^{1,5,6}$
\\  
$^{1}$Center for Computational Astrophysics, Division of Science, National Astronomical Observatory of Japan, 2-21-1, Osawa, Mitaka, Tokyo 181-8588, Japan \\
$^{2}$LMU M\"unchen, Universit\"atssternwarte, Scheinerstr. 1, 81679 M\"unchen, Germany \\ 
$^{3}$National Institute of Technology, Kochi College, 200-1 Monobe, Nankoku, Kochi 783-8508, Japan \\
$^{4}$Penn State Scranton, 120 Ridge View Drive, Dunmore, PA 18512, USA \\ 
$^{5}$Astronomical Science Program, Graduate Institute for Advanced Studies, SOKENDAI 2-21-1 Osawa, Mitaka, Tokyo 181-8588, Japan\\
$^{6}$Department of Physics, Faculty of Science and Engineering, Konan University, 8-9-1 Okamoto, Kobe, Hyogo 658-8501, Japan\\
}
\date{Accepted . Received ; in original form 
}

\pubyear{2024}

\begin{document}
\label{firstpage}
\pagerange{\pageref{firstpage}--\pageref{lastpage}}
\maketitle

\begin{abstract}
At metallicities lower than that of the Small Magellanic Cloud, it remains essentially unexplored how fossil magnetic fields, forming large-scale magnetospheres, could affect the evolution of massive stars, thereby impacting the fundamental building blocks of the early Universe. 
We extend our stellar evolution model grid with representative calculations of main-sequence, single-star models with initial masses of 20 and 60~M$_\odot$, including appropriate changes for low-metallicity environments ($Z = 10^{-3}$--$10^{-6}$). 
We scrutinise the magnetic, rotational, and chemical properties of the models. When lowering the metallicity, the rotational velocities can become higher and the tendency towards quasi-chemically homogeneous evolution increases. While magnetic fields aim to prevent the development of this evolutionary channel, the weakening stellar winds lead to less efficient magnetic braking in our models. Since the stellar radius is almost constant during a blueward evolution caused by efficient chemical mixing, the surface magnetic field strength remains unchanged in some models.
We find core masses at the terminal-age main sequence between 22 and 52~M$_\odot$ for initially 60~M$_\odot$ models. This large difference is due to the vastly different chemical and rotational evolution.
We conclude that in order to explain chemical species and, in particular, high nitrogen abundances in the early Universe, the adopted stellar models need to be under scrutiny. The assumptions regarding wind physics, chemical mixing, and magnetic fields will strongly impact the model predictions.

\end{abstract}

 

\section{Introduction} \label{sec:intro}

Understanding the physics of massive stars ($M_{\rm ini} > 8$~M$_\odot$) is of paramount importance for studies of the early Universe \citep[e.g.,][]{garcia2021,eldridge2022,klessen2023,vink2023a}. The role of massive stars in cosmic re-ionisation, chemical enrichment, and stellar feedback makes them crucial building blocks over various spatial and temporal scales \citep[e.g.,][]{Chiaki2018, Chiaki2020, Chiaki2019}.
%
In low-metallicity environments, from here on defined as metallicity lower than that of the Small-Magellanic Cloud (SMC), that is, $Z < 2.6\cdot10^{-3}$ or $Z/Z_\odot < 20\%$ \citep[e.g.,][]{mokiem2007,dopita2019}, the detailed observational characterisation of massive stars is still in its early stages.
%
The first large sub-SMC catalogue of OB stars was recently compiled by \cite{lorenzo2022} and a census of OBe stars in three low-metallicity dwarf galaxies was studied by \cite{schootemeijer2022}. 
Several other studies scrutinise individual or small number of metal-poor massive stars, nevertheless with important implications for physical processes \citep[e.g.,][]{ji2024,mardini2024}.

Currently, the source of high nitrogen content in distant galaxies --e.g., a factor of 3 nitrogen enrichment compared to the solar value in GN-z11 observed by the James Webb Space Telescope -- remains elusive, with Wolf-Rayet type stars, very massive stars and super massive stars suggested as their possible origin \citep[e.g.,][]{watanabe2024, vink2023,charbonnel2023,marques2024}. Nitrogen enrichment from stars is also reported for several other low-metallicity dwarf galaxies \citep[e.g.,][]{yarorova2023}.

Stellar evolution models are commonly utilised to interpret spectroscopic observations and guide galactic and cosmological-scale studies.
Various codes and assumptions have been employed to calculate grids of low and zero-metallicity massive star evolutionary models \citep[e.g.,][]{meynet2002,heger2010,takahashi2014, szecsi2015,limongi2018,martins2021,groh2019,murphy2021,tsai2023,volpato2024,roberti2024}. 
Insofar, some physical ingredients in massive star evolutionary models, namely, stellar winds and rotation, are often extrapolated from constraints available in Galactic environments. 
Moreover, until recently evolutionary models of massive stars --both at high and low metallicity -- seldom considered surface magnetic fields. Therefore, the impact of this physical mechanism remains mostly without quantification.
For Population III stars with zero metallicity, \cite{haemmerle2019} calculated supermassive star models with magnetic braking and \cite{lou2022} investigated magnetic equilibria and radial pulsations. \cite{urpin2019} studied the thermal induction of magnetic fields. For low metallicities of $Z/Z_\odot \sim 1/10$ and $1/50$, \cite{li2023} utilised stellar evolution calculations, including a branch of magnetic models with a 1 kG field strength, to study supernova progenitors. 
Nevertheless, until now, a coherent extension of typical massive star models with surface magnetic field effects at sub-SMC metallicities is missing. This underpins a major uncertainty in evolutionary model calculations since both stellar winds and rotation are drastically impacted by large-scale magnetic fields, which form a magnetosphere around the star \citep[e.g,][]{ud2002,ud2009,townsend2005,owocki2016}.

%
%
%
%
This paper is part of a series in which we aim to explore the effects of surface fossil magnetic fields on massive star evolution (\citealt{keszthelyi2019, keszthelyi2020, keszthelyi2021, keszthelyi2022}, \PaperI--\PaperIV from here on). 
\PaperI, \PaperII, and \PaperIII focused on solar metallicity models, while in \PaperIV we computed a large grid for metallicities representative of the solar neighbourhood and the Magellanic Clouds. For a very brief summary of the series, we refer the reader to the introduction of \PaperIV.
Given that stellar magnetism is known to largely impact the evolution of stars at high-metallicity environments, it is natural to wonder how stellar model predictions at low and zero metallicity are impacted by magnetism.
In this work, we utilise our implementation of surface fossil magnetic field effects to study typical massive star evolutionary models with initial masses of 20 and 60~M$_\odot$ at a domain of $Z = 10^{-3} - 10^{-6}$, corresponding to $Z/Z_\odot \approx 6\cdot10^{-2} - 6 \cdot 10^{-5}$.
In this metallicity range the incidence rate and general magnetic properties of massive stars are not yet known observationally since spectropolarimetric observations remain limited to detections in a Galactic environment \citep[e.g.,][]{bagnulo2020}. 
In addition to the parametrisation and schemes introduced in \PaperIV, we consider appropriate changes for low-metallicity models. This includes two branches to calculate the applied mass-loss rates. This is because stellar wind physics becomes more uncertain and observationally unconstrained in sub-SMC metallicities. In fact, when lowering the metallicity a critical limit may be reached, leading to effectively no mass loss. We discuss these modelling ingredients in Section \ref{sec:two}.

The paper is organised as follows. In Section~\ref{sec:two}, we detail the main assumptions and general model setup. In Sections~\ref{sec:three}, \ref{sec:four}, and \ref{sec:five} we present and scrutinise the stellar structure and evolution models from our computations. In Section~\ref{sec:six}, we discuss the implications, and in Section~\ref{sec:future} we outline future avenues. Then, we conclude our findings in Section~\ref{sec:concl}.

%
%

\section{Methods}\label{sec:two}

%
%
%
%
\subsection{General model setup}

We use the software instrument Modules for Experiments in Stellar Astrophysics \textsc{mesa} release 22.11.01 \citep[][]{paxton2011,paxton2013,paxton2015,paxton2018,paxton2019} and Software Development Kit (\textsc{SDK}) version 22.6.1 \citep[][]{sdk}. The \textsc{mesa} microphysics are summarised in Appendix~\ref{sec:micro}. 
The models computed in this work are single-star models covering the main sequence evolution. Here we summarise some main points of the model setup but refer the reader to \PaperIV for further details.

%
%
%
%
\begin{table}\label{tab:models}
    \centering
    \caption{Summary of computed models.}
    \begin{tabular}{p{3cm}p{4cm}}
        \toprule
        \toprule
        Parameter/Branch & Values/Schemes \\
        \midrule
        $Z_{\mathrm{ini}}$ &
         $10^{-3}$, $10^{-4}$, $10^{-5}$, $10^{-6}$  \\     
        $M_{\mathrm{ini}}$ [M$_{\odot}$] & 20, 60 \\
        $B_{\mathrm{eq, ini}}$ [kG] & 0, 0.5, 1, 10, 50 \\
        Braking scheme & INT, SURF \\
        AM transport scheme & INT, SURF, NOMAG \\
        Chemical mixing scheme & Mix1, Mix2 \\
        Wind scheme & W1, W2 \\
        \bottomrule
        \bottomrule
    \end{tabular}
\end{table}

%
%
\subsubsection{Initial mass}
In this study, we consider initial masses of $M_{\rm ini} = \{20, 60 \} \, \mathrm{M}_{\odot}$. 
We choose this mass range for four reasons. 
i) It is consistent with the upper mass range presented in \PaperIV, where the effects of winds and rotation are the most pronounced.
ii) Stable fossil fields are anchored in the radiative envelope and are most likely expelled from the convective cores of main-sequence massive stars \citep{keszthelyi2023}. For initial masses above 60~M$_\odot$, the convective core starts to occupy a much larger fraction of the total mass of the star. Due to strong stellar winds that remove much of the radiative envelope, the core boundary will reside much closer to the stellar photosphere. In addition, the envelopes of Very Massive Stars (VMSs) may reach super-Eddington luminosities. In those layers, convective turbulence could develop to transport the luminosity (\citealt{quataert2012,quataert2016} \footnote{See also \citet{cantiello2011,cantiello2019,jermyn2021} for a discussion on subsurface convection in massive stars.}). The stability of fossil fields is unclear under those conditions (see Section~\ref{sec:future}).
iii) Given the initial effective temperatures ($\approx 50$~kK), the models are most representative of typical O-type stars. Therefore, these models may be directly compared with observations of stars in metal-poor galaxies.
iv) VMSs and supermassive stars (SMSs) might dominate the luminosity budget and chemical evolution of stellar populations and have received considerable interest at near zero-metallicity conditions. 
The "typical" massive stars that we model in this work would be just as frequent in number as more exotic ones if the initial stellar mass function (IMF) was completely flat for high masses.
However, the IMF is likely not flat but top-heavy close to zero metallicity, supported by star formation studies (e.g., \citealt{Chiaki2022, Chon2022}, see also \citealt{bastian2010}). In addition, protostellar feedback could also play a role in setting the upper mass limit of a forming star \citep{hosokawa2011}. These imply that typical massive stars are possibly more numerous than VMSs and SMSs and therefore could have an important contribution to the luminosity budget and chemical evolution. 

%
%
\subsubsection{Initial magnetic field strength}
We consider initial equatorial surface magnetic field strengths $B_{\rm eq, ini} = \{0, 0.5, 1, 10, 50 \} $~kG. The upper limit is the same as in \PaperIV and is chosen as an exploration of the parameter space. While observationally inferred magnetic field strengths of Galactic massive stars are in the kG range \citep{grunhut2017,shultz2018}, it is presently unclear if fossil fields in low-metallicity environments would have different strengths.
Star-forming molecular clouds have more than sufficient magnetic flux to produce magnetised stars \citep[e.g.,][and references therein]{keszthelyi2023}. At high metallicity, efficient dissipation via non-ideal magnetohydrodynamic processes need to be invoked to reduce the magnetic flux from order of $10^{33}$~G\,cm$^{2}$ in star-forming cores to around $10^{27}$~G\,cm$^{2}$ that is observed in stars (\citealt{petit2017}, \PaperII). So far, it remains unclear what initial magnetic flux and dissipation could characterise molecular clouds at low metallicities.

%
%
\subsubsection{Initial rotation}
Consistently with \PaperIV, rotation is initialised assuming a flat angular velocity profile and we use the \textsc{mesa} control option $\Omega/\Omega_{\rm crit}$ (set to 0.5), where the critical angular velocity is defined in \textsc{mesa} as $\Omega_{\rm crit}~=~\sqrt{(1 - \Gamma) GM / R } $ with $\Gamma$ the Eddington parameter calculated for all opacity sources at the stellar photosphere. 
This results in initial surface equatorial rotational  velocities in the range of $\approx$~350--550~km\,s$^{-1}$, with increasing values towards higher initial masses and lower metallicity (due to smaller stellar radii).
This assumption is a reasonable approximation since star-formation studies indicate that low-metallicity massive stars may initiate their evolution with high rotational velocities \citep{hirano2018}. We refrain from computing models with various initial $\Omega/\Omega_{\rm crit}$ ratios since to some extent this is substituted by our assumptions regarding chemical mixing.

\subsubsection{Chemical mixing schemes}

As in \PaperIV, we adopt two chemical mixing schemes that account for transporting elements inside the stellar model. The "Mix1" scheme contains the typical description used in \textsc{mesa}, which is based on the works of \cite{endal1978} and \cite{pin1989}. The "Mix2" scheme is our implementation of the \cite{zahn92} formalism into \textsc{mesa}. We found the latter case to be very efficient in chemical mixing at higher-metallicity environments, therefore these scenarios may be considered as boundary cases for weak and strong chemical mixing. We further elaborate on this, particularly, in Section~\ref{sec:three}.

\subsubsection{Angular momentum transport and braking schemes}
In \PaperIV, the models were split into 3 main branches to account for uncertainties in applying angular momentum transport and magnetic braking.
"INT" models are assumed to be solid-body rotating. Angular momentum loss accounting for magnetic braking and stellar winds is distributed through all layers of these models. "SURF" models on the other hand have a rotation profile, which allows for radial differential rotation. In the SURF models, specific angular momentum is removed from the envelope and surface layers in one time-step. It is assumed that this region contains 20 per cent of the total mass of the star. In the "NOMAG" models, we assume that only hydrodynamic instabilities transport angular momentum in \textsc{mesa}'s diffusive scheme. In this case, angular momentum loss follows the default \textsc{mesa} procedure. Specific angular momentum is subtracted from those near surface layers where mass is removed to account for stellar winds.

%
%
\begin{figure}
\includegraphics[width=0.48\textwidth]{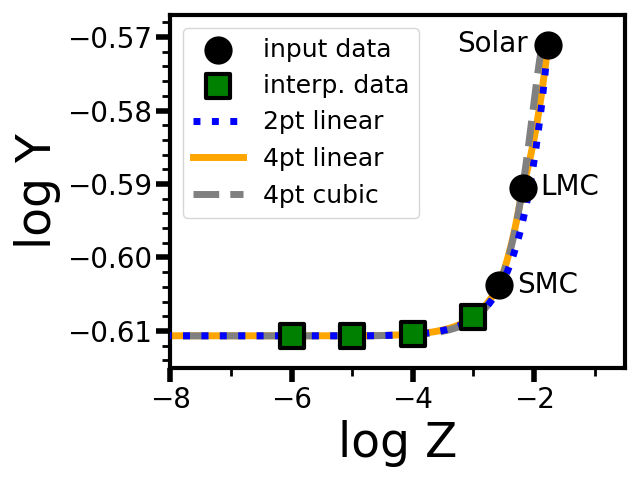}
\caption{Helium abundances as a function of metallicity. Symbols with black circles show the input anchoring values (Table~\ref{tab:t2}). Zero metallicity is omitted from the figure. Interpolation is performed for the normal values. The 2-point interpolation uses the Solar and zero metallicity points, the 4-point interpolations include data from the Magellanic Clouds. The blue dotted, yellow full, and grey dashed lines show the interpolation methods. Symbols with green squares indicate our obtained initial helium abundances for a given metallicity, using the 4-point linear interpolation.}
\label{fig:hecalc}
\end{figure}
%
%
%

%
%
\begin{table}
\centering
\caption{Mean helium and metal abundances from the literature.}
\begin{threeparttable}
\label{tab:t2}
\begin{tabular}{lllll}
\toprule
\toprule
& primordial & LMC & SMC & proto-solar \\
\midrule
$Y_{\rm anchor}$ & 0.245075 $^{[1,2,3,4]}$ &   0.24900 $^{[5]}$ &  0.25671 $^{[5]}$ & 0.26848 $^{[6]}$ \\ 
$Z_{\rm anchor}$ & 0                       &   0.00260 $^{[5]}$ &  0.00644 $^{[5]}$ & 0.0169 $^{[6]}$   \\
\bottomrule
\bottomrule
\end{tabular}
\begin{tablenotes}
\item References: [1] \cite{peimbert2007}, [2] \cite{planck2016}, [3] \cite{matsumoto2022}, [4] \cite{kurichin2022}, [5] \cite{dopita2019}, [6] \cite{magg2022}
\end{tablenotes}
\end{threeparttable}
\end{table}
%
%

%
%
\begin{table}
\centering
\caption{Initial hydrogen, helium and metal abundances in our models.}
\label{tab:t3}
\begin{tabular}{llll}
\toprule
\toprule
$X_{\mathrm{ini}}$ & $Y_{\mathrm{ini}}$ & $Z_{\mathrm{ini}}$ & $Z_{\mathrm{ini}}/Z_\odot$\\
\midrule
0.75242 & 0.24658 & $10^{-3}$ & $6\cdot10^{-2}$ \\ 
0.75467 & 0.24523 & $10^{-4}$ & $6\cdot10^{-3}$ \\
0.75490 & 0.24509 & $10^{-5}$ & $6\cdot10^{-4}$ \\
0.75492 & 0.24508 & $10^{-6}$ & $6\cdot10^{-5}$ \\
\bottomrule
\bottomrule
\end{tabular}
\end{table}

\subsubsection{Differences compared to \PaperIV}

While the models are generally consistent with those of \PaperIV, we should highlight that in the present work we use a newer \textsc{mesa} release and consider a more extended 21-isotope "approx21.net" nuclear network compared to the 8-isotope "basic.net" used in \PaperIV.

\subsection{Main assumptions for low metallicity}

To consider models at low metallicity we rely on some further assumptions. We now discuss changing the initial chemical composition, as well as the consideration of magnetic braking and stellar-wind driven mass loss.

%
%
\subsubsection{Metallicity}

We adopt initial metallicities in our models from $Z = 10^{-3}$ to $10^{-6}$, which is approximately from $\approx$1/17 to $\approx$1/17,000 of solar metallicity, considering $Z_{\odot} \approx 0.014-0.017$ \citep{asplund2009,magg2022}. This range encompasses the presently known most metal-poor massive stars observed in Sextans A and IZw18 and also extends to the high-redshift universe. 
To obtain the corresponding initial hydrogen ($X_{\rm ini}$) and initial helium ($Y_{\rm ini}$) abundances, we test a few different interpolation methods. To this extent, we use anchoring values of metals and helium of the primordial abundance, the initial abundance in the Large and Small Magellanic Clouds, and the initial solar abundance. Table~\ref{tab:t2} lists the mean values, where the study of \cite{dopita2019} used the mean values of nine different investigations and the study of \cite{magg2022} considered the mean of six values from different studies and methods. Generally, the helium abundances at a given metallicity are consistent between various studies within a 5\% level. 
A usual method is to perform a two-point linear interpolation, typically using the proto-solar and primordial helium abundances. We additionally consider whether four-point linear and cubic interpolations (including data from the Magellanic Clouds) could result in differences and find that in the metallicity range of our interest, the three methods agree very well with deviations below a 0.1\% level (Figure~\ref{fig:hecalc}). We opt to use the results obtained from the four-point linear interpolation. 

%
%

%
The resulting initial helium and hydrogen ($X = 1 - Z - Y$) abundances are listed in Table~\ref{tab:t3}.
We assume that all initial hydrogen is in the $^{1}$H isotope. 
For simplicity, we make the assumption that the initial helium abundance is distributed between its two isotopes as $^{3}\mathrm{He}/^{4}\mathrm{He} \sim 10^{-4}$ across all metallicities (see, e.g., \citealt{meynet2002}). We tested models with initially only $^{4}\mathrm{He}$ and found that they generally lead to small differences in the main model characteristics. Nonetheless, such an assumption can indeed be important when extrapolating to zero metallicity environments (c.f. \citealt{west2013,takahashi2018}).
Although it is likely that the distribution of individual metal abundances over cosmic times does not follow a simple scaling down from measured solar values, the uncertainty in describing individual metal elemental abundances makes it impossible to tailor this input in our models. For this reason, we choose to use the initial distribution of metals with the option \texttt{initial\_zfracs = 8} in \textsc{mesa}, corresponding to the \cite{asplund2009} data, modified by \cite{nieva2012} and \cite{prz2013}, considering B-type stars in the solar neighbourhood\footnote{In \PaperIV, we use the same mixture of initial metal abundances for our solar metallicity models, with $Z=0.014$. For the LMC and SMC models, we use the mixture of metals as described by \cite{dopita2019}.}. 
In all cases, we use the \cite{lodders2003} isotopic ratios and the metallicity adopted for the chemical evolution is the same value that is used for the opacity calculations. 
%

%
%
\subsubsection{Magnetic braking}\label{sec:magbraking}

Large-scale magnetic fields, which form a magnetosphere around the star, are efficient in transporting angular momentum along the field lines independently of the plasma flow. As the star rotates, this Maxwell stress and associated Poynting flux can transfer angular momentum from the magnetic field to the surrounding wind material. The angular momentum loss from the star leads to rapid spin down. This is well evidenced by the generally slow surface rotation of magnetic stars \citep[e.g.,][]{shultz2018,shultz2019b}.

Two-dimensional magnetohydrodynamic simulations of rotation-aligned dipole fields of massive stars by \cite{ud2009} indicated that the angular momentum carried out by a magnetically torqued stellar wind follows the same simple, split-monopole scaling law derived for the Sun by \cite{weber1967}, $\dot J =\frac{2}{3} {\dot M} \, \Omega \, R_A^{2}$ -- with, however, the Alfv\'en radius $R_A$ now given by the dipole scaling $R_A \sim \eta_\star^{1/4}$, instead the oft-quoted, stronger scaling $R_{A} \sim \eta_\star^{1/2}$ for a split monopole. Here $\eta_\star$ is the `wind magnetic confinement parameter' and represents the ratio of magnetic energy density and wind kinetic energy density \citep{ud2002}. 
For typical confinement by a dipole field ($\eta_\star \approx 10 - 100$), $\eta_\star^{1/2}  \propto B  \dot{M}^{-1/2} \propto R_A^2$, with $B$ the magnetic field strength and $\dot{M}$ the mass-loss rate. Therefore, the angular momentum loss $\dot{J}$ is proportional to $B  \dot{M}^{1/2}$.
This scaling forms the basis for modelling the rotational evolution of magnetic massive stars (\PaperI, \PaperII, \PaperIII, \PaperIV). 

Thus far, it is little explored how magnetic braking operates in low wind density conditions. The key point to achieve the stellar spin-down is that the magnetic field at the distance of the Alfv\'en radius has to be able to transfer angular momentum to some surrounding material that is detached from the star. In principle, this could also be the interstellar medium. In our models, we assume that magnetic braking operates for (very) low mass-loss rates; however, we do not apply it in the case of $\dot{M}=0$.

%
%

\subsubsection{Stellar wind physics}\label{sec:winds}

Radiative-line driving is a well understood mechanism that leads to significant mass loss from high-metallicity, hot, massive stars  \citep[e.g.,][for reviews]{puls2008,vink2022}. However, the nature of stellar winds at metallicities lower than that of the SMC is highly uncertain. Radiative-line driving is expected to become less efficient when lowering the metallicity, primarily due to the decrease in the iron abundance, which is the major driver at near-solar metallicity conditions. 

At low wind densities (as predicted for low metallicity environments and/or for stars of comparatively low luminosity), a potential de-coupling of ions from the bulk wind plasma could lead to a runaway effect. The corresponding frictional heating might dominate the energy balance, thus even enforcing the de-coupling \citep{springmann1992}. In a series of publication, \cite{krticka2000,krticka2001a,krticka2001b,krticka2002,krticka2006a} studied two- and three-component stellar wind models (partly including frictional heating), and surprisingly found steady-state solutions even for low wind densities. They argued that the de-coupling of the accelerated ions from the bulk H+He plasma becomes quite difficult, if not impossible, under typical conditions. We note here that in contrast to metals, hydrogen and helium have only a few lines available, therefore any form of wind driving from H+He alone would result in very low mass-loss rates \citep{krticka2006b}.

\cite{owocki2002} confirmed the findings by \cite{krticka2002} for stationary assumptions, and discovered a new, very strong and rapidly-propagating "ion separation instability". This could destroy any stationary flow at those densities (or even somewhat higher ones) where the de-coupling may occur. Based on these results, we follow the parametrisation of \cite{springmann1992} to calculate the critical mass-loss rate $\dot{M}_{\rm crit}$ as a function of metallicity where the metal ions are predicted to de-couple from the H/He bulk wind. Below this limit, the accelerated metal ions \textit{might} de-couple from the bulk wind, such that mass loss either terminates at all, or continues at extremely low rates (if driving by H and He alone was possible). We have also compared with the alternative approach by \cite{owocki2002}, which results in slightly different values, though at maximum by a factor of two. For simplicity, and to obtain a lower limit for $\dot{M}_{\rm crit}$, we neglect the impact of frictional heating.

We start with the condition for the boundary between 
coupling and de-coupling, by equating 
\begin{equation}
\Gamma_{\rm B} = \Gamma_{\rm L},
\end{equation}
where $\Gamma_{\rm B}$ is the $\Gamma$-value ($= g_{\rm
rad}(r)/g_{\rm grav}(r)$) for which the drift between ions and bulk plasma reaches the thermal speed. For larger drifts, the collisional coupling decreases rapidly, and de-coupling becomes (principally) possible in a runaway fashion. $\Gamma_{\rm L}$ is the radiative
line force scaled to the total wind density, divided by the
gravitational acceleration. In the supersonic approximation and
neglecting the acceleration by electrons (this is one of the
differences compared to  the study by \citealt{owocki2002}),
 \begin{equation}
\Gamma_{\rm L} \approx v(r) \frac{\mathrm{d} v}{\mathrm{d} r}\frac{1}{g_{\rm grav}(r)}
\end{equation}
Following Equations.~(12) and (13) of \cite{springmann1992}, i.e.,
neglecting the number density of metals compared to the number density of the bulk plasma (with a Helium content\footnote{For a plasma with very large Helium content (compared to hydrogen), e.g., a wind of a classical WR-star, this formulation needs to be adapted, since otherwise $Y_{\rm He}
\rightarrow \infty$ (though compensated by $Y_i$
$\rightarrow \infty$ as well).} $Y_{\rm He}=n_{\rm
He}/n_{\rm H}$, and a number of $I_{\rm He}$ free electrons per
He nucleus), $\Gamma_{\rm B}$ can be approximated by
\begin{equation}
\Gamma_{\rm B} \approx Y_i \frac{\dot M}{r^2 v(r)} \frac{\ln \Lambda Z_i^2
e^4}{m_{\rm H}^2 k_{\rm B} T} \Bigl(\frac{1+I_{\rm He}Y_{\rm He}}
{1+4 Y_{\rm He}}\Bigr)^2 \frac{1}{1+Y_{\rm He}} \frac{G_{\rm
max}}{g_{\rm grav}(r)},
\end{equation}
with $Y_i$ the metallic abundance (expressed as a number fraction,
$Y_i=n_i/n_{\rm H}$), $\ln \Lambda$ the Coulomb logarithm ($\approx 20)$, $Z_i$ the average charge of the metallic ions in units of the electron charge $e$ ($Z_i \approx$ between 2
and 3 for OB stars), $m_{\rm H}$ the hydrogen mass, $k_{\rm B}$ the Boltzmann constant, $G_{\rm max} = 0.214$ the maximum value of the Chandrasekhar function, and all other quantities as conventionally
designated.

\begin{figure}
\includegraphics[width=0.48\textwidth]{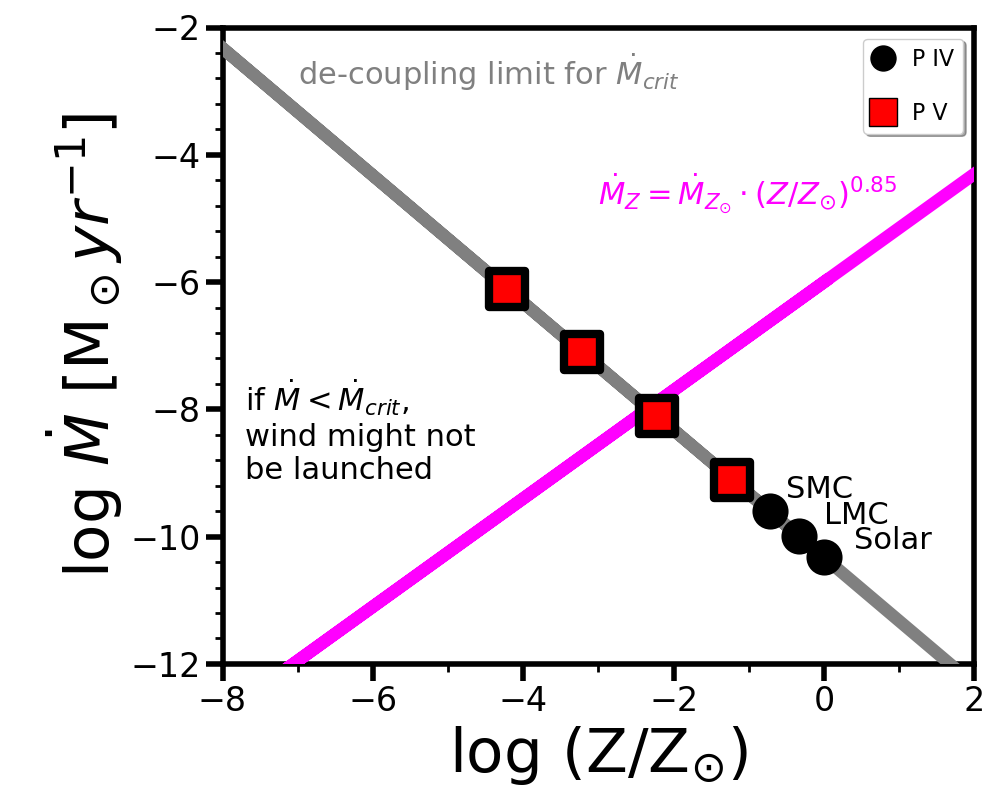}
\caption{Critical de-coupling mass-loss rates as a function of metallicity for parameters of a typical 60 M$_\odot$ star near the ZAMS. The symbols show the metallicities considered in our previous (P IV) and current (P V) work, respectively. The magenta regression line shows the expected mass-loss rate if the rates from solar metallicity decreased with an exponent of 0.85. The representative value for solar metallicity mass-loss rate is assumed to be $10^{-6}$~M$_\odot$~yr$^{-1}$. } 
\label{fig:mdotcrit}
\end{figure}
%

The velocities and velocity gradients can be expressed
by the canonical $\beta$ law (e.g., \citealt{pauldrach1986}), which for the prototypical value $\beta = 1$ results in
\begin{equation}
\Gamma_{\rm L} \approx v_{\infty}^2 (1-\frac{R_\star}{r})
\frac{R_\star}{r^2 g_{\rm grav}(r)},
\end{equation}
with $v_{\infty}$ the terminal speed of the (coupled) wind, and
$R_\star$ the stellar radius. Equating the relations for $\Gamma_{\rm B}$ and $\Gamma_{\rm L}$, and requiring a completely de-coupled wind (i.e., setting $v(r) =: v_{\rm crit} \approx 0.1 v_{\infty}$ as a prototypical value for the speed of a line-driven wind at its critical point, see \citealt{cak1975,pauldrach1986}), we finally
obtain
\begin{align}
\dot M_{\rm crit} \approx v_{\infty}^3 \Bigl(\frac{v_{\rm
crit}}{v_{\infty}}\Bigr)^2  & 
\frac{R_\star m_{\rm H}^2 k_{\rm B} T}{Y_i \ln \Lambda Z_i^2 e^4 G_{\rm max}} \nonumber \\
&
\Bigl(\frac{1+4Y_{\rm He}}{1+I_{\rm He}Y_{\rm He}}\Bigr)^2 (1+Y_{\rm He}).
\end{align}
Overall, the critical mass-loss rate (w.r.t. de-coupling) scales as
\begin{equation}
\dot M_{\rm crit} \propto \frac{v_{\infty}^3  R_\star T}{Y_i}, 
\end{equation}
where the temperature might be replaced by the effective one.
Obviously, the lower the metallicity, the larger $\dot M_{\rm crit}$.
Accounting for the fact that (observed) terminal velocities of OB stars scale with $T_{\rm eff}$ \citep{hawcroft2023}, the (effective) temperature has a large impact on the critical mass-loss rate. 

Our results for a typical 60 M$_\odot$ star near the ZAMS are shown in Figure \ref{fig:mdotcrit}. If the actual mass-loss rate is below the de-coupling limit, then the radiative force might be insufficient to drive a wind. Here, the actual mass-loss rate is assumed with a typical value for Galactic massive stars and then scaled down by a 0.85 exponent for lower metallicity \citep{vink2001,mokiem2007}. This results in our first-order estimate shown with the magenta line.
From solar metallicity to a metallicity of $Z/Z_\odot \sim 6 \cdot 10^{-3}$, the critical mass-loss rates for O-type dwarfs with effective temperatures of 60~kK are on the order of $10^{-11}$ and $10^{-8}~$M$_\odot$~yr$^{-1}$, respectively. Typical mass-loss rates calculated from wind prescriptions or obtained from observational diagnostics are above this limit at solar metallicity. However, at around $Z/Z_\odot \sim 6 \cdot 10^{-3}$ and at even lower metallicities, the calculated rates will tend to fall below the de-coupling limit. 
Thus, according to our estimates, reaching the de-coupling limit will take place in the metallicity range considered in our study. For this reason, we implement the calculation of the critical de-coupling limit in the \textsc{mesa} \texttt{run\_star\_extras} extension and systematically compare whether \textit{the mass-loss rate calculated from a prescription} is above or below this limit at each point during the evolution of a stellar model. In the following, we use the terminologies of \textit{calculated} and \textit{applied} mass-loss rates. By calculated rates, we refer to theoretical predictions from a wind prescription. By applied rates, we refer to the rates used to reduce the mass of the stellar model.

%
\begin{figure*}
\includegraphics[width=0.95\textwidth]{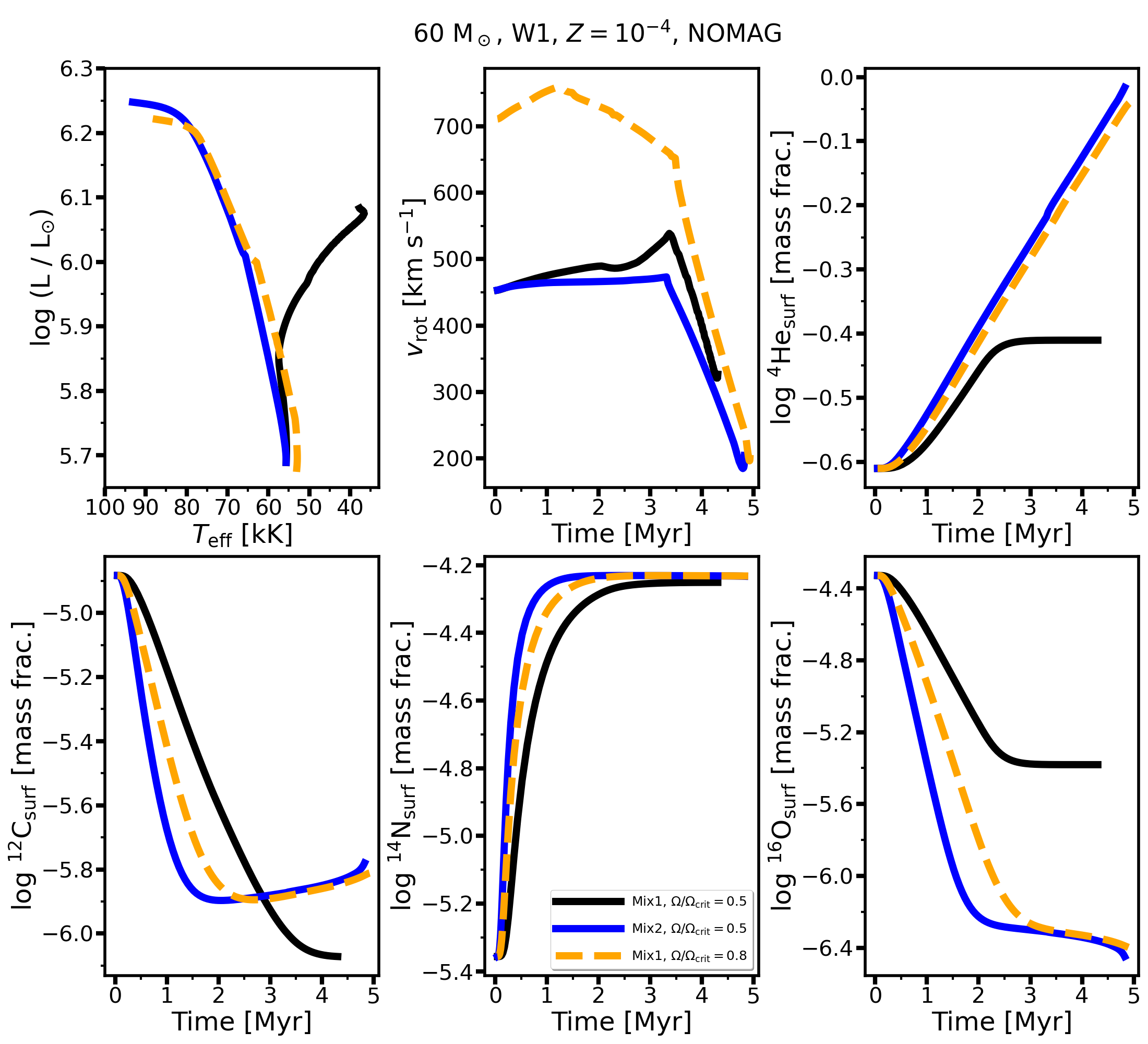}
\caption{HRD, surface equatorial rotational velocity, and surface He and CNO abundances (in mass fractions) as a function of time. The models are in the W1 wind scheme (mass-loss rates are applied even if the de-coupling limit is reached) at $Z=10^{-4}$ in the non-magnetic angular momentum transport scheme with $M_{\rm ini}=60$~M$_\odot$. The black and blue lines show our standard grid models in the Mix1 and Mix2 chemical mixing schemes, respectively. The orange dashed line shows the track, which demonstrates the effect of increasing the initial rotation from 50 percent to 80 percent of critical rotation in a Mix1 model.}
\label{fig:mixdemo}
\end{figure*}

%

%
%
%
\subsubsection{Mass loss schemes}\label{sec:winds_evol}

Given the above considerations, we distinguish between two modelling scenarios. In the first case (W1), even if the calculated mass-loss rates are below the critical value, we assume that some other form of wind-driving might take place, thus we apply non-zero mass loss. 
Even if radiative-line driving becomes completely inefficient, winds from low and zero metallicity massive stars might be launched by other mechanisms (which mechanisms might not depend on metallicity).
The prime candidates are pulsations-driven mass loss \citep[e.g.,][]{appenzeller1970,shiode2012}, Super-Eddington winds \citep[e.g.,][]{vanmarle2008,sanyal2015,owocki2017}, or other non-steady state eruptive mass-loss episodes. Though, these mechanism might require specific conditions (evolutionary stage, high luminosity, etc). 
For the sake of simplicity, we apply the originally calculated mass-loss rates in the W1 branch. 
In the second case (W2), we assume that the models experience no mass loss at all if their calculated rates would be below the de-coupling limit by applying $\dot{M}= 0$. Note however that this is time dependent and may not characterise the entire evolution of a model.

Rapid rotation could enhance (the existing) mass loss.
We include this effect and calculate rotational enhancement on the mass-loss rates following \cite{maeder2000}. 
In the case of models approaching critical rotation (which threshold we arbitrarily set to $\Omega/\Omega_{\rm crit} = 0.9$), we apply a numerical scheme built into \textsc{mesa} to artificially increase the mass-loss rates. Insofar, the application of such numerical techniques is seldom verified and we will refrain from discussing physical details near critical rotation, which is a largely unexplored and unresolved issue.

In consistency with \PaperIV, we adopt the \cite{vink2001} rates reduced by a factor of 2 for hot, hydrogen-rich stars ($T_{\rm eff} > 10$~kK, $X>0.4$), and the \cite{nugis2002} rates for hot, helium-rich stars ($T_{\rm eff} > 10$~kK, $X<0.4$). If chemical mixing is very efficient in the models (or the initial rotation is high), they can become helium rich and meet this condition already on the main sequence.  
We tested the new \cite{sander2020} rates (developed for $2.0 > Z/Z_{\odot} > 0.02$); however, the exponential decrease in the predicted mass-loss rates below SMC metallicity leads to prohibiting numerical issues. We use the metallicity scaling included in the \cite{vink2001} prescription, that is $\dot{M} \sim (Z/Z_\odot)^{0.85}$. In this case, we use the initial metallicity of the models. The \cite{nugis2002} prescription also includes a metallicity scaling ($\dot{M} \sim Z(t)^{0.5}$), which however uses the actual (time-dependent) metallicity of the stellar model (see discussion by \citealt{higgins2021}).

%
%
%
%
\section{Degeneracy between chemical mixing efficiency and initial rotational velocity}\label{sec:three}

In this section, we briefly demonstrate how varying either chemical mixing or the rotational velocity can lead to similar effects. 
Traditionally, uncertainties in chemical mixing have been less explored in evolutionary models (however, see, e.g., \citealt{meynet2013,nandal2024}), instead the initial rotation rates were used to study various scenarios. These findings indicated that when rotation is rapid enough, the star may have a structure that is almost perfectly mixed in composition \citep{maeder1987, maeder1998,yoon2006,cantiello2007}. Such quasi-chemically homogeneous evolution (QCHE) is expected to be more common at metallicities lower than that of the SMC \citep[e.g.,][]{langer2012,kubatova2019}. In \PaperIV, we fixed the initial rotation rates and tested two cases for chemical mixing, the less efficient Mix1 scheme and the more efficient Mix2 scheme. 

In Figure \ref{fig:mixdemo}, we demonstrate that the Mix1 case tends toward the Mix2 case if the initial angular velocity is increased. The models considered here are without magnetic field effects and in the W1 wind scheme (mass-loss rates are applied even if the de-coupling limit is reached). Keeping the initial rotation the same ($\Omega/\Omega_{\rm crit}=0.5$), the Mix1 model evolves redward, whereas the Mix2 model evolves blueward in the HRD (top left panel). The latter evolutionary trajectory is indeed associated with QCHE. A model within the Mix1 chemical mixing scheme but with higher initial rotation ($\Omega/\Omega_{\rm crit}=0.8$) has a Zero Age Main Sequence (ZAMS) that is slightly shifted to a lower effective temperature. This is because the centrifugal force adds support against gravity and allows for a larger stellar radius. This track (orange dashed line) follows an evolution that is almost identical to the Mix2 model which is initially less rapidly rotating but has a higher efficiency of  transporting chemical species. Besides the obvious quantitative difference in rotational velocities, the evolution of the surface abundances of He and CNO elements in the Mix1 $\Omega/\Omega_{\rm crit}=0.8$ model also tends toward the Mix2 $\Omega/\Omega_{\rm crit}=0.5$ case.
For this reason, in the following it may be considered that the Mix2 scheme with initially $\Omega/\Omega_{\rm crit}=0.5$ gives an indication for results where, instead of the mixing efficiency, the rotational velocity was increased in a Mix1 model. With our approach, we wish to emphasise that chemical mixing is largely unresolved and as such remains a key physical aspect to calibrate in evolutionary models.

%
%
%
%
\section{Results within the W1 wind scheme}\label{sec:four}

In this section, we present our findings when utilising the assumption of the W1 wind scheme. This results in gradually weaker stellar winds as a function of decreasing metallicity. In the next section, we investigate the impact of this choice by discussing the W2 wind scheme, where the applied mass-loss rates can become zero.

%
%
%
%

\begin{figure*}
\includegraphics[width=0.45\textwidth]{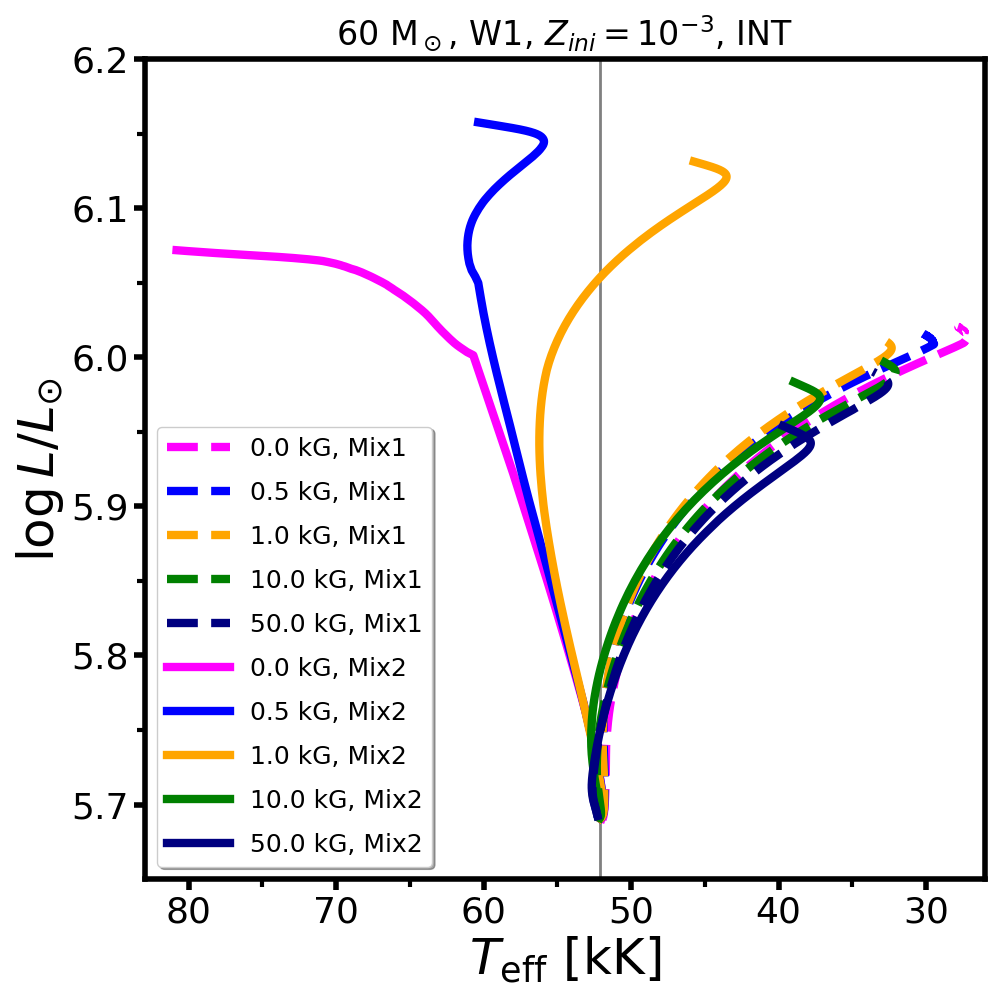}\includegraphics[width=0.45\textwidth]{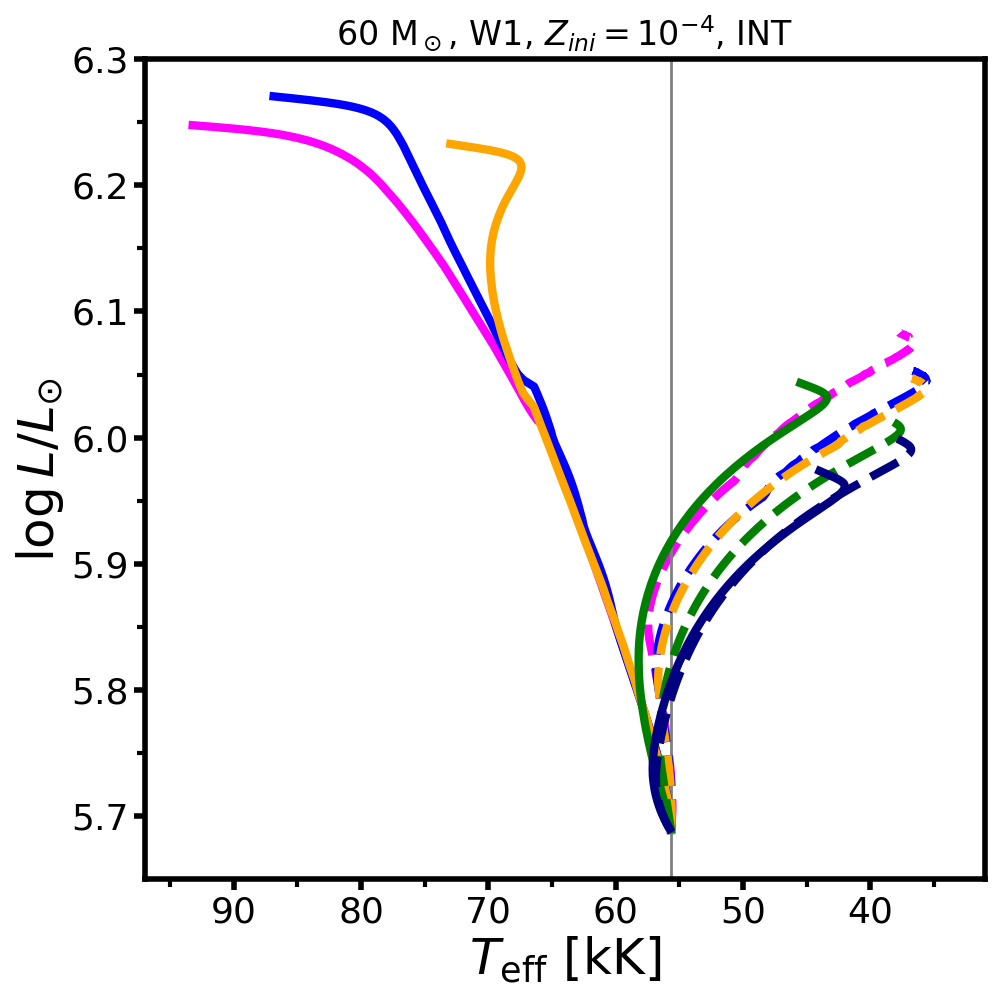}
\includegraphics[width=0.45\textwidth]{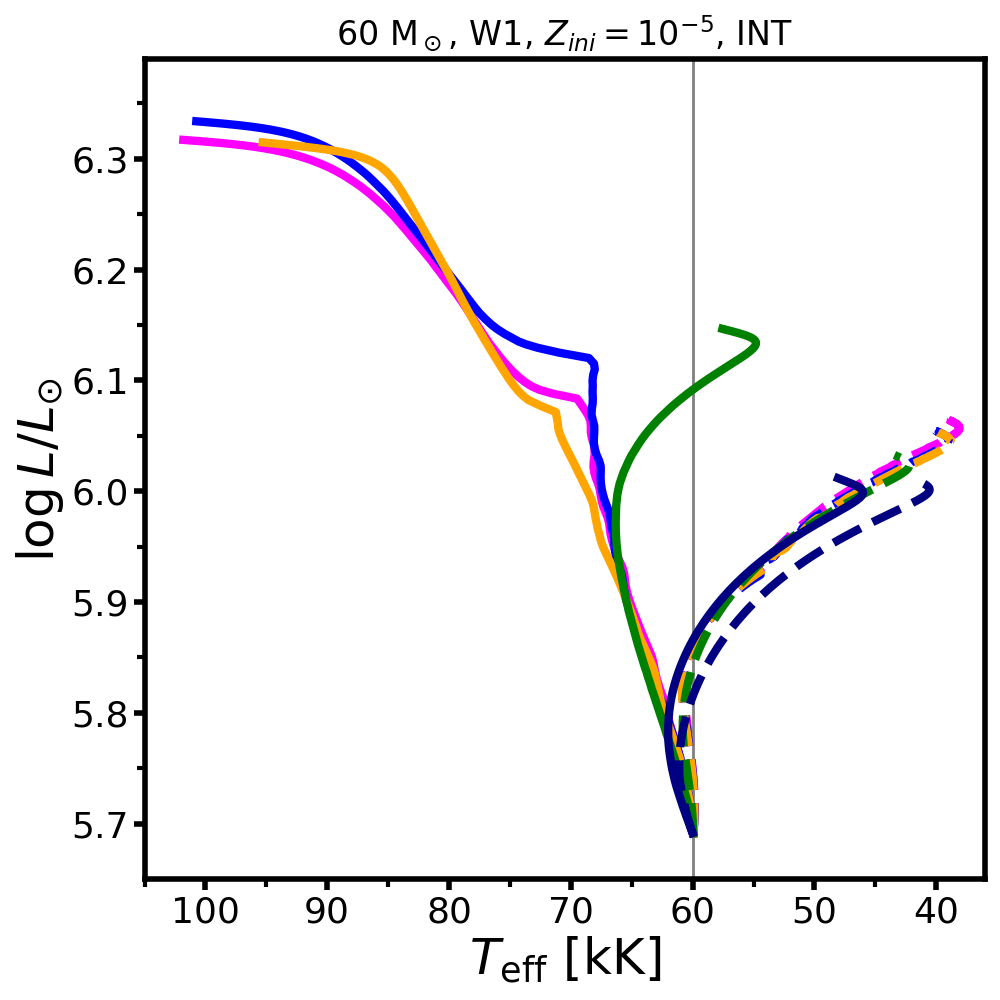}\includegraphics[width=0.45\textwidth]{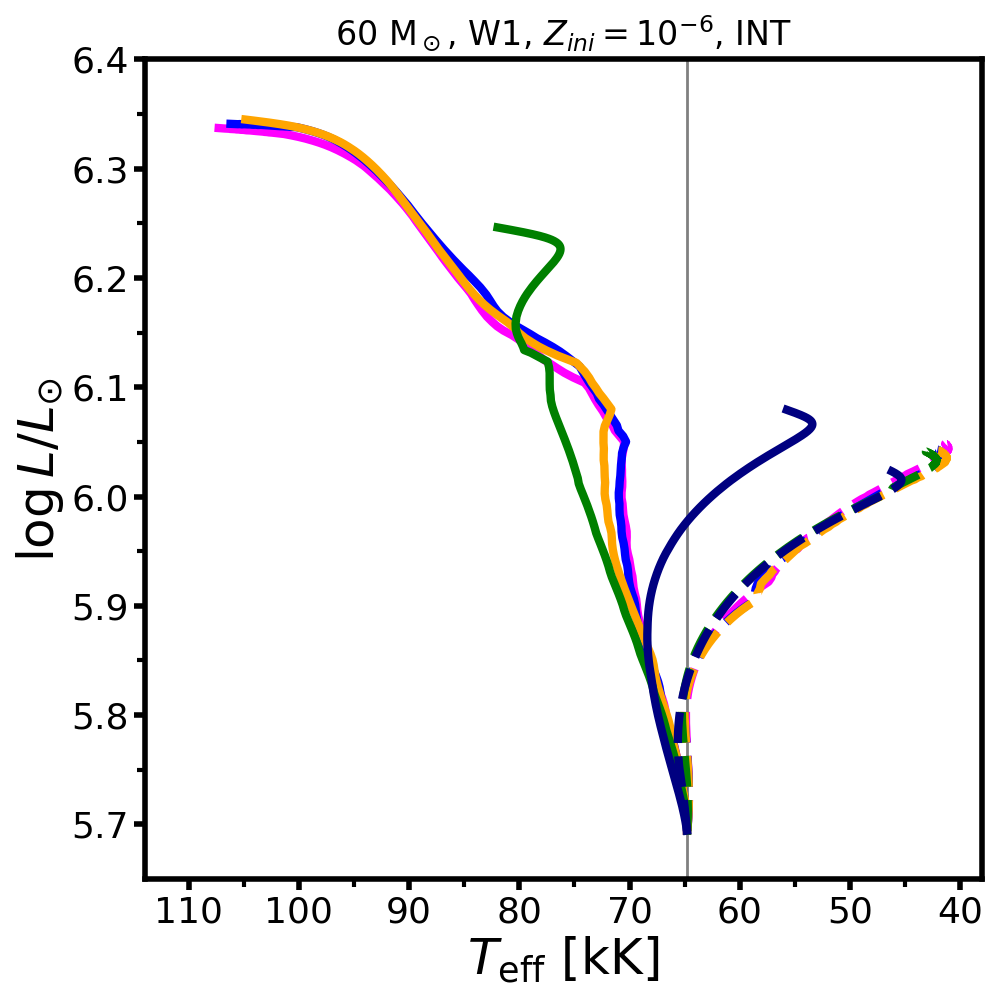}
\caption{HRD of initially 60~M$_\odot$ main-sequence stellar evolution models in the W1 wind scheme at various metallicities, as indicated in the panel titles. The magnetic models are in the INT scheme. The initial magnetic field strength is colour-coded, the chemical mixing is line-coded. Note the different axis scales. The grey vertical line is placed at the ZAMS effective temperature.}
\label{fig:hrd1}
\end{figure*}

\begin{figure*}
\includegraphics[width=0.45\textwidth]{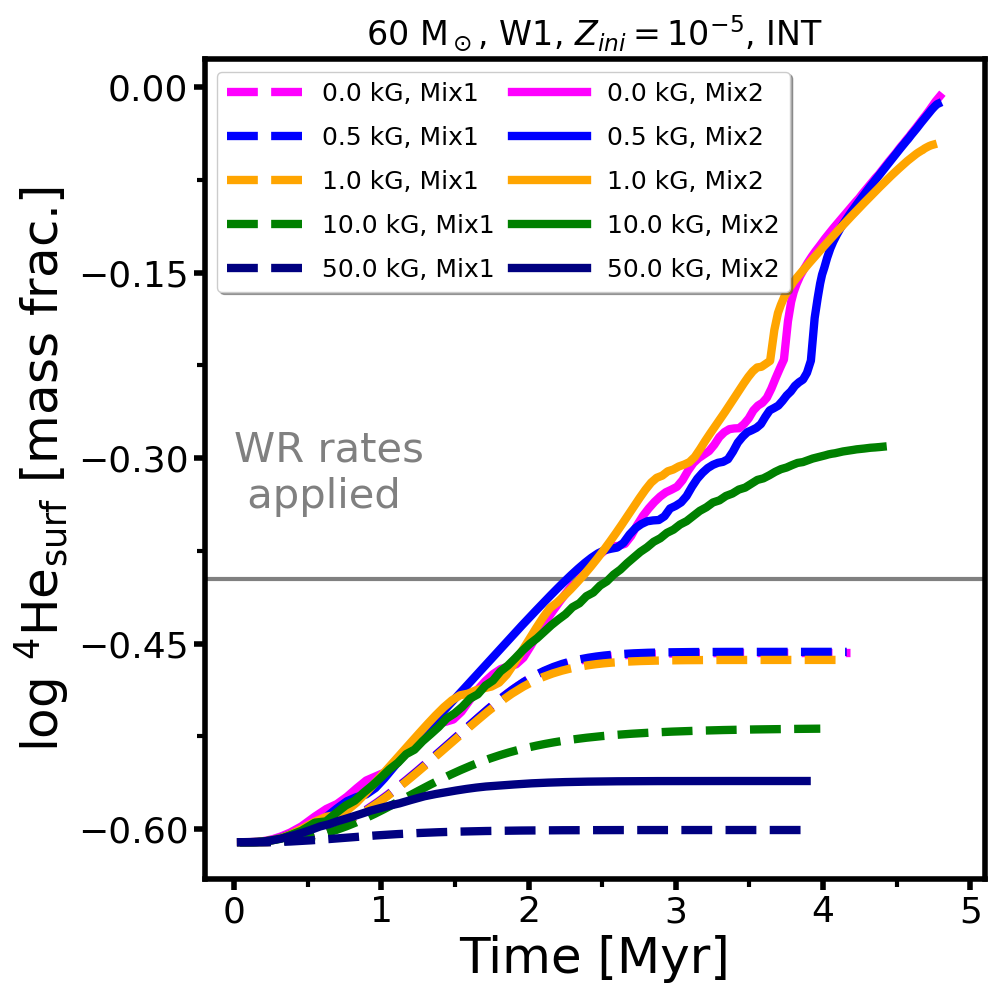}\includegraphics[width=0.45\textwidth]{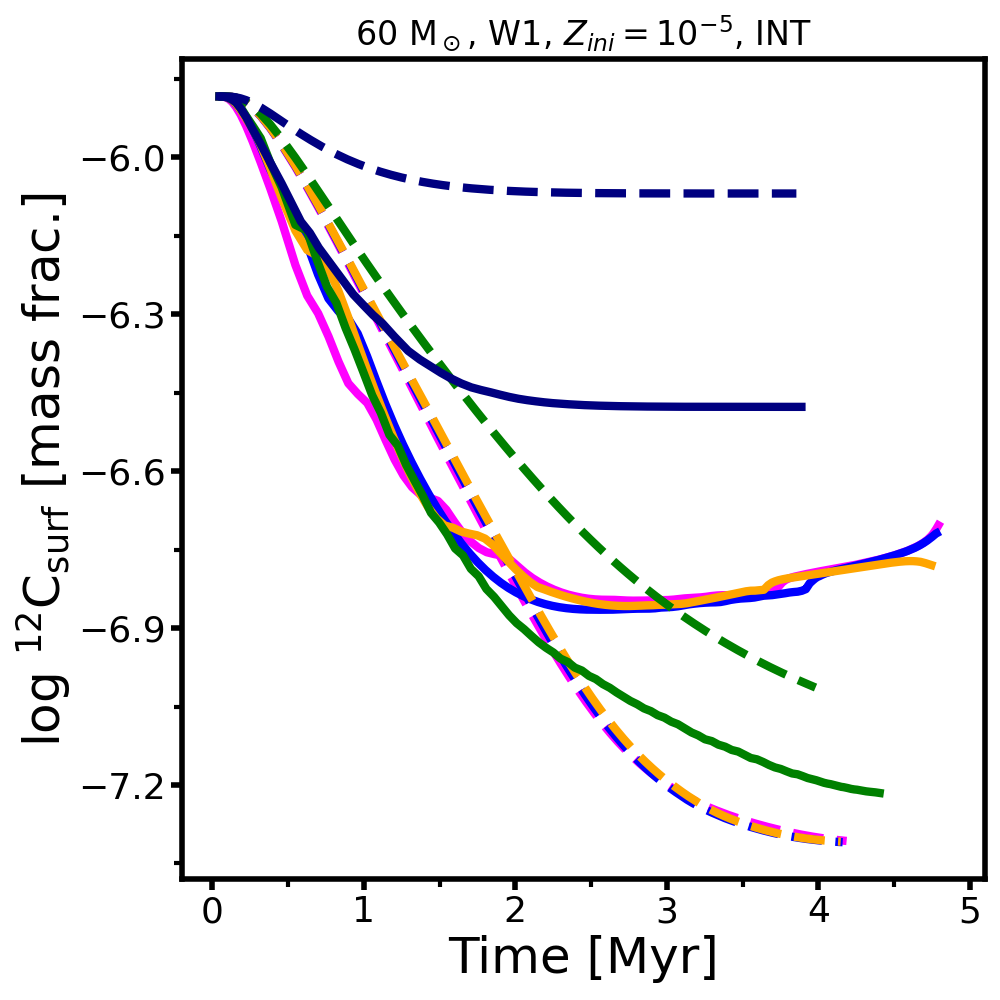}
\includegraphics[width=0.45\textwidth]{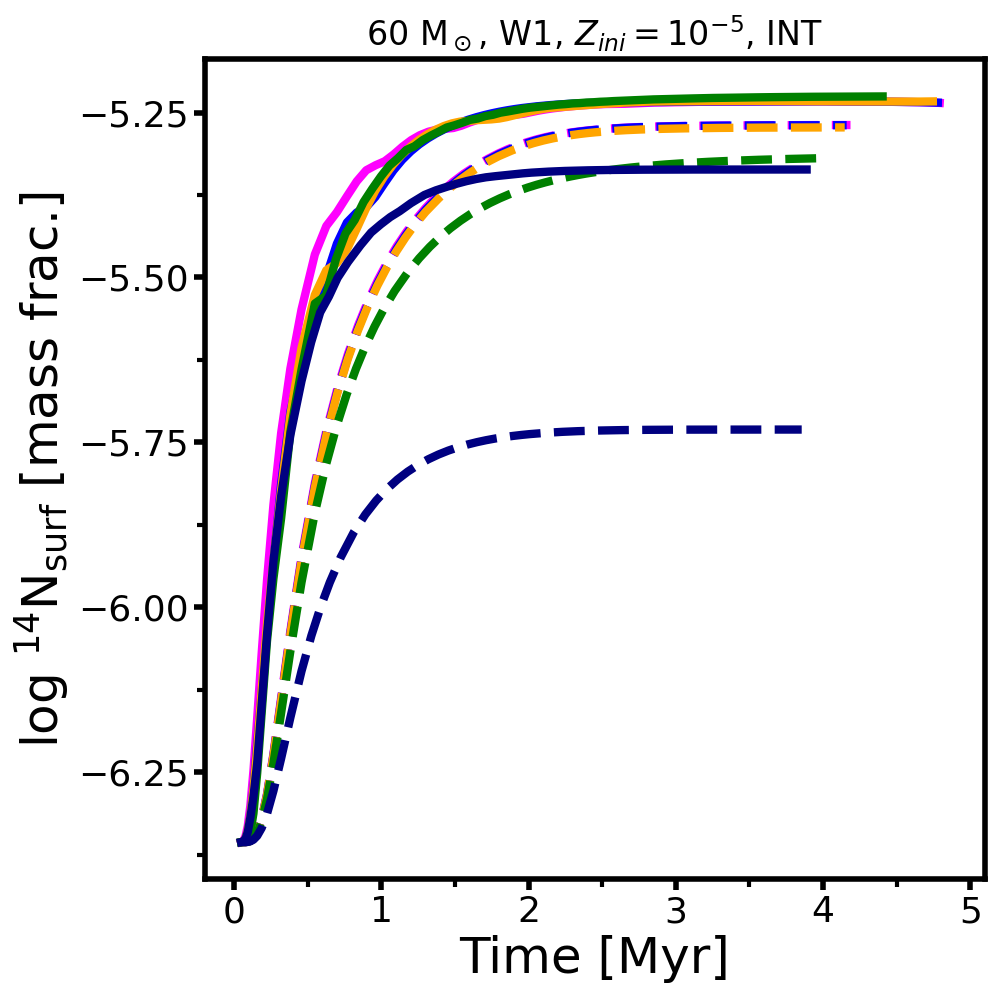}\includegraphics[width=0.45\textwidth]{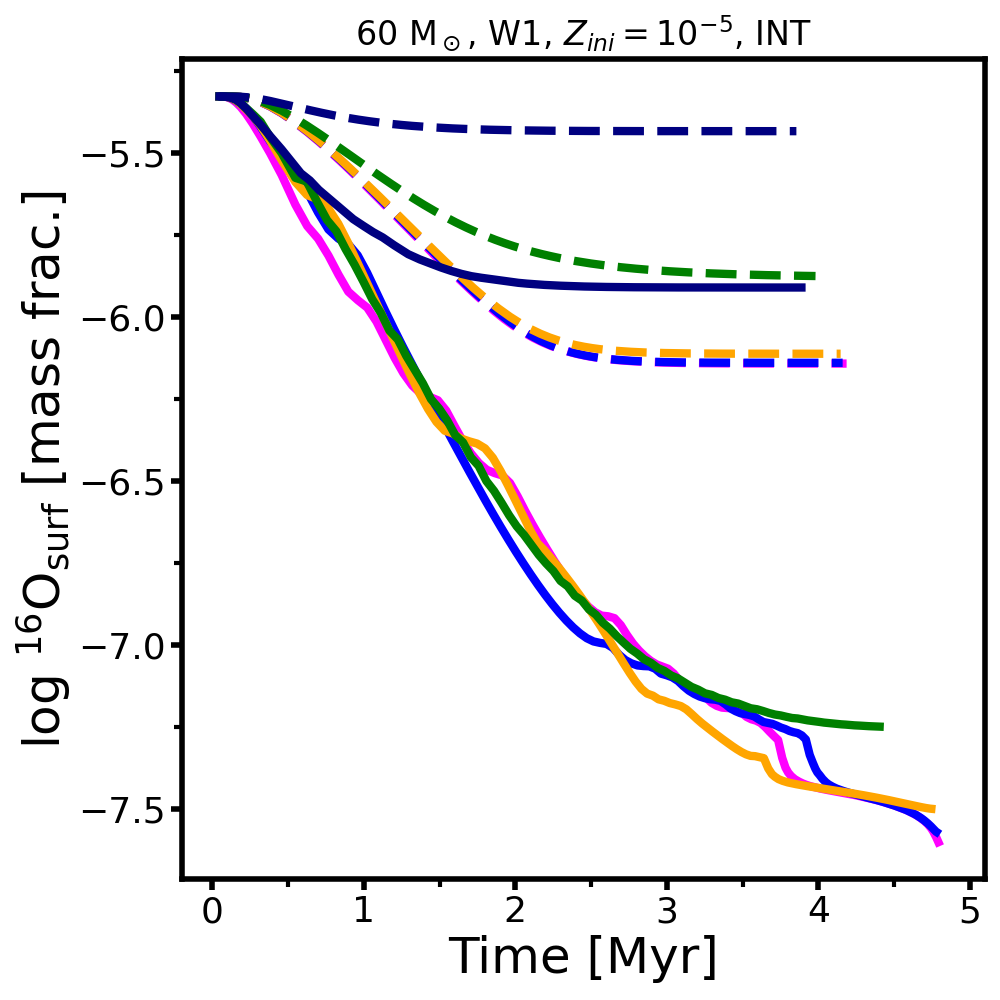}
\caption{Time evolution of the surface mass fraction of helium, carbon, nitrogen, and oxygen for models with $M_{\rm ini}=$~60~M$_\odot$ at a metallicity of $Z = 10^{-5}$ within the W1 wind scheme. The magnetic models are in the INT scheme. The initial magnetic field strength is colour-coded, the chemical mixing is line-coded. Note the different axis scales. The grey horizontal line indicates the surface helium abundance above which Wolf-Rayet type mass-loss rates are applied. }
\label{fig:hecno}
\end{figure*}

\begin{figure*}
\includegraphics[width=0.45\textwidth]{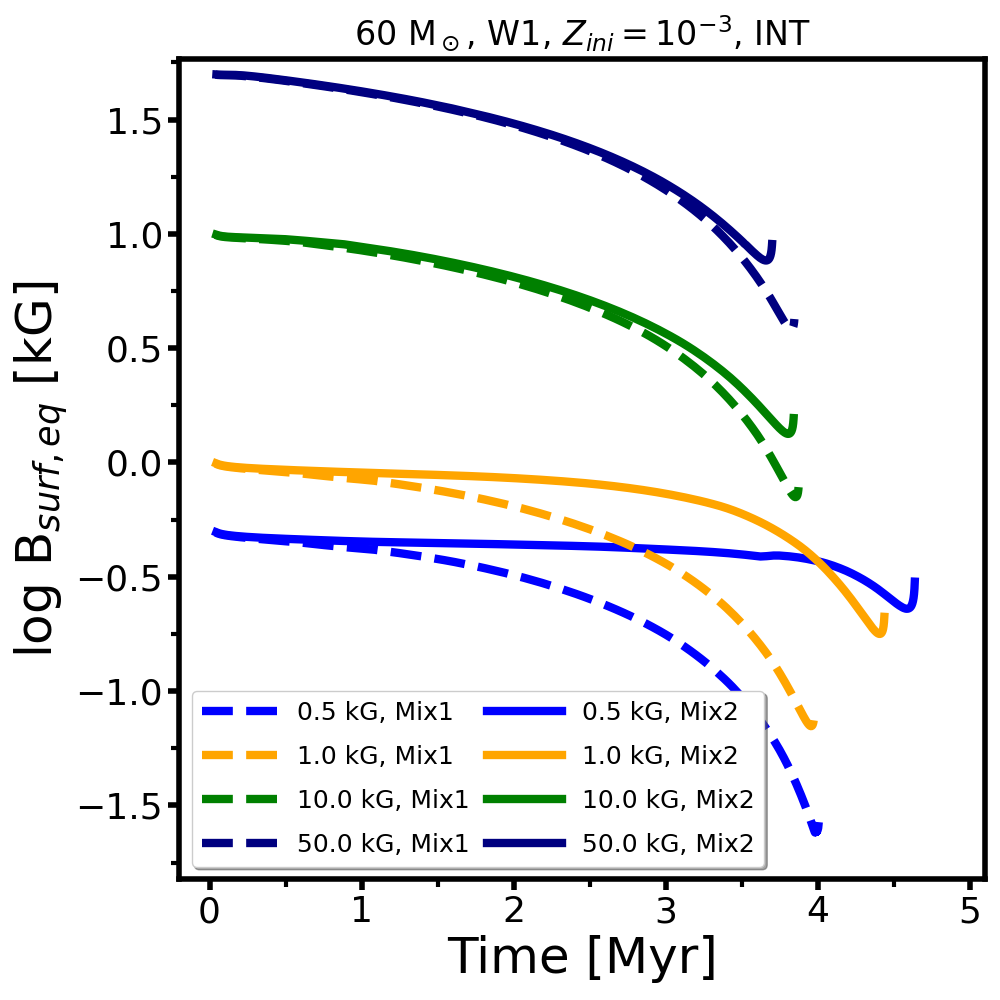}\includegraphics[width=0.45\textwidth]{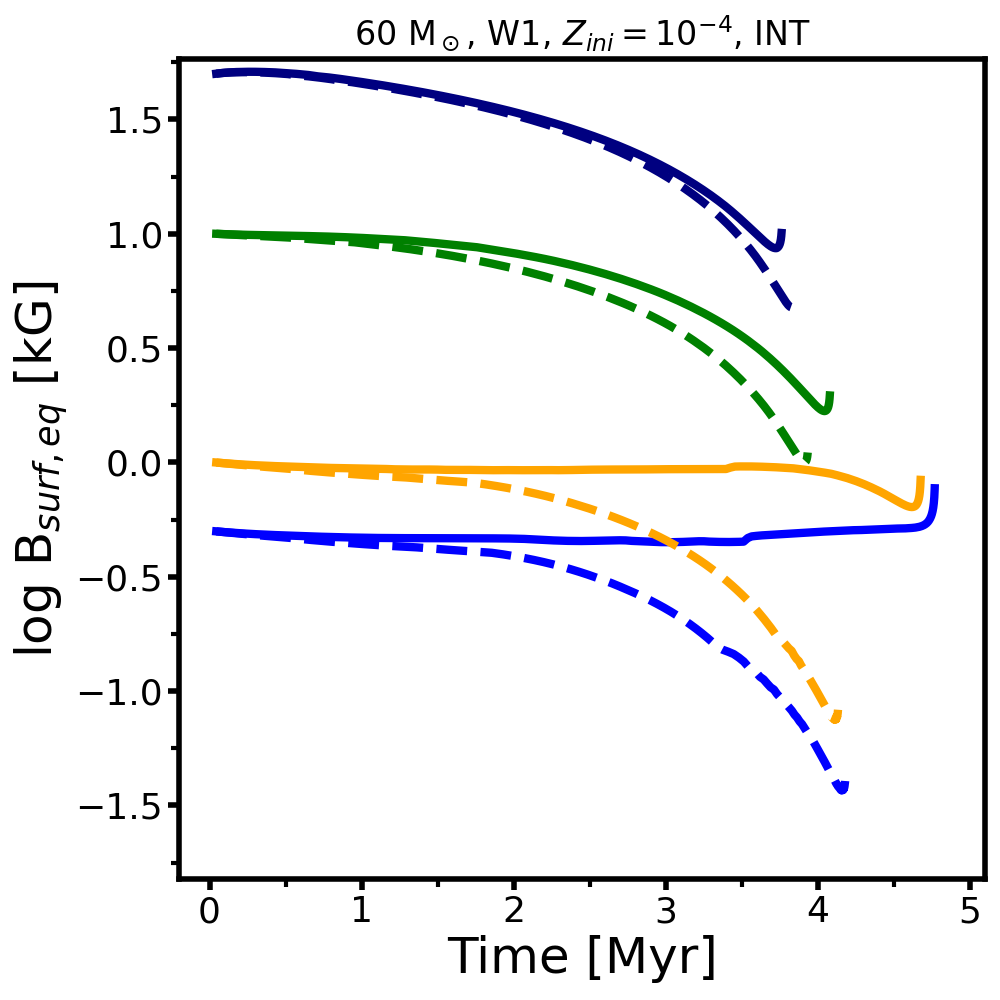}
\includegraphics[width=0.45\textwidth]{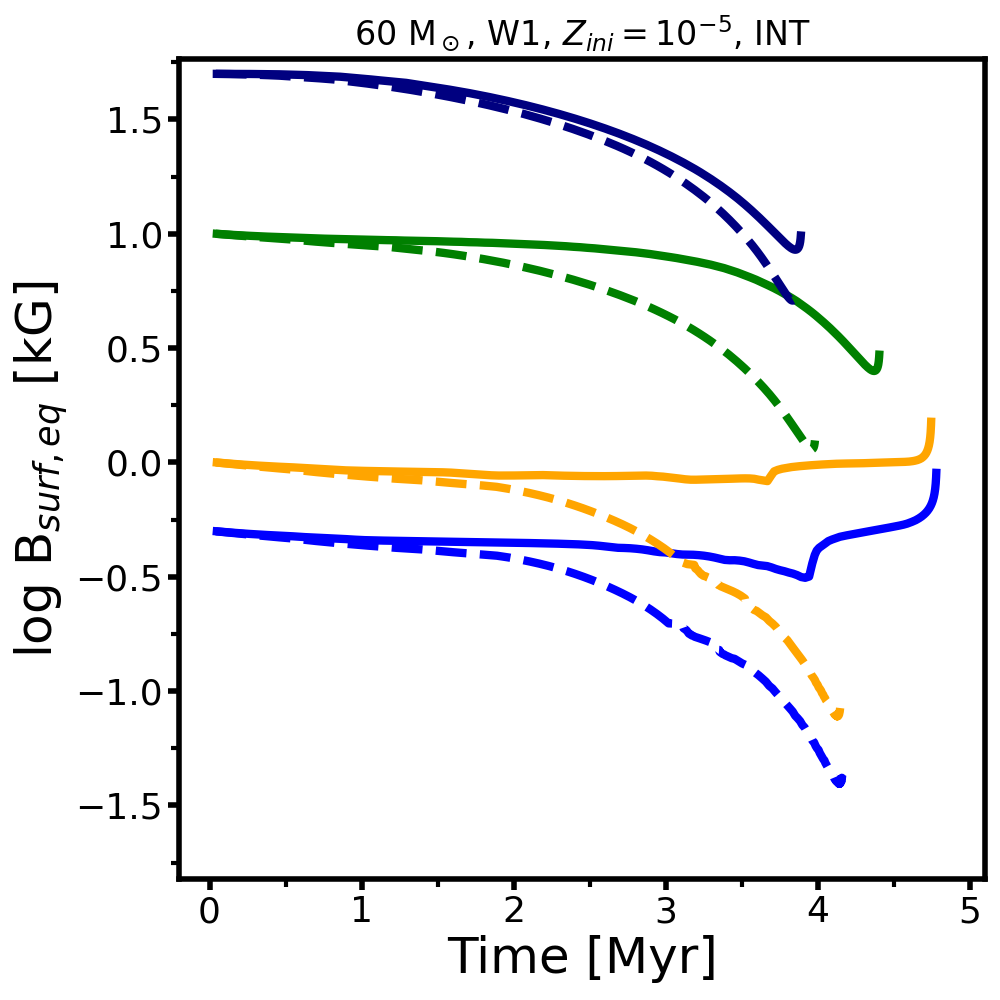}\includegraphics[width=0.45\textwidth]{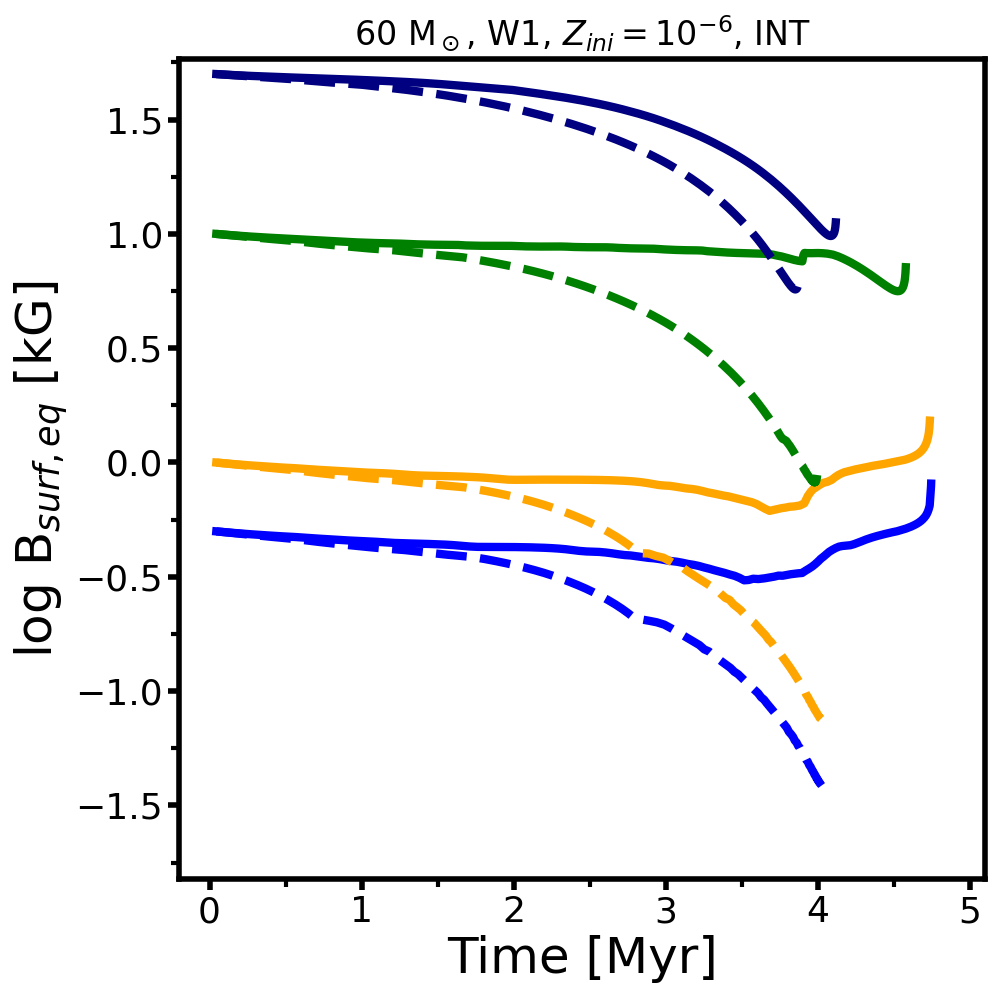}
\caption{Surface equatorial magnetic field evolution in our models with initial mass of 60~M$_\odot$ in the W1 wind scheme and in the INT magnetic braking and angular momentum transport scheme. The panels show different initial metallicities.}
\label{fig:bevol}
\end{figure*}

\begin{figure*}
\includegraphics[width=0.45\textwidth]{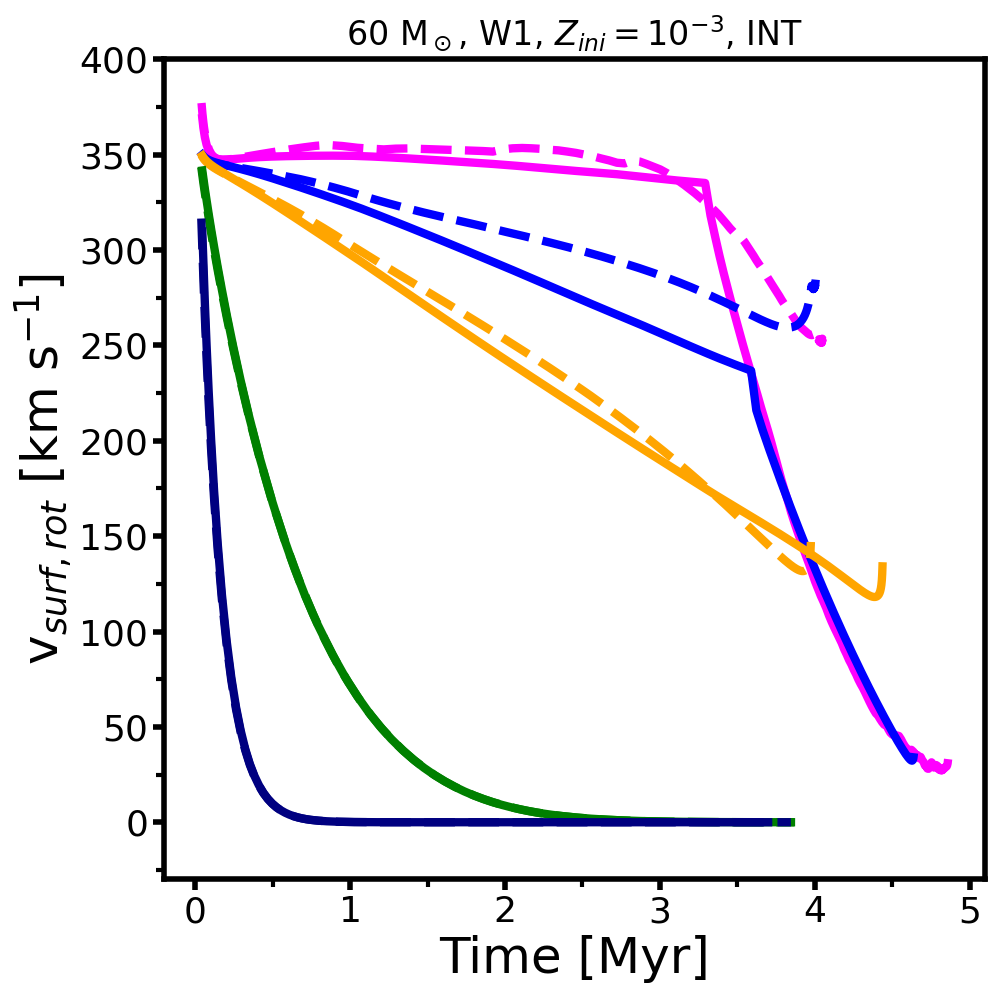}\includegraphics[width=0.45\textwidth]{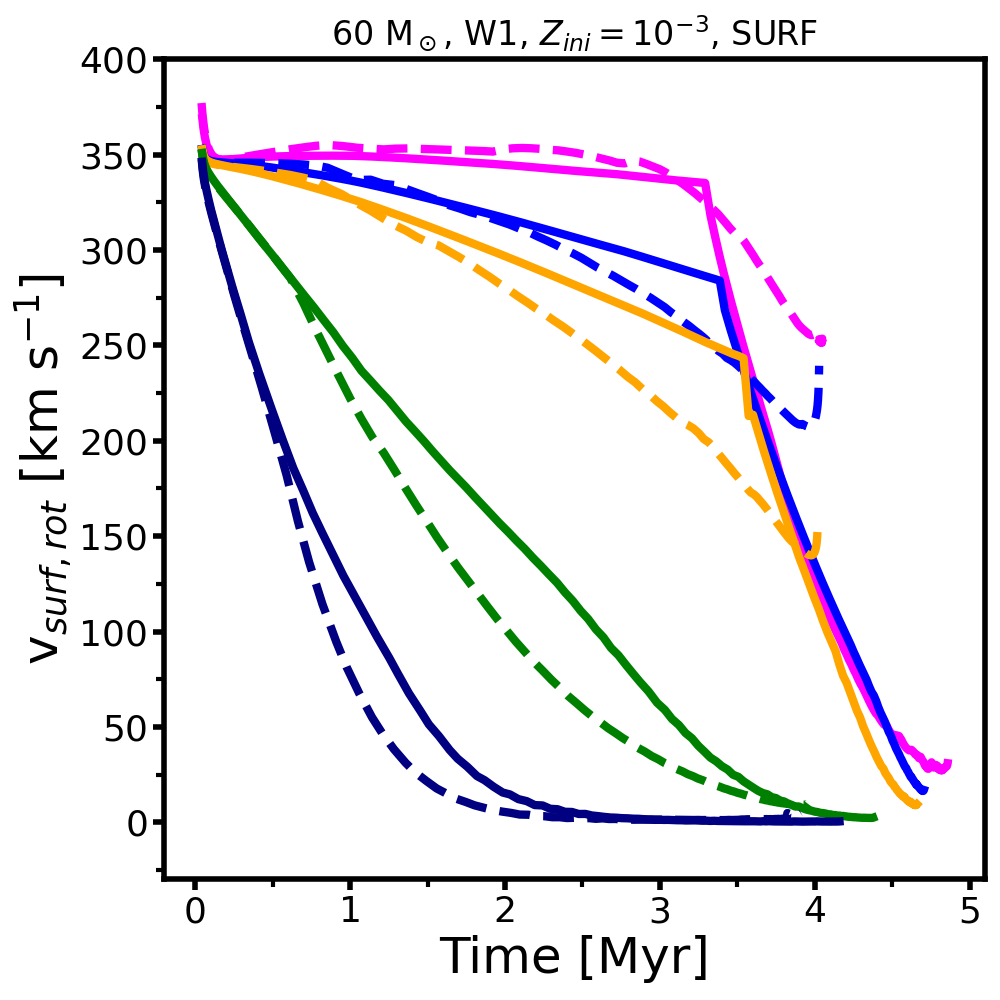}
\includegraphics[width=0.45\textwidth]{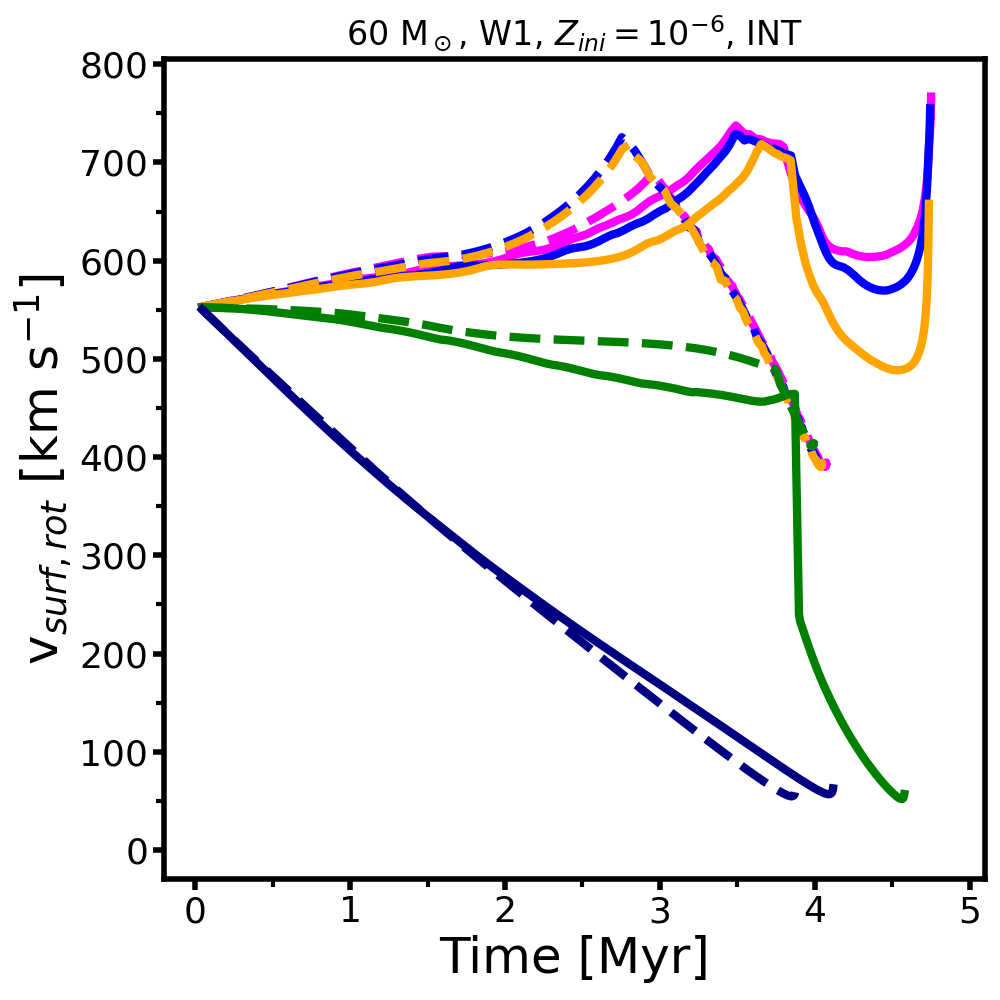}\includegraphics[width=0.45\textwidth]{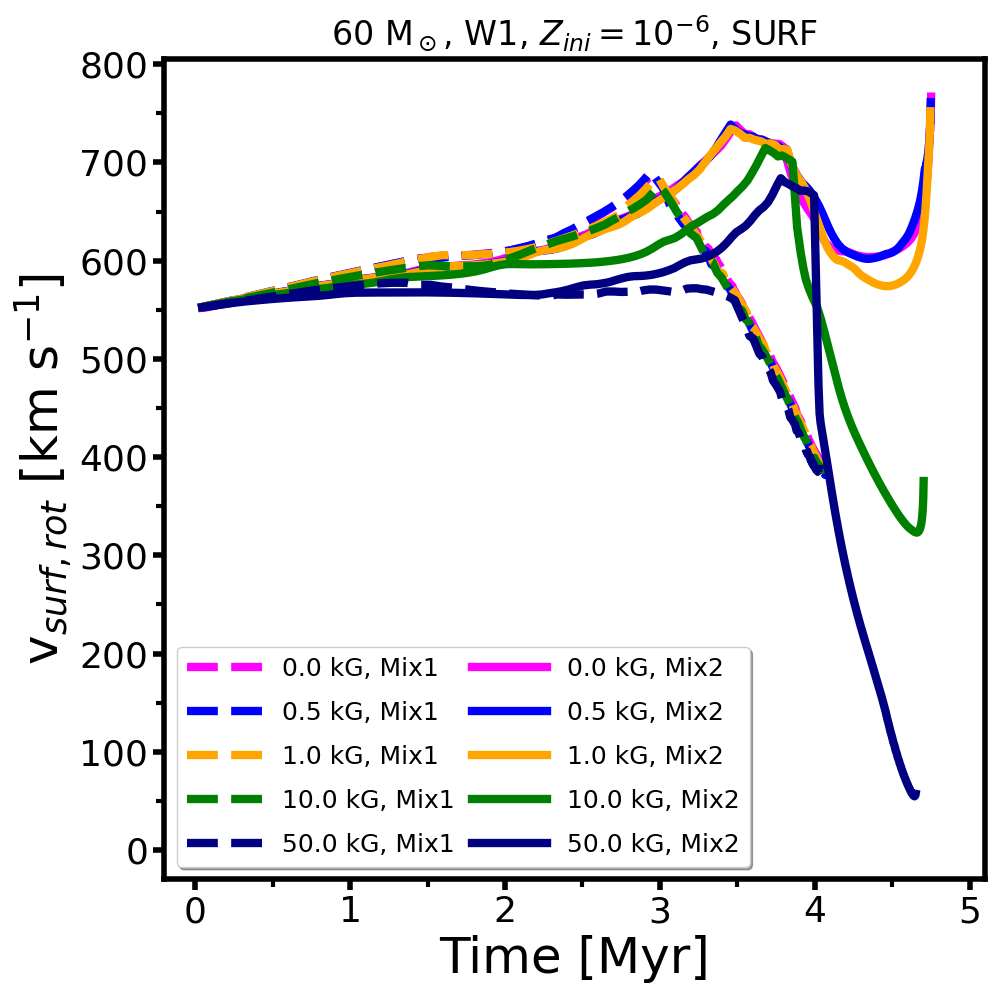}
\caption{Evolution of surface equatorial rotational velocity for models with initial mass of 60~M$_\odot$ in the W1 wind scheme. Different metallicities (top: $Z=10^{-3}$, lower: $Z=10^{-6}$) and different magnetic braking schemes (left: INT, right: SURF) are shown in the panels. Note the different vertical axis scales.}
\label{fig:revol}
\end{figure*}

%
%

\subsection{Evolution of main stellar parameters and surface abundances}\label{sec:r1}

In contrast to high-metallicity environments, stellar mass loss is modest, whereas rotation and chemical mixing lead to much more pronounced effects at low metallicity \citep[e.g.,][]{meynet2002b}. We proceed with discussing our models with the assumption of a fixed initial rotation rate and two boundary cases (Mix1, Mix2) for chemical mixing. 

%
%
%
Figure~\ref{fig:hrd1} shows the HRD for models with an initial mass of 60~M$_\odot$ at the initial metallicities considered in our study ($Z=10^{-3} $--$10^{-6}$, from top left to lower right panels), using the W1 wind scheme. The initially 20~M$_\odot$ models follow similar qualitative trends. Here, we only focus on the magnetic models in the INT scheme (solid-body rotators). The SURF models shows similar evolutionary trajectories.
Following \cite{szecsi2015}, we may distinguish between three evolutionary channels: i) a "classical", redward trajectory in the HRD, ii) a transitional, intermediate situation, in which some part of the evolution is blueward but the tracks return to a redward trajectory, and iii) a completely blueward evolution. The blueward evolution may be associated with QCHE, that is, the star is almost completely mixed in composition. 
In the following, we simply use the initial, ZAMS effective temperature of the models as a reference and compare whether the tracks follow an evolution towards lower (redward) or higher (blueward) effective temperatures on the main sequence. 

The initial metallicity plays a large role in the development of the QCHE channel. Lowering the metallicity leads to the well-known effect of shifting the ZAMS effective temperatures to higher values, which is due to the smaller stellar radius. For this reason, and consequently due to the higher rotational velocities, lower metallicity also favours a more extended blueward evolution towards high effective temperatures and luminosities.

Within the Mix1 chemical mixing scheme, the models tend to evolve along a classical path. However, when lowering the metallicity, the evolutionary trajectory can remain vertical in the HRD for up to half of the main sequence lifetime ($\sim 2$~Myr), or even with a slightly increasing $T_{\rm eff}$.
In the Mix1 case, the assumption about initial magnetic field strength or angular momentum transport scheme lead to relatively small differences in the evolutionary tracks, although the exact width of the main sequence, the lifetimes, and other physical quantities vary in the models (c.f. \PaperII, \PaperIV). 

Models with the more efficient Mix2 chemical mixing scheme tend to evolve blueward in the HRD. The stronger the magnetic field, the slower the rotation becomes, resulting in less efficient mixing. Therefore, magnetic fields work to prevent the QCHE channel (c.f. \PaperIV). Compared to a non-magnetic track, the magnetic tracks start to deviate from the QCHE channel and turn towards a classical evolution. However, while in higher metallicity environments weaker magnetic fields are sufficient to brake the rotation and suppress chemical mixing to prevent this evolutionary channel, in low-metallicity, a stronger magnetic field is required. In particular, at $Z=10^{-3}$ (top left) only the 10 and 50 kG models return to lower effective temperatures than the ZAMS value and follow a classical evolution. The model with 1 kG field strength follows a transitional evolution, initially evolving bluewards and returning to a redward trajectory towards the TAMS. At $Z=10^{-6}$, only the 50 kG model in the INT angular momentum transport and magnetic braking scheme departs from the QCHE channel. This highlights that at lower metallicity environments even strong magnetic fields are insufficient to prevent the efficient chemical mixing.

%
%
The surface abundances of He and CNO elements are shown in Figure \ref{fig:hecno} for the models with initial mass of 60~M$_\odot$ at \nobreak{$Z=10^{-5}$}. QCHE leads to helium-rich stars on the main sequence in the Mix2 case, except the 50~kG model which brakes its rotation and returns to a classical redward evolution (top left panel, c.f. also Figure~\ref{fig:hrd1} and discussion above). In both Mix1 and Mix2 models, the stronger the magnetic field, the less helium enrichment is predicted.
During the main sequence, as set by the CNO cycle, first the surface carbon and then the oxygen abundances decrease, while the surface nitrogen abundance increases rapidly. These changes are more prominent in the initially 60~M$_\odot$ models, but the initially 20 M$_\odot$ models also show similar trends.
The magnetic models systematically deviate from the non-magnetic models. Stronger magnetic fields lead to less surface enrichment of nitrogen, whereas the depletion of surface carbon and oxygen is more moderate. This is because magnetic braking leads to slower spin, which decreases the efficiency of chemical mixing inside the star.
We find that typical, order of kG magnetic fields tend to moderately impact the evolution of surface abundances, within about 0.2 dex. Very strong magnetic fields (50~kG) are needed to significantly inhibit chemical mixing.

%
%
\subsection{Magnetic field evolution}\label{sec:r2}

Utilising the assumption of magnetic flux conservation, the evolution of surface equatorial magnetic field strength only depends on the stellar radius (see \PaperII and \PaperIII and references therein). 
%
%
%
In Figure~\ref{fig:bevol}, we compare the time evolution of the surface equatorial magnetic field strength of models with an initial mass of 60~M$_\odot$ at the metallicities considered in our study.
At $Z = 10^{-3}$, all models within this specific setup follow a decline in their surface equatorial magnetic field strength. 
At $Z = 10^{-4}$, the magnetic field evolution of the models with initial field strength of 0.5 and 1~kG in the Mix2-SURF and Mix2-INT schemes remains practically constant. 
At $Z = 10^{-5}$ and $Z = 10^{-6}$, also some models with initially 10 kG and 50~kG field strength display a nearly constant magnetic field evolution in the Mix2 chemical mixing scheme.

This might seem surprising; however, the reason for these trends is the competition between chemical mixing and surface magnetism. Strong chemical mixing favours blueward evolution in the HRD with nearly no change in the stellar radius, whereas a strong magnetic field favours a classical evolution with increasing radius due to slow rotation and less efficient mixing. 
Consequently, our results have interesting implications. i) At low metallicity, if chemical mixing is efficient (or initial rotation is rapid, c.f. Section~\ref{sec:three}), magnetic fields could maintain the same strength over the course of their entire main sequence evolution. This is unlike in observed Galactic samples, where a decline in magnetic field strength over the main sequence has been identified \citep{landstreet2007,landsreet2008,fossati2015,shultz2019b}. Approximately an order of magnitude decline in the surface equatorial magnetic field strength is also expected from our low-metallicity model predictions in the Mix1 scheme with less efficient chemical mixing.
ii) High-mass, main-sequence stars are likely to become Wolf-Rayet type stars during the post-main sequence. In fact, several of our models are also Wolf-Rayet type-like already on the main sequence in terms of their surface helium enrichment as shown in Figure~\ref{fig:hecno}. Therefore, the initial equatorial magnetic field strengths of 0.5, 1, and 10 kG (implying a 1, 2, and 20 kG dipole field strength), which are predicted to remain quasi constant during the main sequence evolution in the Mix2 models at $Z=10^{-5}$ and $Z=10^{-6}$ (lower row of Figure~\ref{fig:bevol}), might be assumed to be the magnetic field strength of a Wolf-Rayet type star inherited from its progenitor.

%
%

\begin{figure*}
\includegraphics[width=0.95\textwidth]{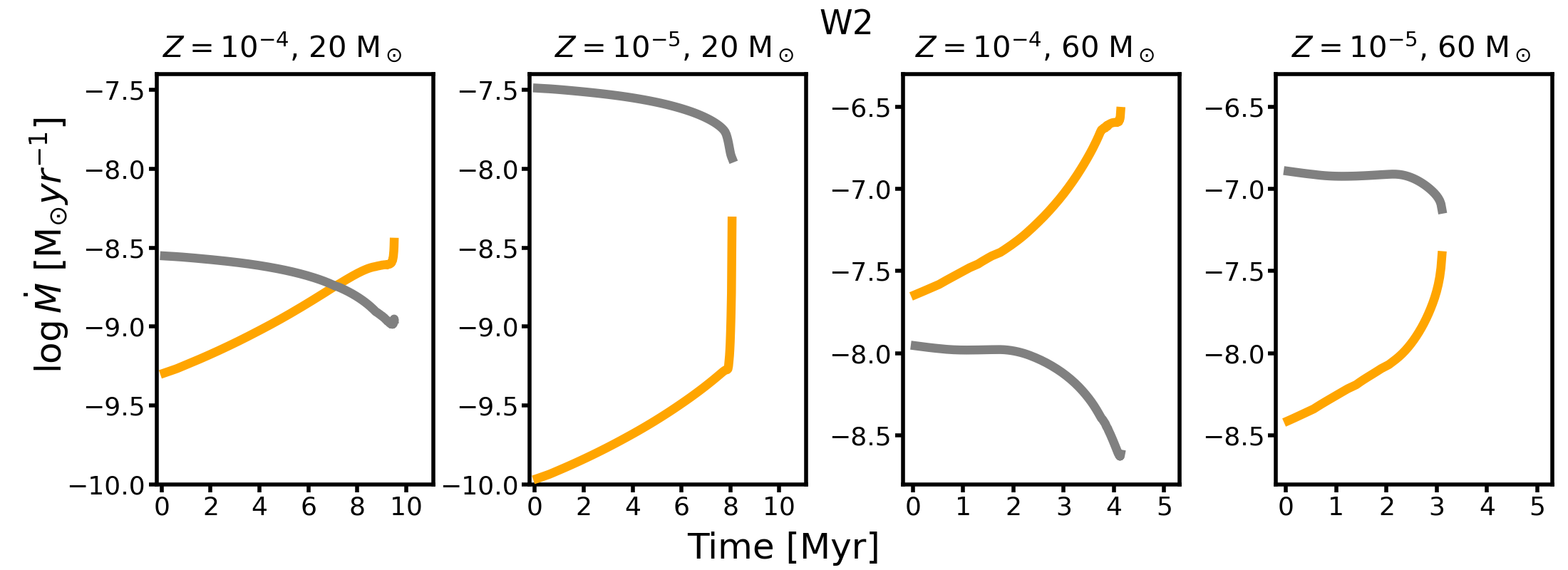}
\caption{Mass-loss rates as a function of age. The panels show models with $M_{\rm ini}=20$~M$_\odot$ (left two) and $M_{\rm ini}=60$~M$_\odot$ (right two) at metallicities of $Z=10^{-4}$ and $Z=10^{-5}$. 
In each case, the selected model is in the INT-Mix1 scheme with initial field strength of 1 kG. 
Note the different scales on the axes. The mass-loss rates calculated from a wind prescription are shown with the orange line for a given model. The corresponding de-coupling limit is shown with the grey line. If the calculated rates are below the de-coupling limit, the wind might not be launched. In the W1 scheme nonetheless we assume non-zero applied rates in such a case, but in the W2 scheme, we apply zero mass loss.}
\label{fig:mdotdec}
\end{figure*}

%
%
\subsection{Evolution of surface rotation}\label{sec:r3}

Stellar rotation plays a major role in the evolution of low-metallicity massive stars. In diffusive angular momentum schemes as in \textsc{mesa}, the speed of rotation is the main driver of chemical mixing (primarily via meridional currents) in radiative envelopes\footnote{In advective schemes, not only the speed of rotation but rather the degree of radial differential rotation is important in driving chemical mixing. While shear mixing can develop in diffusive schemes, it is not sustained because diffusion aims to erase radial differential rotation. In contrast, shear can not only be maintained but even enhanced in advective schemes. Therefore, even for a slow surface rotational velocity, chemical mixing can be efficient in the stellar interiors if there is strong radial differential rotation.}. Therefore it is important to quantify how magnetic fields impact rotation.

For this reason, we scrutinise the evolution of the surface equatorial rotational velocity as predicted in our models (Figure \ref{fig:revol}). If the star has a magnetosphere, most of the angular momentum loss is driven by the magnetic field \citep{weber1967,ud2009}.
At $Z = 10^{-3}$, the non-magnetic Mix1 models maintain the highest rotational velocity on the main sequence. The non-magnetic Mix2 model follows a similar rotational evolution until about 3.5~Myr. At that point, the increasing surface helium abundance leads to Wolf-Rayet type mass-loss rates. These rates are much higher than the optically-thin wind prescriptions, therefore an enhanced rotational braking proceeds. 
In general, the Mix1 models can inflate their envelopes more, which also contributes to lowering their surface rotation due to angular momentum conservation. The Mix2 models tend to remain more compact, which favours maintaining higher surface rotational velocities.
In some cases, the models can experience a spin up. This is because the timescale to transport and replenish angular momentum in the surface layers is shorter than the timescale to remove it.
The magnetic models follow the trend evidenced in high-metallicity, namely, the stronger the magnetic field, the faster the surface rotation brakes (e.g. \PaperII and \PaperIV). 
At $Z = 10^{-5}$ and $Z = 10^{-6}$, the rotational evolution becomes more complex due to the tendency towards QCHE. In these cases too, the sudden drop (by hundreds of km\,s$^{-1}$) in surface rotational velocity close to the end of the main sequence is attributed to switching to Wolf-Rayet type mass-loss rates. Overall, we can see that when lowering the metallicity, progressively stronger magnetic fields are needed to brake rotation. 
The difference between the INT and SURF magnetic braking schemes is the most prominent for strong magnetic fields. For 10 and 50 kG fields, the INT models display lower surface rotational velocities over their evolution than the SURF models. This is because efficient chemical mixing remains in the SURF models, which allows them to evolve on a blueward evolutionary trajectory. In turn, these models are more compact and therefore can maintain higher surface rotational velocities than the INT models.

%
%
%
%

\begin{figure*}
\includegraphics[width=0.25\textwidth]{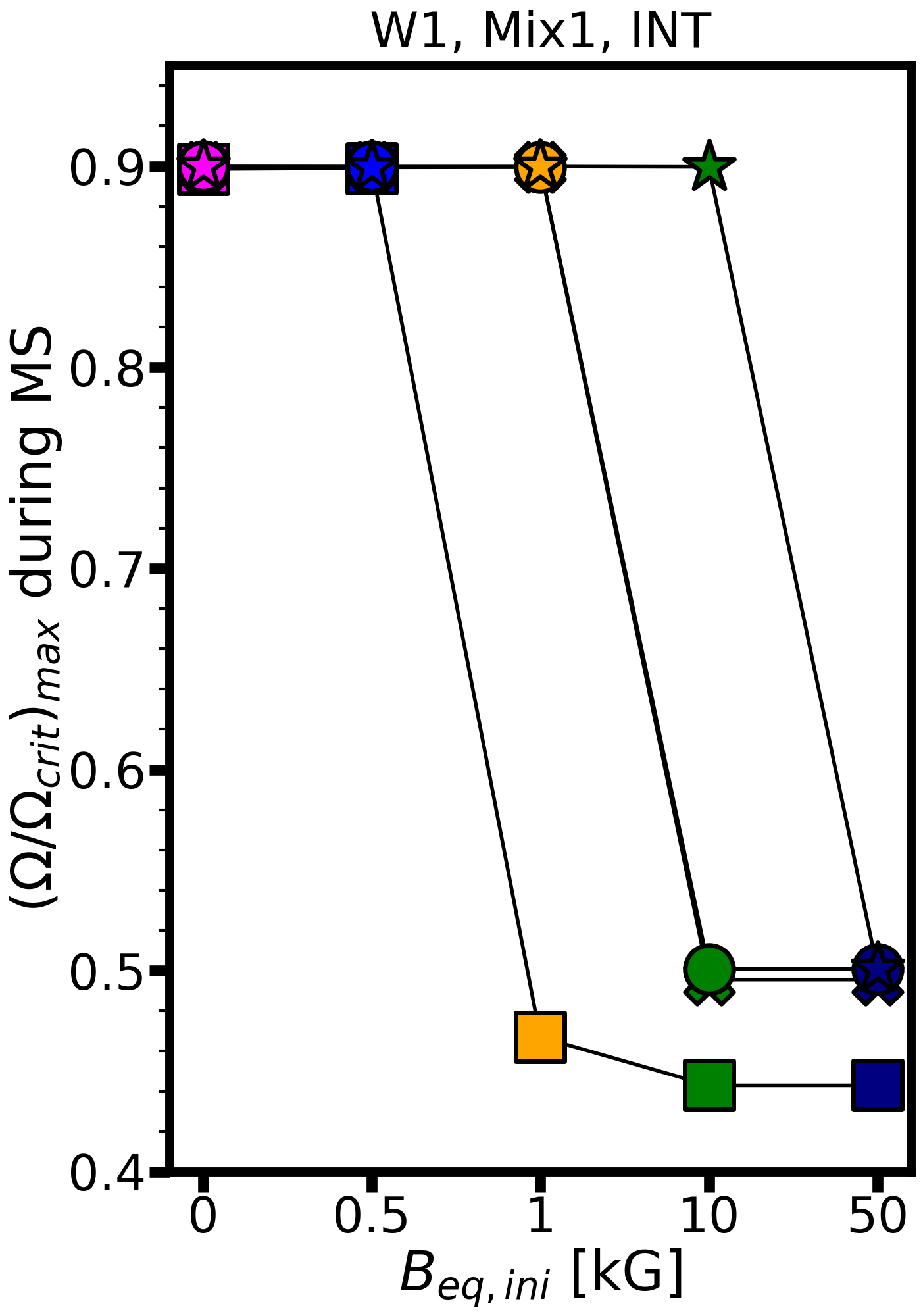}\includegraphics[width=0.25\textwidth]{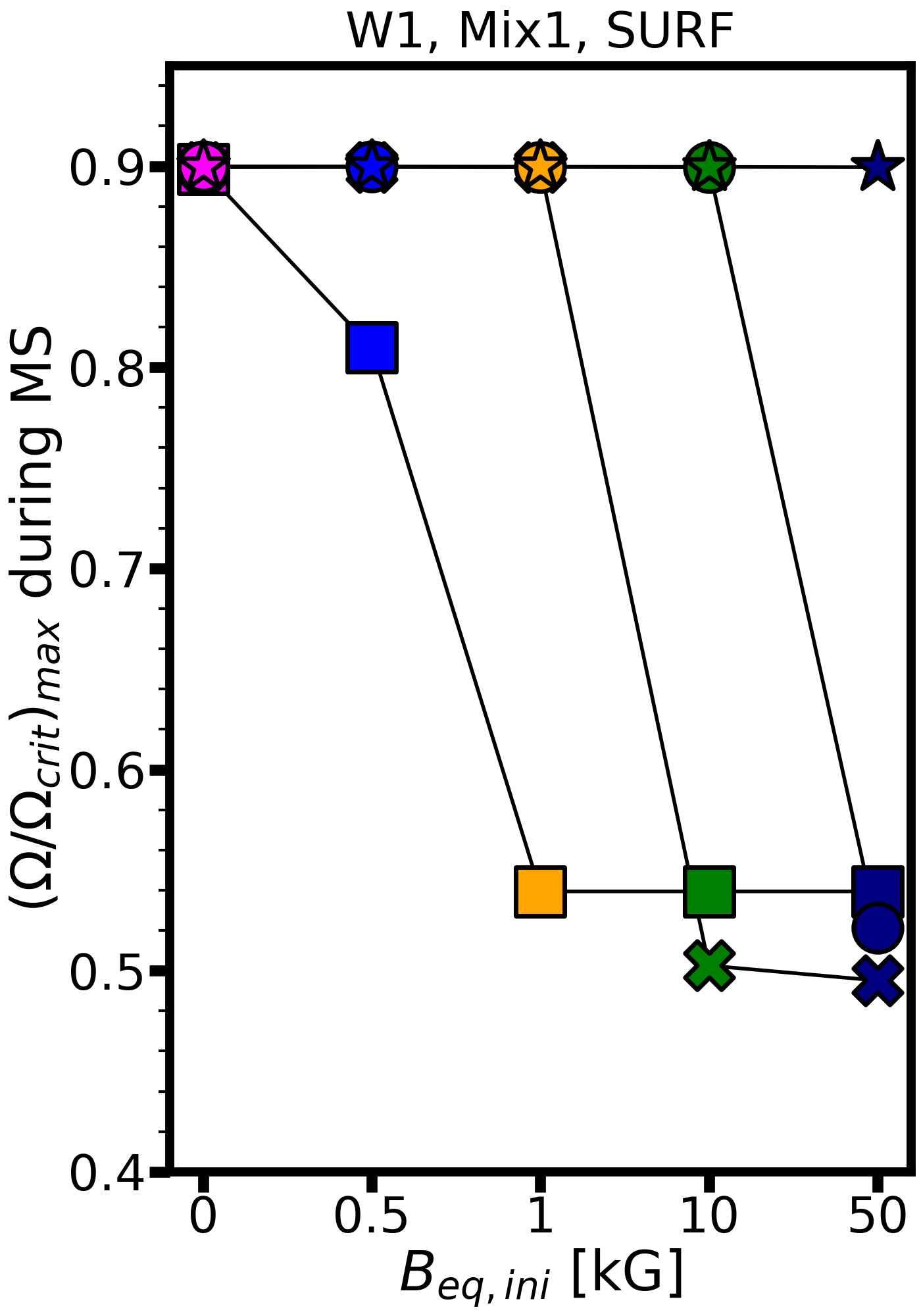}\includegraphics[width=0.25\textwidth]{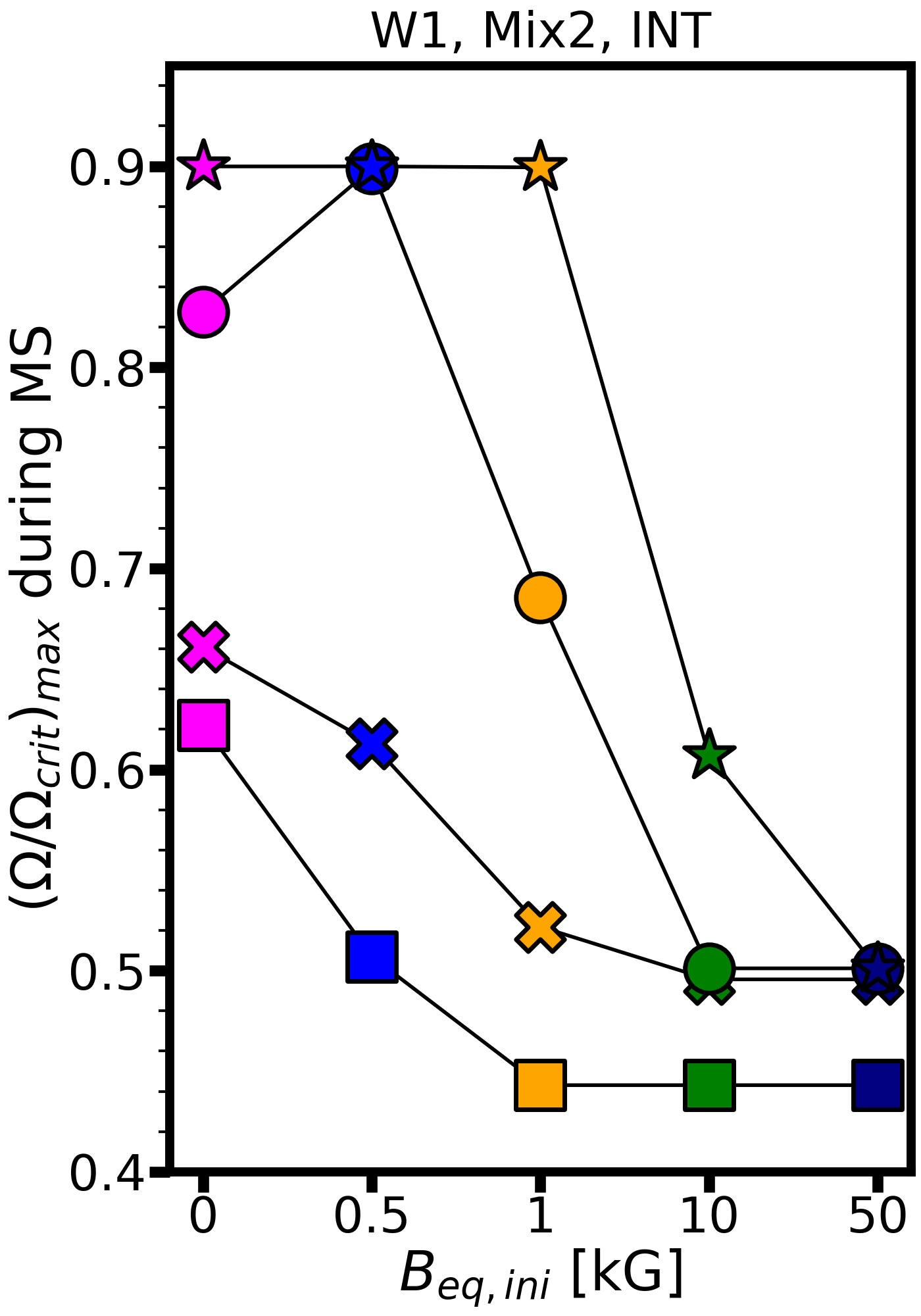}\includegraphics[width=0.25\textwidth]{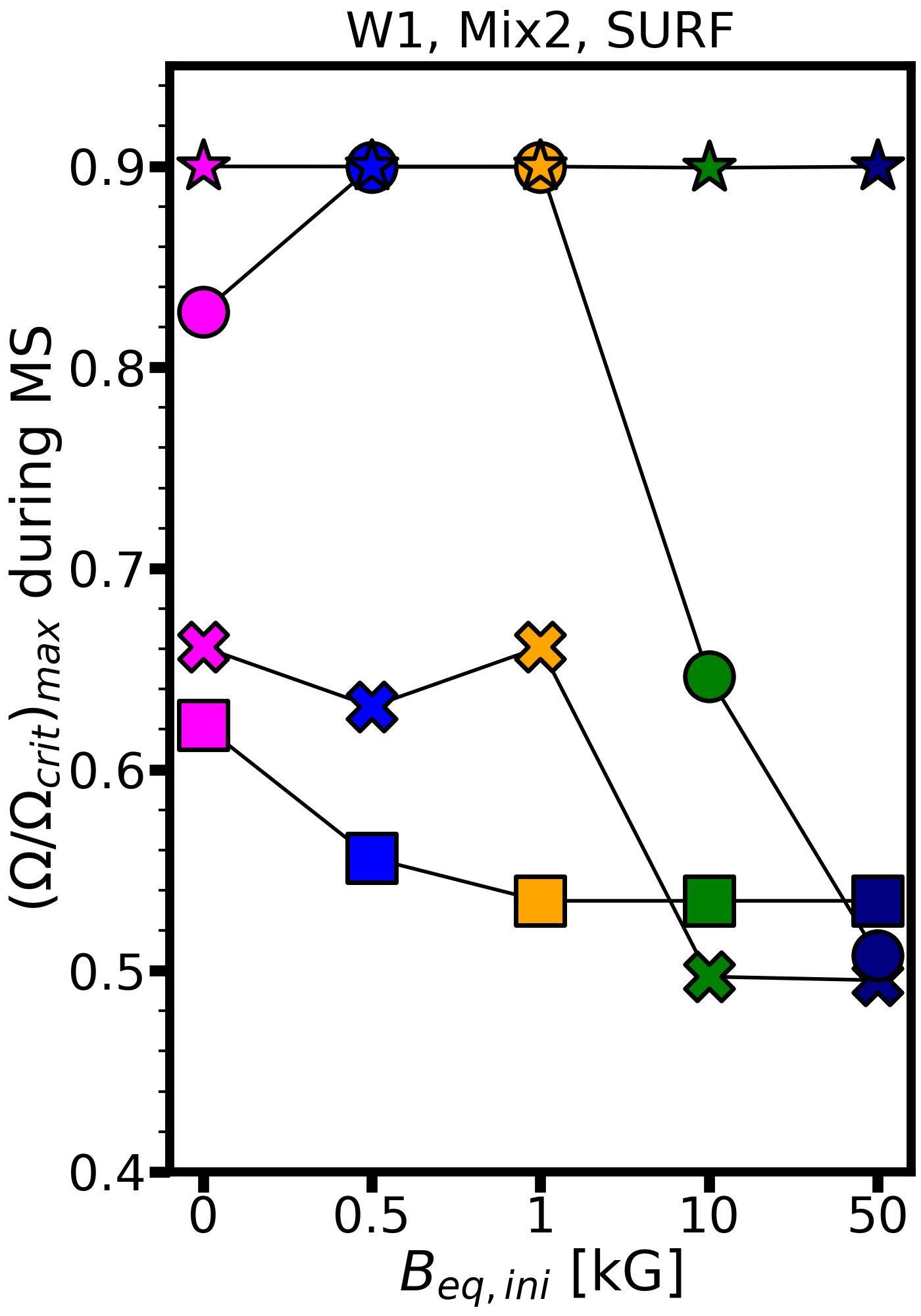}
\includegraphics[width=0.25\textwidth]{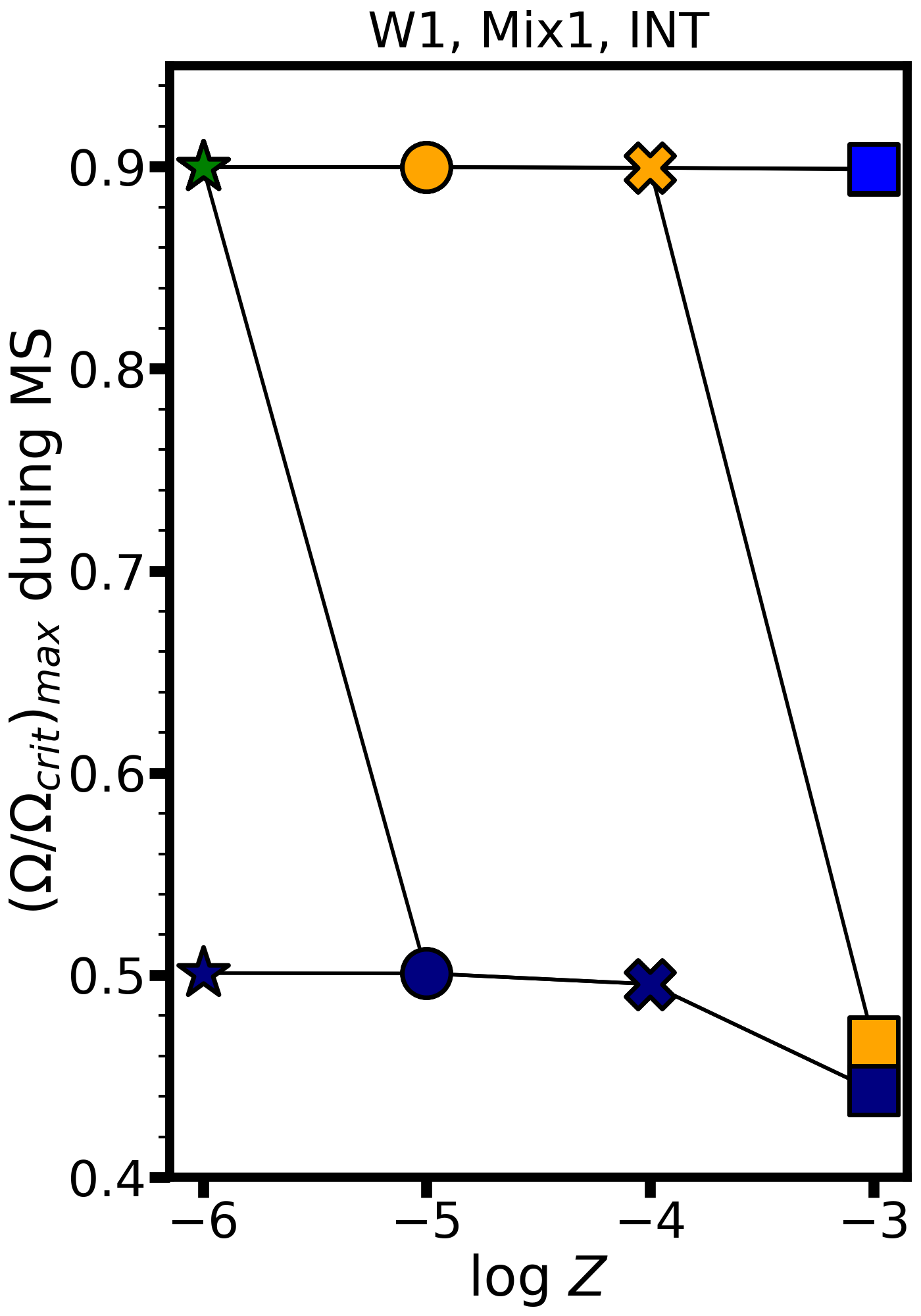}\includegraphics[width=0.25\textwidth]{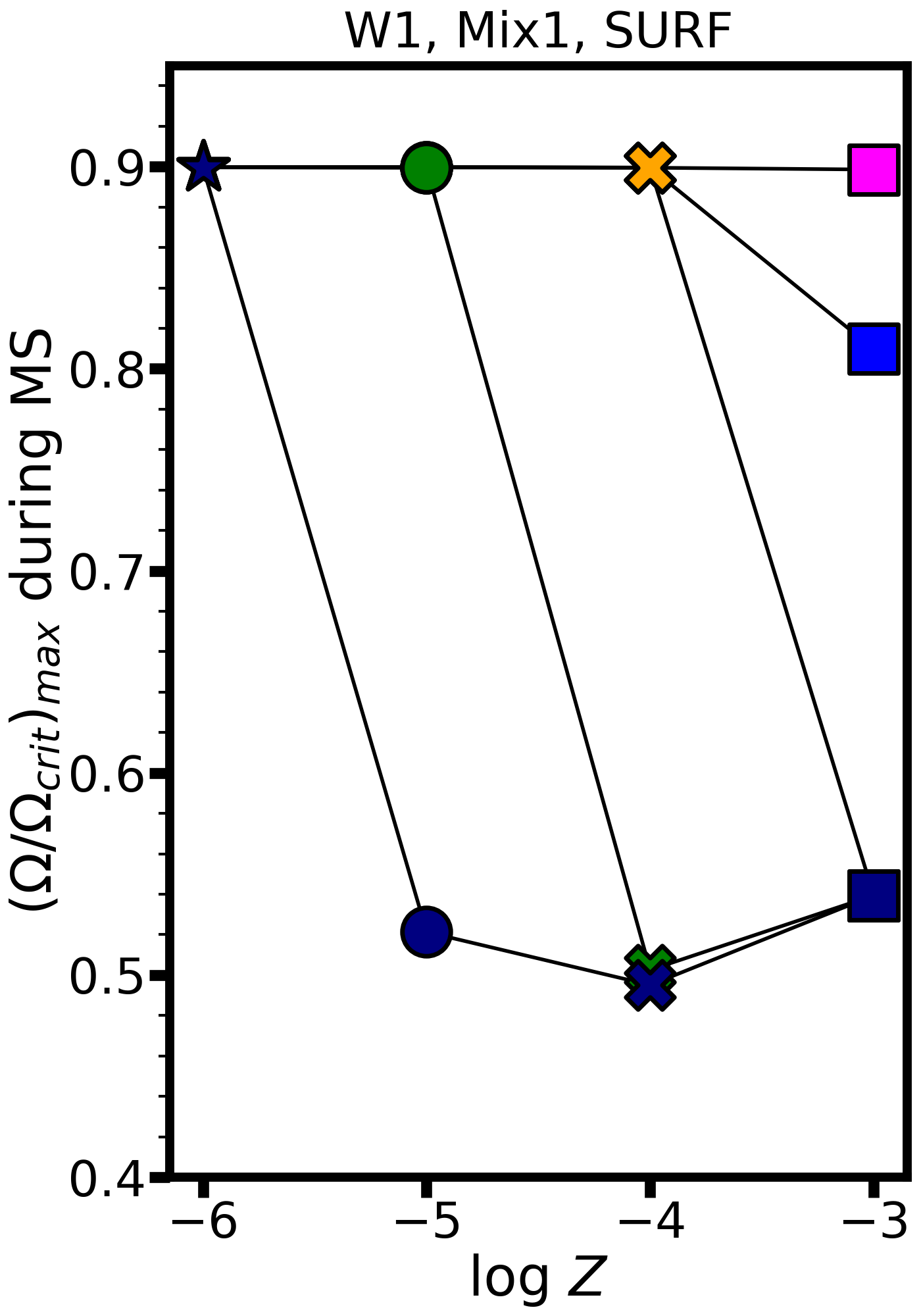}\includegraphics[width=0.25\textwidth]{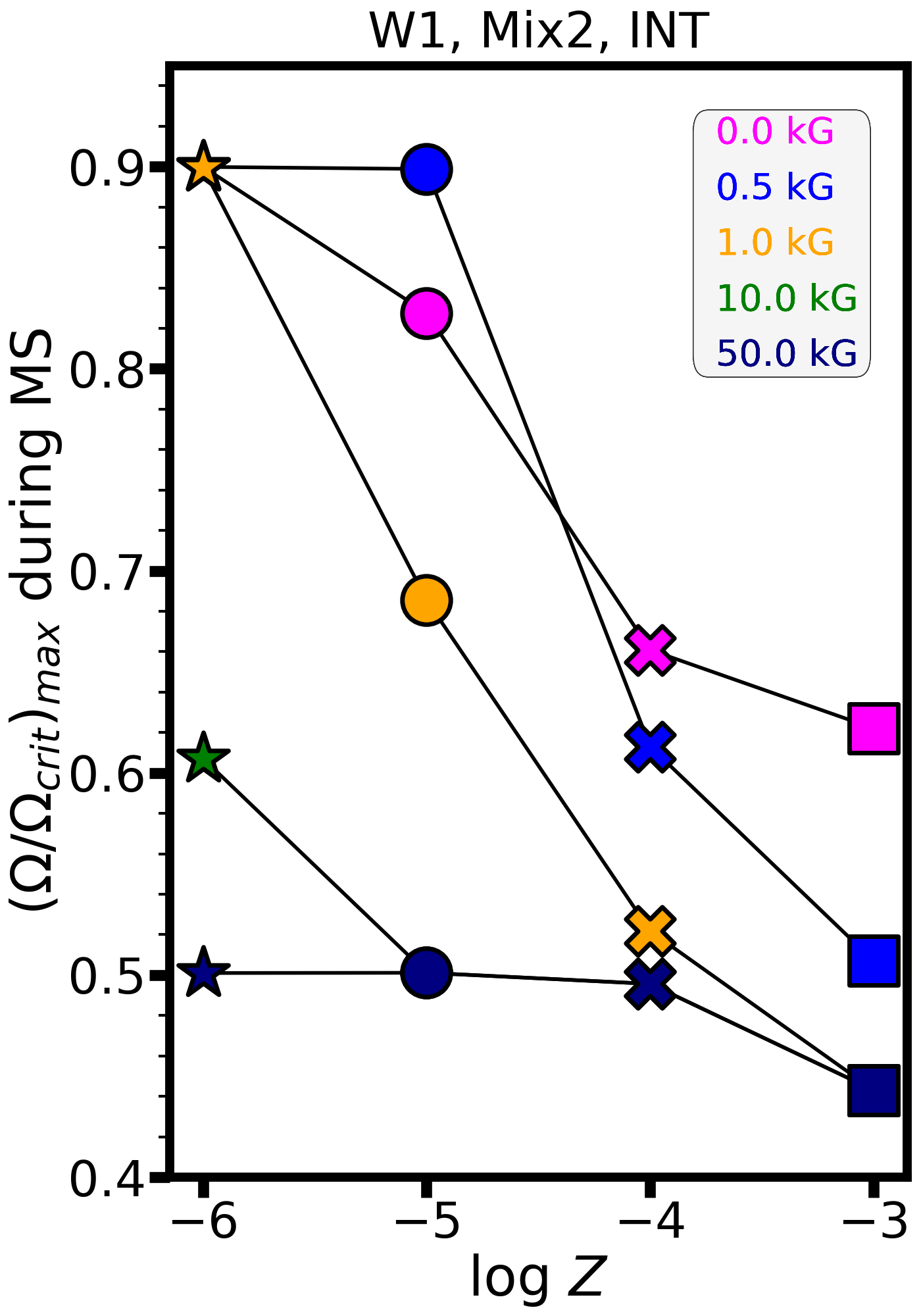}\includegraphics[width=0.25\textwidth]{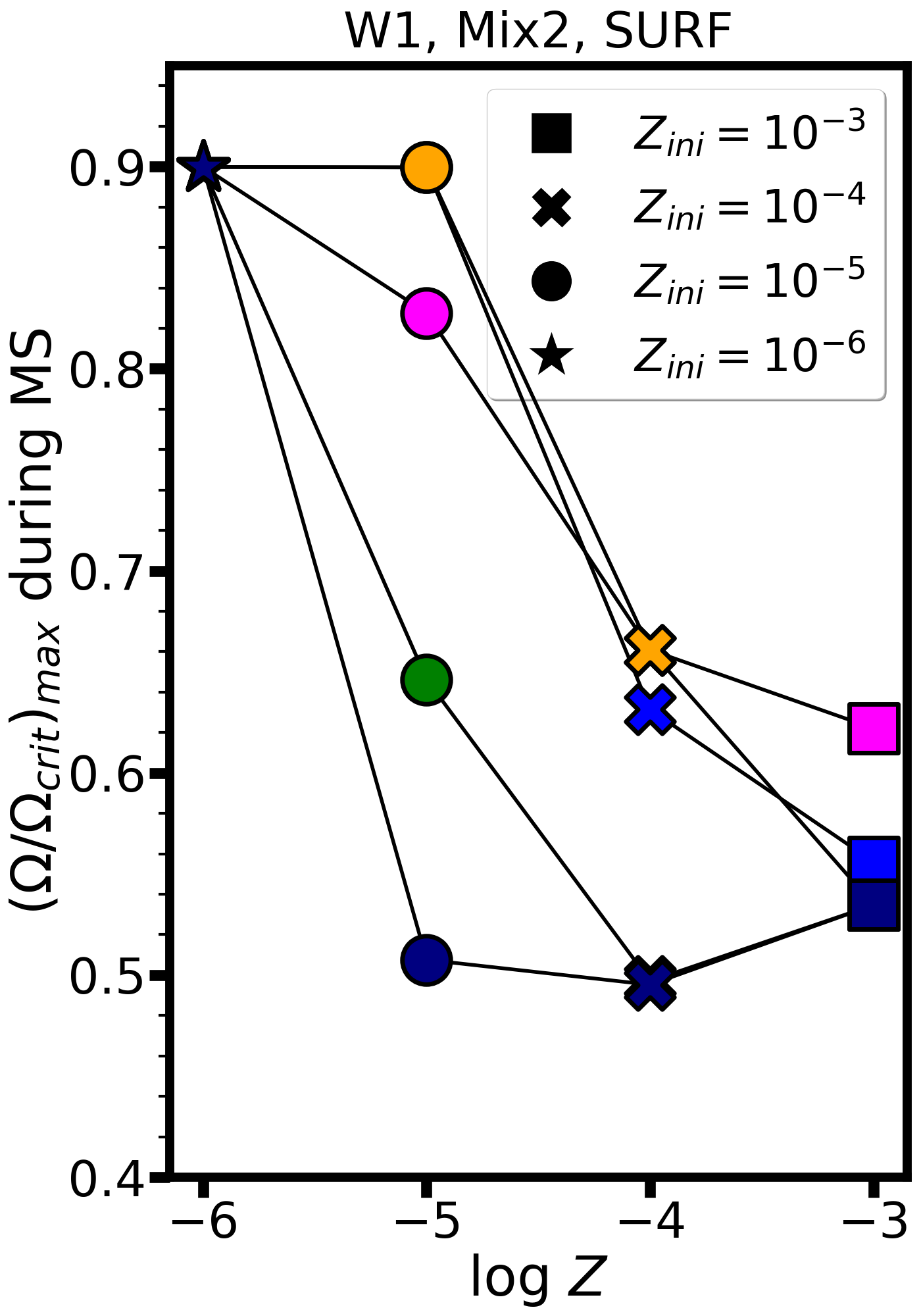}
\caption{Maximum of the angular velocity normalised by its critical value during the main-sequence evolution as a function of initial equatorial magnetic field strength (top) and metallicity (lower) for $M_{\rm ini}=60$~M$_\odot$ models in the W1 wind scheme. The panel title indicates the wind, chemical mixing and magnetic braking schemes. The symbol shapes and colours denote the initial metallicity and magnetic field strength, respectively. The lines connect symbols with the same metallicity (top) or magnetic field strength (lower).}
\label{fig:critrotW1}
\end{figure*}

\begin{figure*}
\includegraphics[width=0.25\textwidth]{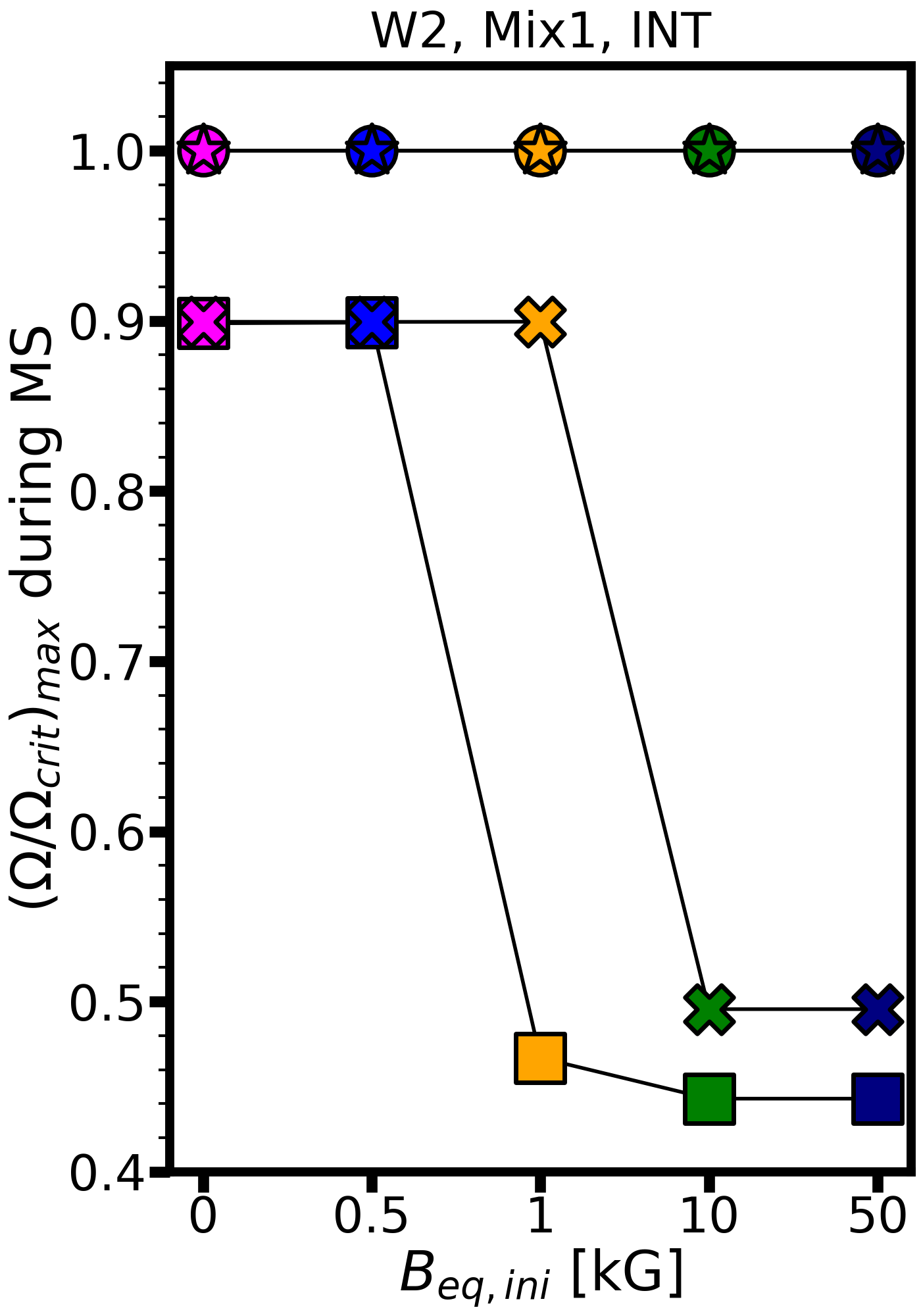}\includegraphics[width=0.25\textwidth]{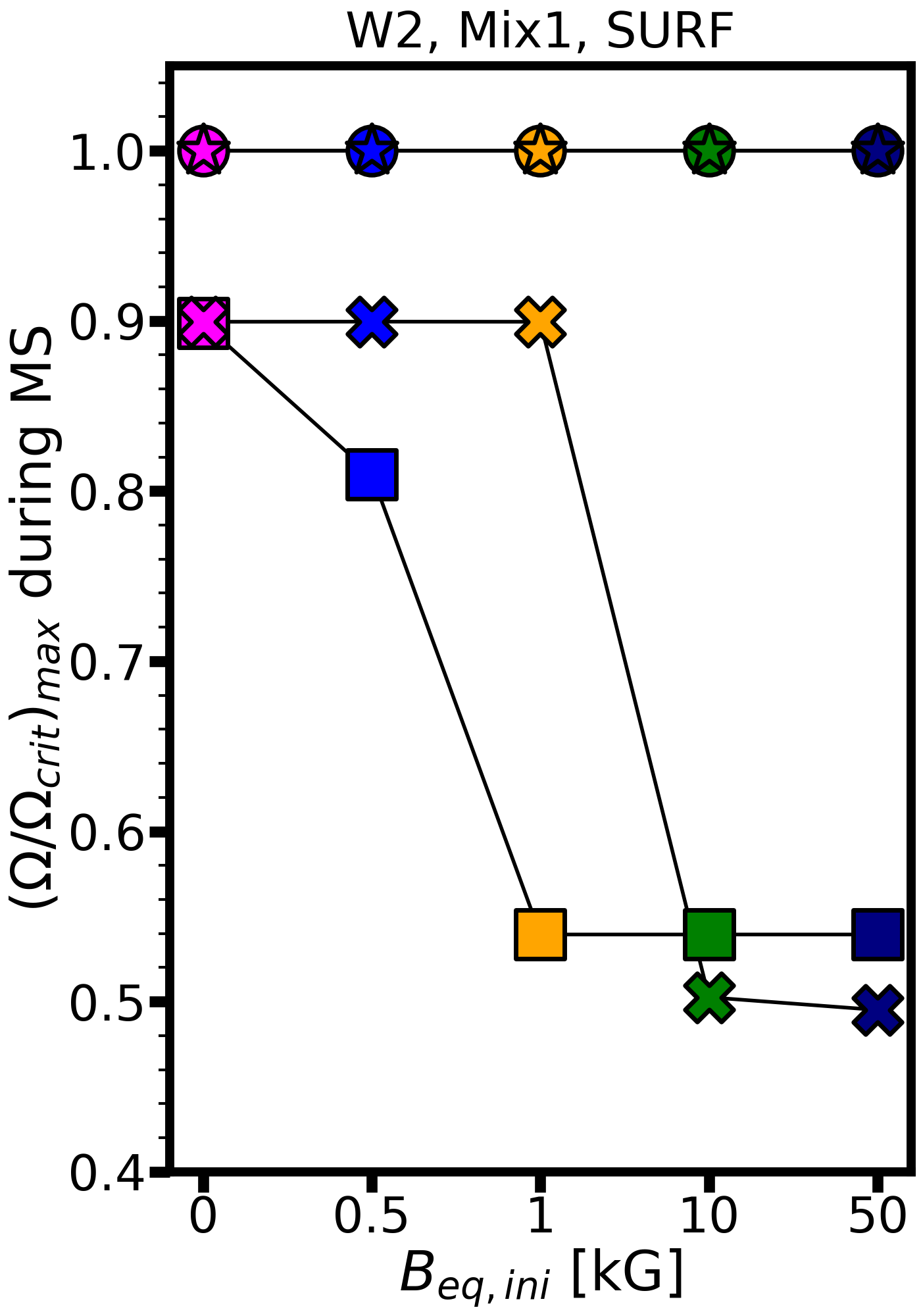}\includegraphics[width=0.25\textwidth]{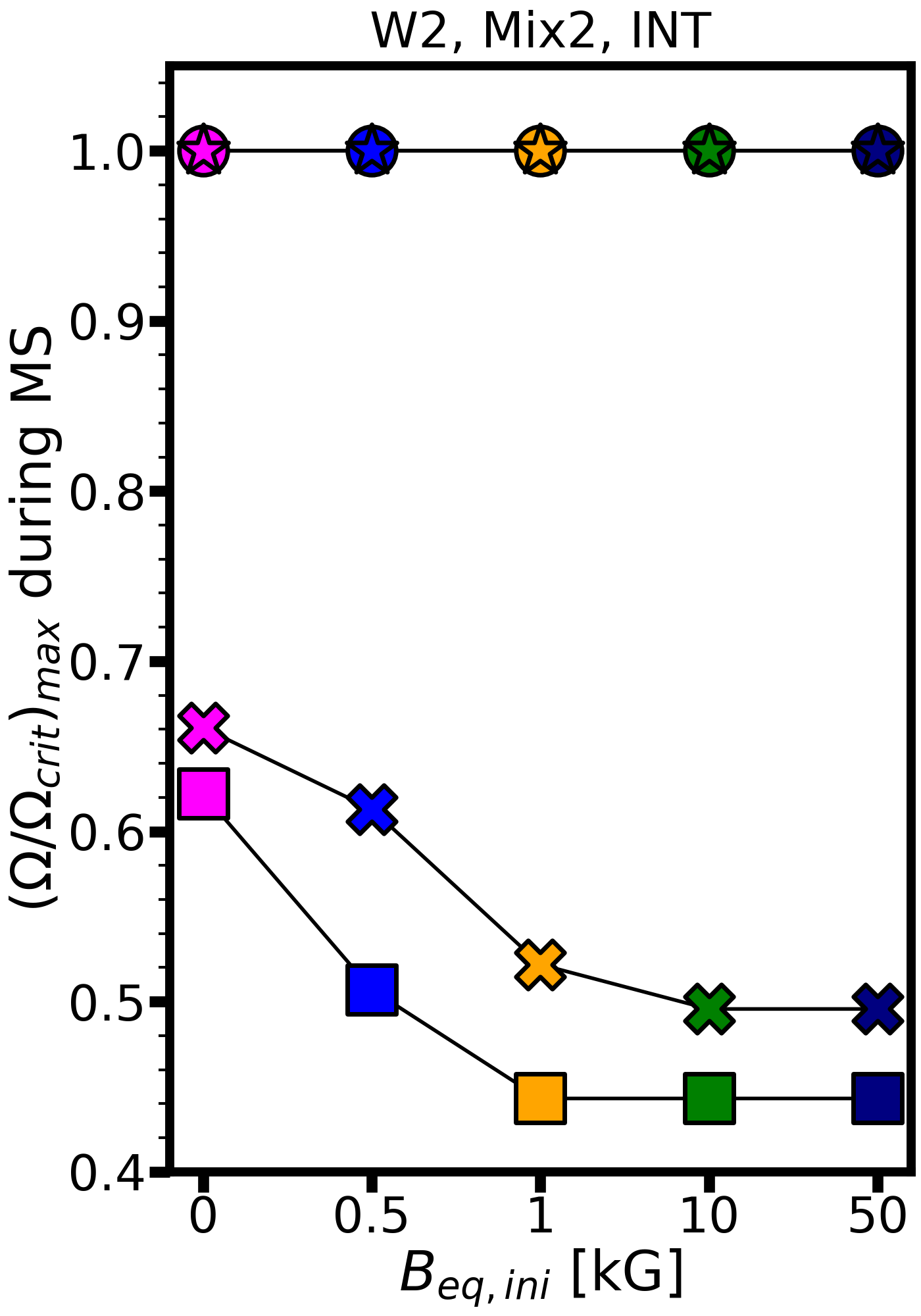}\includegraphics[width=0.25\textwidth]{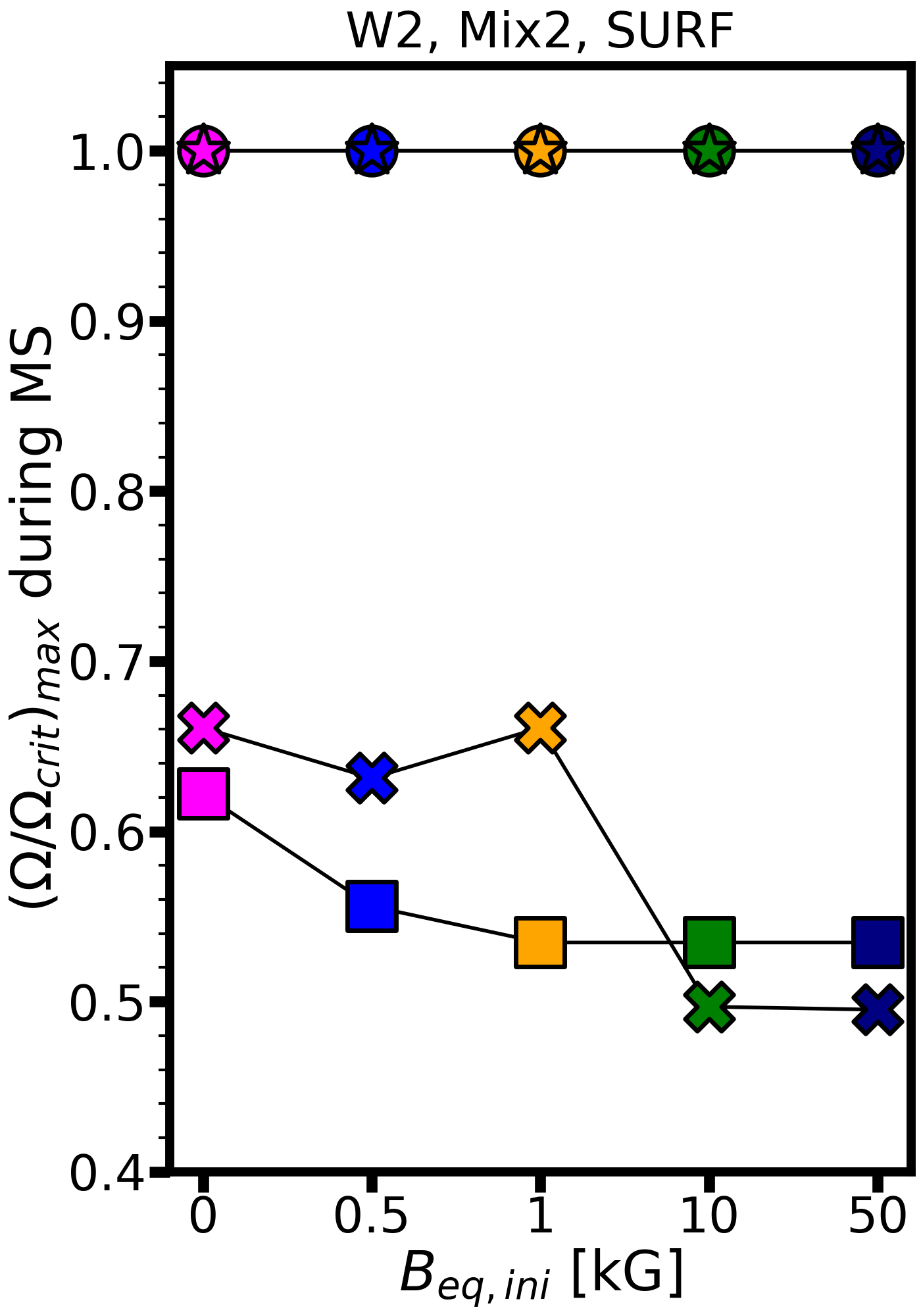}
\includegraphics[width=0.25\textwidth]{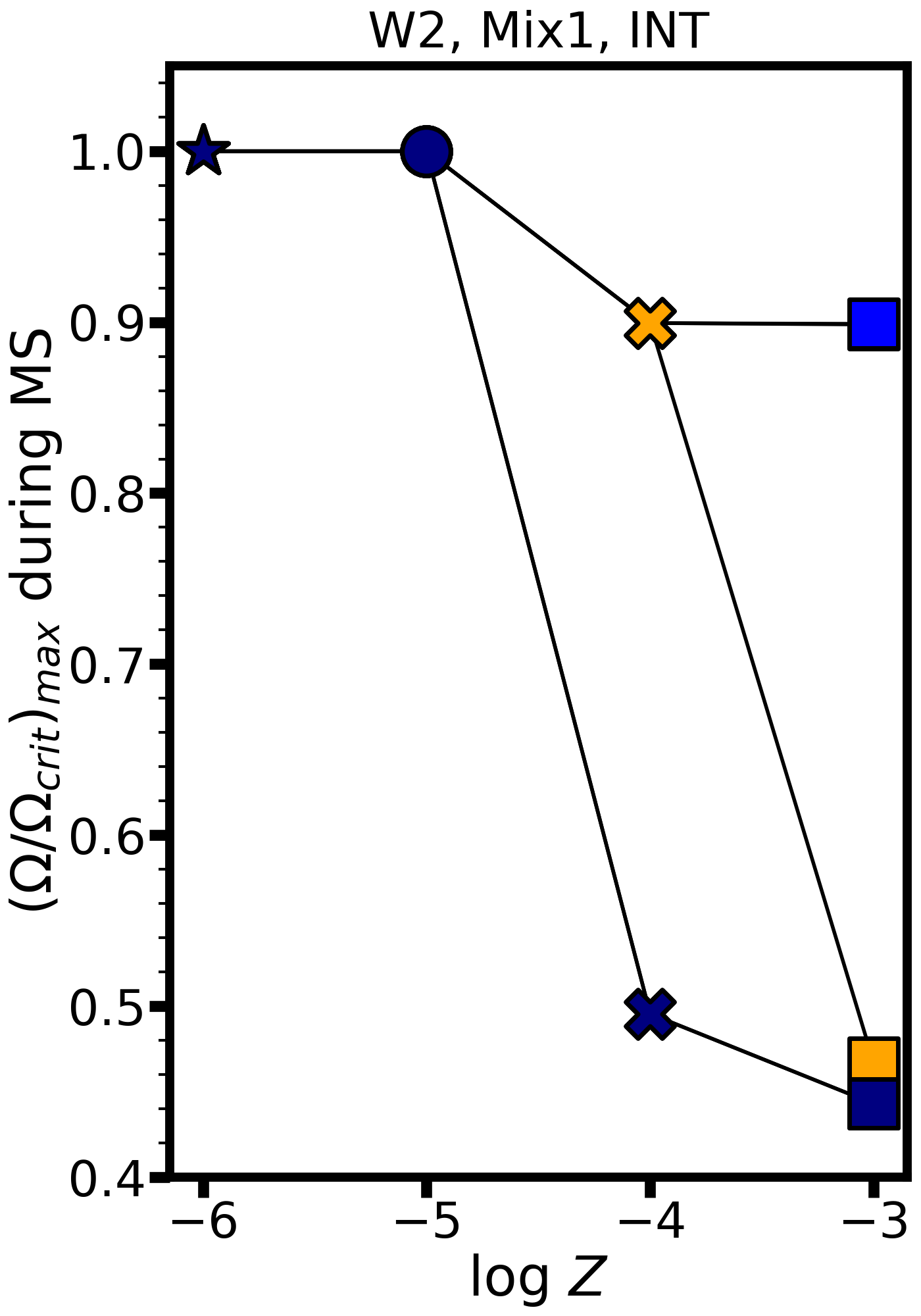}\includegraphics[width=0.25\textwidth]{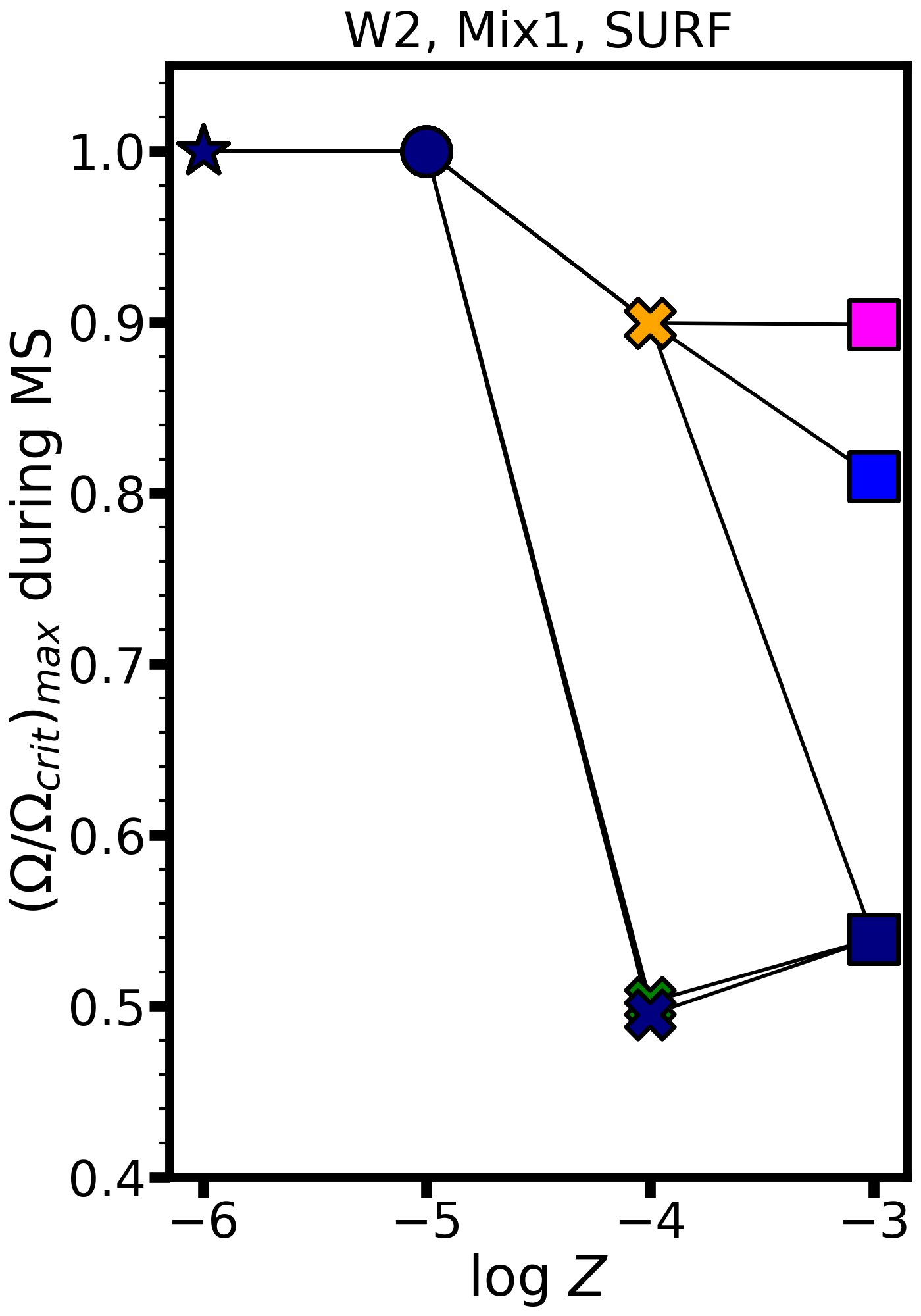}\includegraphics[width=0.25\textwidth]{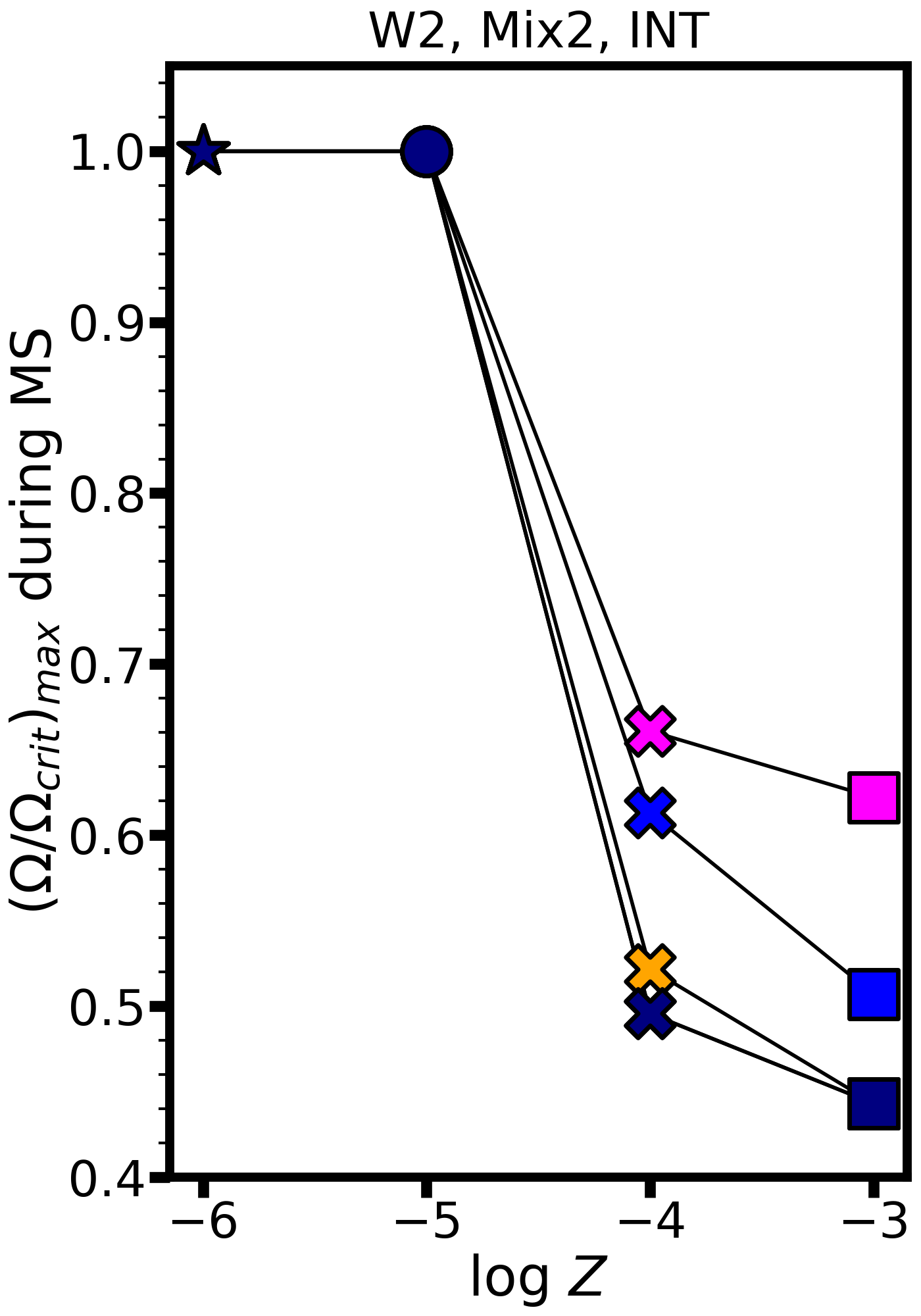}\includegraphics[width=0.25\textwidth]{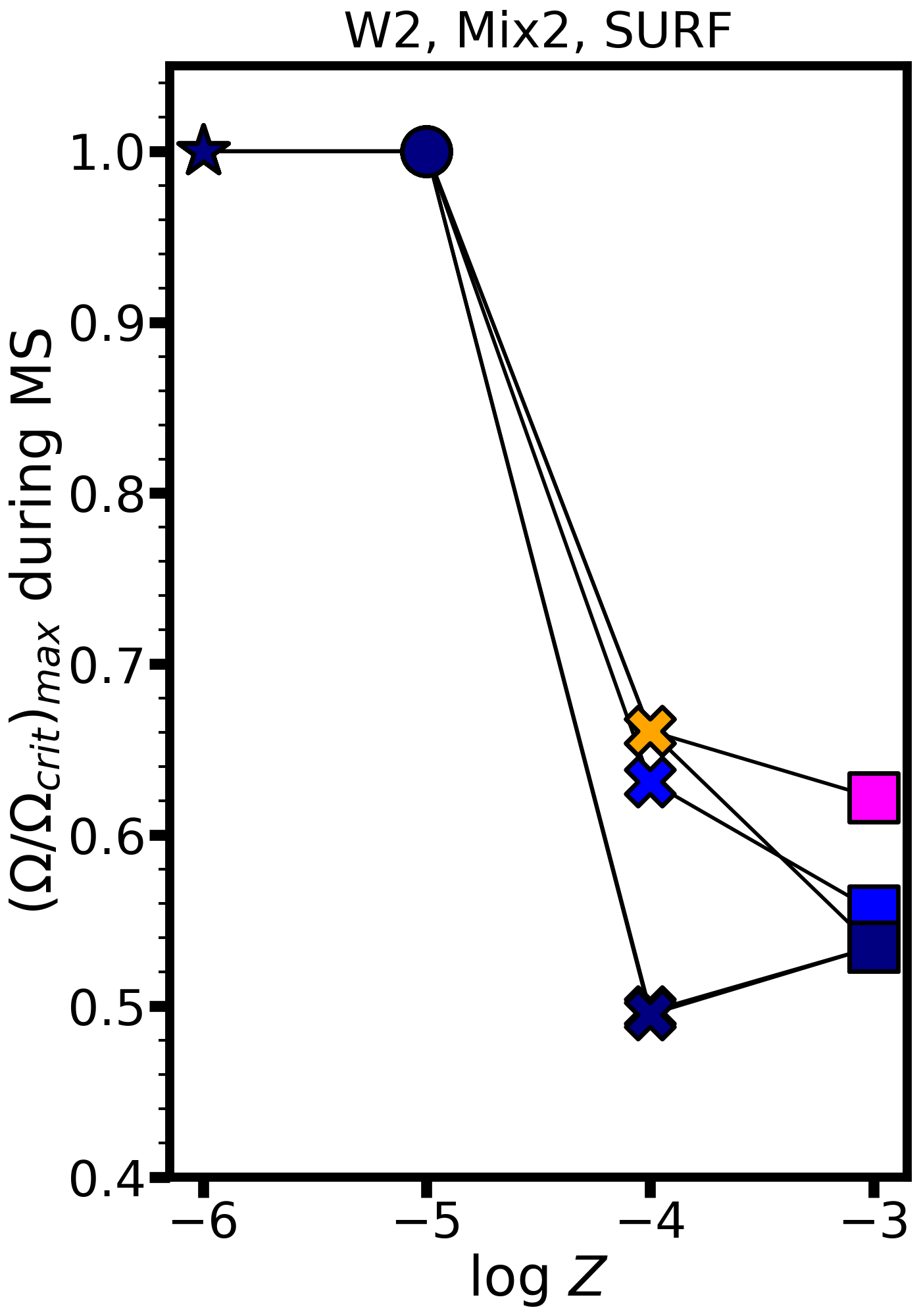}
\caption{Same as Figure~\ref{fig:critrotW1} but for models in the W2 wind scheme.}
\label{fig:critrotW2}
\end{figure*}

%
%
%
%
\section{Results within the W2 wind scheme}\label{sec:five}

In this section, we use a few selected examples to highlight how the assumption on stellar wind impacts the results. We begin with discussing the scenarios, in which the de-coupling limit is reached and zero mass-loss rates are applied in the W2 scheme.

%
%
\subsection{Mass-loss rates}\label{sec:t1}

As shown in Section~\ref{sec:winds}, the de-coupling limit concerns the metallicities of $Z = 10^{-5}$ and $Z = 10^{-6}$ in the case of an initial mass of 60~M$_\odot$. When decreasing the initial mass, the lower luminosity results in weaker winds. Consequently, the calculated mass-loss rate of the 20~M$_\odot$ model falls below the de-coupling limit already at $Z = 10^{-4}$. 
However, the critical mass-loss rate to reach the de-coupling of ions and bulk plasma in the stellar wind is dependent on stellar parameters, which continuously change during the evolution of the star. In Figure~\ref{fig:mdotdec}, we demonstrate this with the specific example of models using the Mix1-INT scheme with $B_{\rm eq, ini}= 1$~kG. 
At $Z = 10^{-4}$, the model with $M_{\rm ini}=20$~M$_\odot$ initiates its evolution with zero applied mass-loss rates since the calculated rates (orange line) are below the critical de-coupling limit (grey line). At around 7~Myr, the calculated rates reach above the de-coupling limit hence the model will experience mass loss from that time onward.
At $Z = 10^{-5}$, the calculated mass-loss rate for the $M_{\rm ini}=20$~M$_\odot$ model is constantly below the de-coupling limit, thus we apply zero mass loss in that case.
Contrary to this, at $Z = 10^{-4}$, the calculated mass-loss rate for the $M_{\rm ini}=60$~M$_\odot$ model is above the de-coupling limit, hence mass loss is applied. In this case, the models in the W1 and W2 branches are completely identical. 
At $Z = 10^{-5}$, the $M_{\rm ini}=60$~M$_\odot$ model in the W2 has zero applied mass-loss rates since the calculated values are systematically below the critical de-coupling limit throughout the main sequence evolution.

%
%
\subsection{Critical rotation}\label{sec:t2}

One of the immediate consequences of applying zero mass-loss rates is that our massive star evolutionary models predict a spin-up. Therefore, these models are much more prone to approach the limit of critical rotation, at which point the centrifugal force balances gravity and radiation pressure \citep{maeder2000}. 

The physics of critical rotation and its implementation in one-dimensional stellar evolution calculations have remained challenging, unresolved problems. Here we simply aim to identify the parameter space that favours the development of critical rotation but we refrain from further discussion of this complex issue.
To this extent, Figures \ref{fig:critrotW1} and \ref{fig:critrotW2} show the tendency towards critical rotation for initially 60~M$_\odot$ models using the W1 and W2 wind schemes. None of the models reach critical rotation when using the W1 scheme. The mass-loss rates are enhanced when reaching $\Omega/\Omega_{\rm crit} \approx 0.9$ and this prevents reaching critical rotation. In general, lower metallicity and weaker magnetic fields favour the development of near critical rotation. 
At $Z = 10^{-3}$ and $Z = 10^{-4}$, the W1 and W2 branches are identical for the 60~M$_\odot$ models since their mass-loss rates are above the de-coupling limit (see Section \ref{sec:t1} above). In the W2 scheme, all models at $Z = 10^{-5}$ and $Z = 10^{-6}$ reach critical rotation.
Thus, we find that critical rotation becomes inevitable below $Z = 10^{-4}$ if we assume zero mass-loss rates when the de-coupling limit is reached. At this point of the evolution, the models generally experience convergence problems, and their further evolution strongly depends on numerical choices. For this reason we do not discuss the model behaviour after reaching critical rotation. 
In principle, stronger magnetic fields work to prevent high rotation. However, we assume that some non-zero mass-loss rate is required for this process to operate. The magnetic field dependence on the maximum angular velocities shows clearly in the W1 case. While strong magnetic fields are very efficient to brake the surface rotation at high-metallicity, when lowering the metallicity, they become less efficient -- even if some (low) mass-loss rates are applied in the models.

%
%
%
\begin{figure*}
\includegraphics[width=0.45\textwidth]{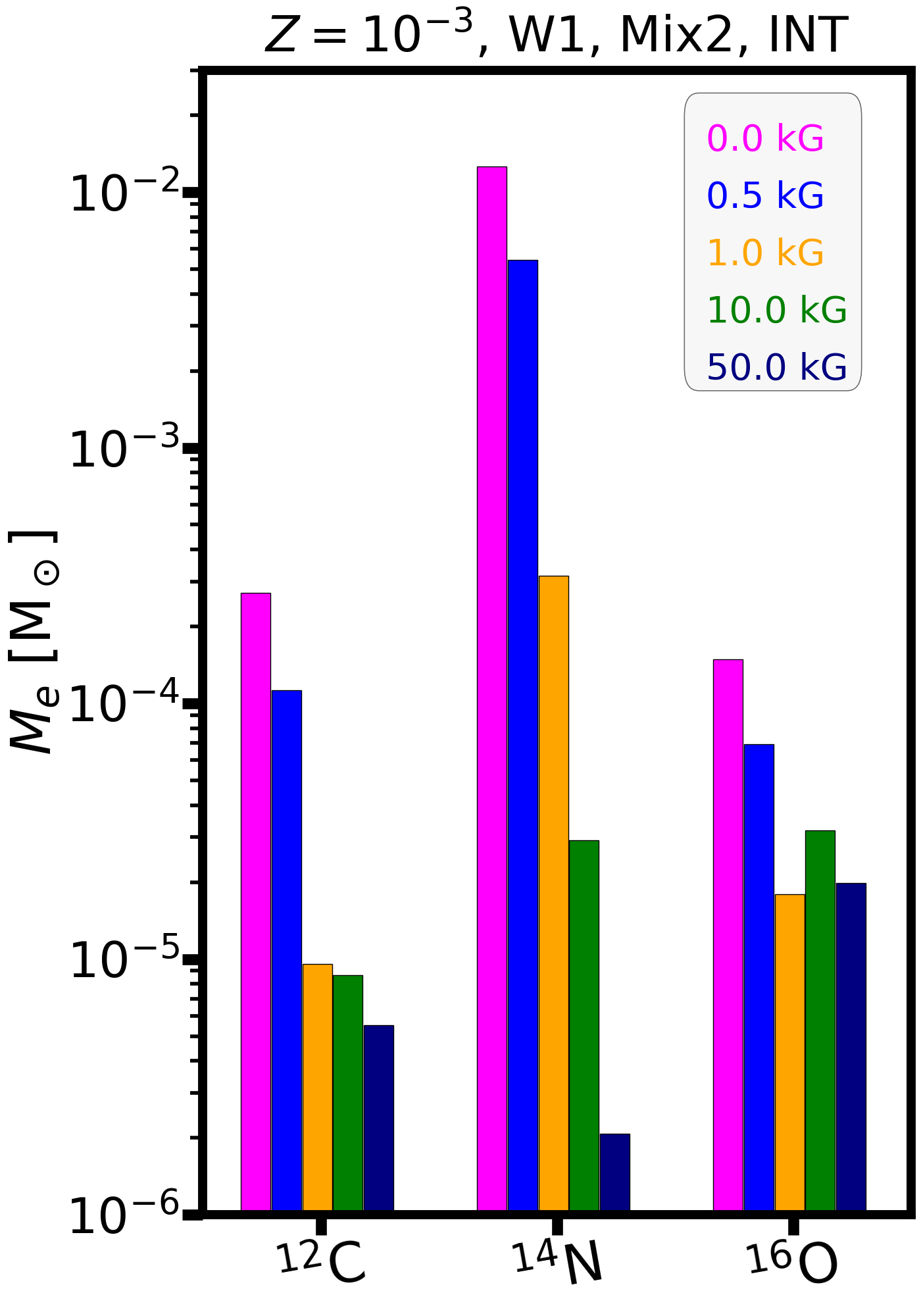}\includegraphics[width=0.45\textwidth]{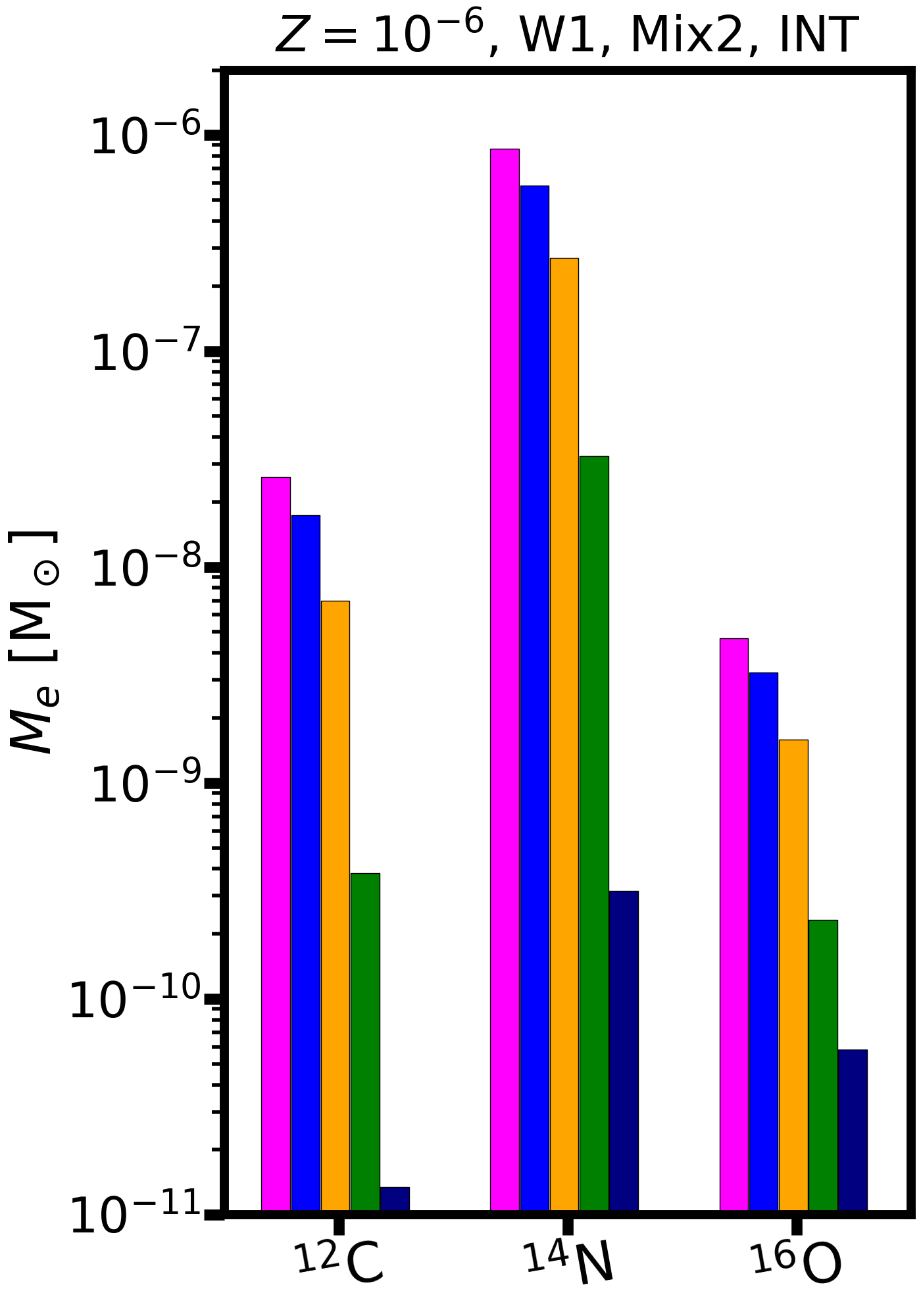}
\caption{Ejected mass of CNO chemical species integrated over the main sequence evolution for initially 60~M$_\odot$ models. The colour-coding corresponds to the magnetic field strength. The panel titles indicate the wind, chemical mixing, and magnetic braking schemes. The left and right panels show metallicities of $Z=10^{-3}$ and $10^{-6}$, respectively. Note the different vertical axis scales.}
\label{fig:yields}
\end{figure*}
%
%
%

\cite{meynet2002} also studied $M_{\rm ini} =60$~M$_\odot$ models at $Z = 10^{-5}$. The main difference in their model is the advective treatment of angular momentum transport, which, without including magnetic fields, results in growing radial differential rotation. Although, this is qualitatively different, it still aids outward angular momentum transport in the model \citep{meynet2001}. Combined with the lower mass-loss rates for lower metallicity (scaled down from solar-metallicity conditions, thus similar to our W1 scheme), their model also predicts a spin up early on in the main sequence and reaches critical rotation close to the end of the main sequence (their Figures 5 and 7). In our case, we mitigate reaching exactly the critical velocity. Overall our results are in very good agreement.

%
%
%
%
\section{Discussion}\label{sec:six}

%
%
\subsection{Ejected mass of chemical species during the main sequence}\label{sec:d1}

Understanding chemical enrichment on local, galactic, and even cosmological scales requires well-constrained stellar models. However, the different assumptions lead to large deviations in model predictions. 

We demonstrated in Figure~\ref{fig:hecno} that the He and CNO abundances can be an order of magnitude different when using various mixing efficiencies and magnetic field strengths. Here, we calculate the ejected mass of chemical species over the main sequence evolution from our models, following the parametrisation of \cite{hirschi2005} and \cite{higgins2023}.
\begin{equation}\label{eq:yields}
    M_{e} = \int_{\rm ZAMS}^{\rm TAMS} \dot{M} \, X_{\rm surf}  \, \mathrm{d} t \, ,
\end{equation}
where $X_{\rm surf}$ is the surface abundance (mass fraction) of a given isotope, and ZAMS and TAMS are the zero age and terminal age main sequence, respectively. 
Our results show that the ejected mass is the most sensitive to metallicity and magnetic field strength (Figure~\ref{fig:yields}). 
In general, the more efficient Mix2 chemical mixing scheme results in higher nitrogen yields than the Mix1 scheme; however the carbon and oxygen yields show more complicated trends with chemical mixing scheme. 
Assuming the W1 wind scheme, the decrease in metallicity results in a gradual decrease in the ejected mass.
At $Z=10^{-3}$, the strongly magnetised models predict the least amount of ejected mass. On the one hand, these models have less efficient mixing due to their slower rotation. This keeps the surface abundances of carbon and nitrogen relatively unchanged, whereas other models allow for carbon depletion and nitrogen enrichment following the changes in the stellar core. The ejected mass of oxygen is also the smallest for the most strongly magnetised model in a given scheme at a given metallicity. 
On the other hand, these models also predict shorter main-sequence lifetimes. Since the yields are integrated over a time period, this also means lower values of $M_{e}$ for the magnetic models. 
At $Z=10^{-6}$, these trends are mostly maintained. In the INT scheme, a very strong, 50 kG magnetic field could have a drastic impact of several orders of magnitude on the ejected mass. The models with different initial magnetic fields strengths in the SURF scheme maintain smaller differences.

In previous works, the rotational properties of low-metallicity massive stars were considered to study the impact on CNO yields by \cite{meynet2002,meynet2002b,meynet2006}. We have now demonstrated that additional physical ingredients in the model calculations are able to change the quantitative results already on the main sequence. Interestingly, even if the star initially rotates fast, the nitrogen abundance is not necessarily enhanced when the magnetic field is strong. 
In particular, the high nitrogen abundance observed in GN-z11 is also accompanied by a relatively low oxygen abundance, which suggests that the source is not a supernova ejecta (which contains oxygen).
We note however that, in general, supernova ejecta also contributes to the chemical evolution of the Universe in addition to stellar winds \citep[e.g.,][]{Chiaki2018}. 
While we only focus on main-sequence models in the present work, we discuss some potential implications for later evolutionary stages in the next section.

%
%

\begin{figure*}
\includegraphics[width=0.25\textwidth]{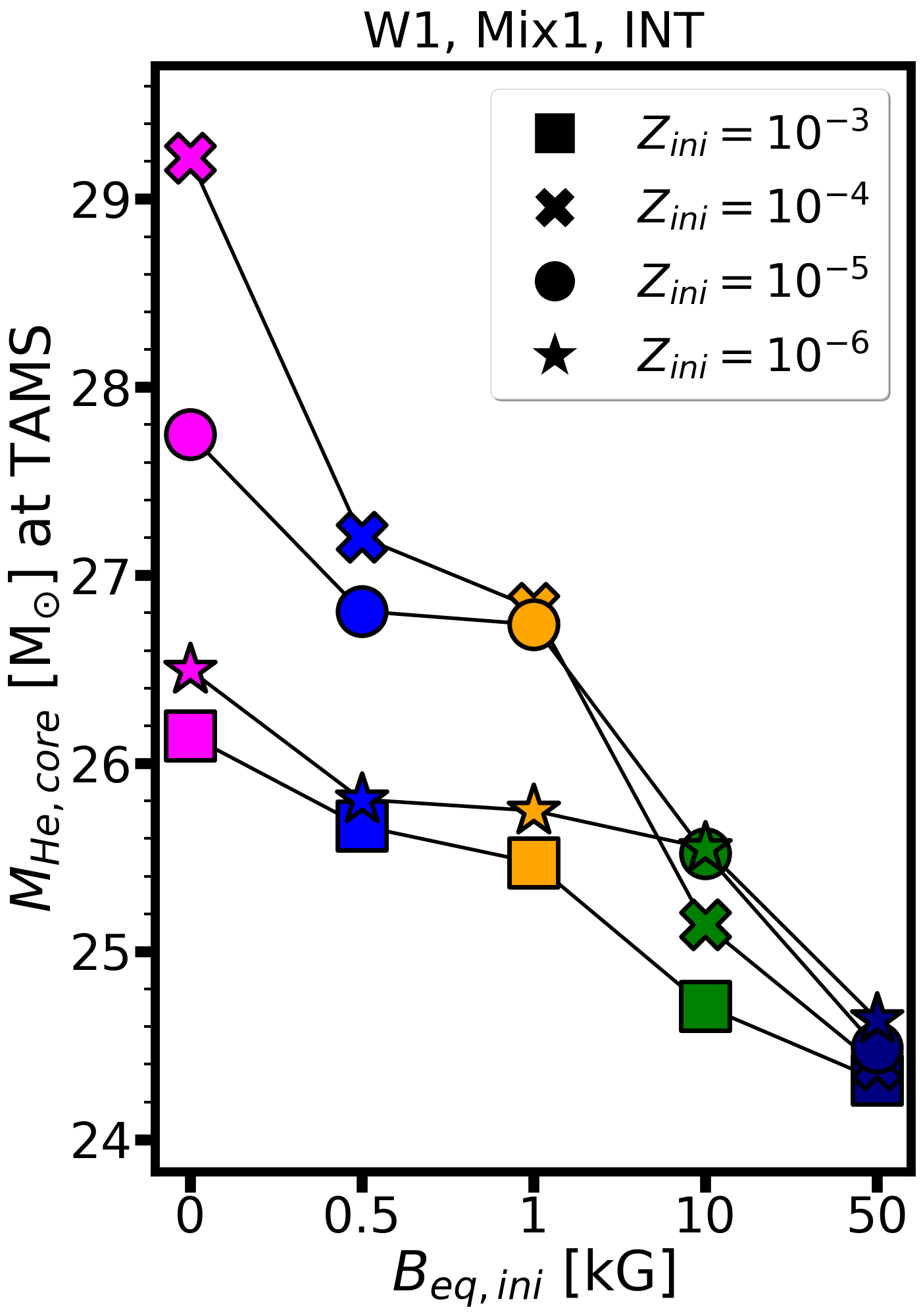}\includegraphics[width=0.25\textwidth]{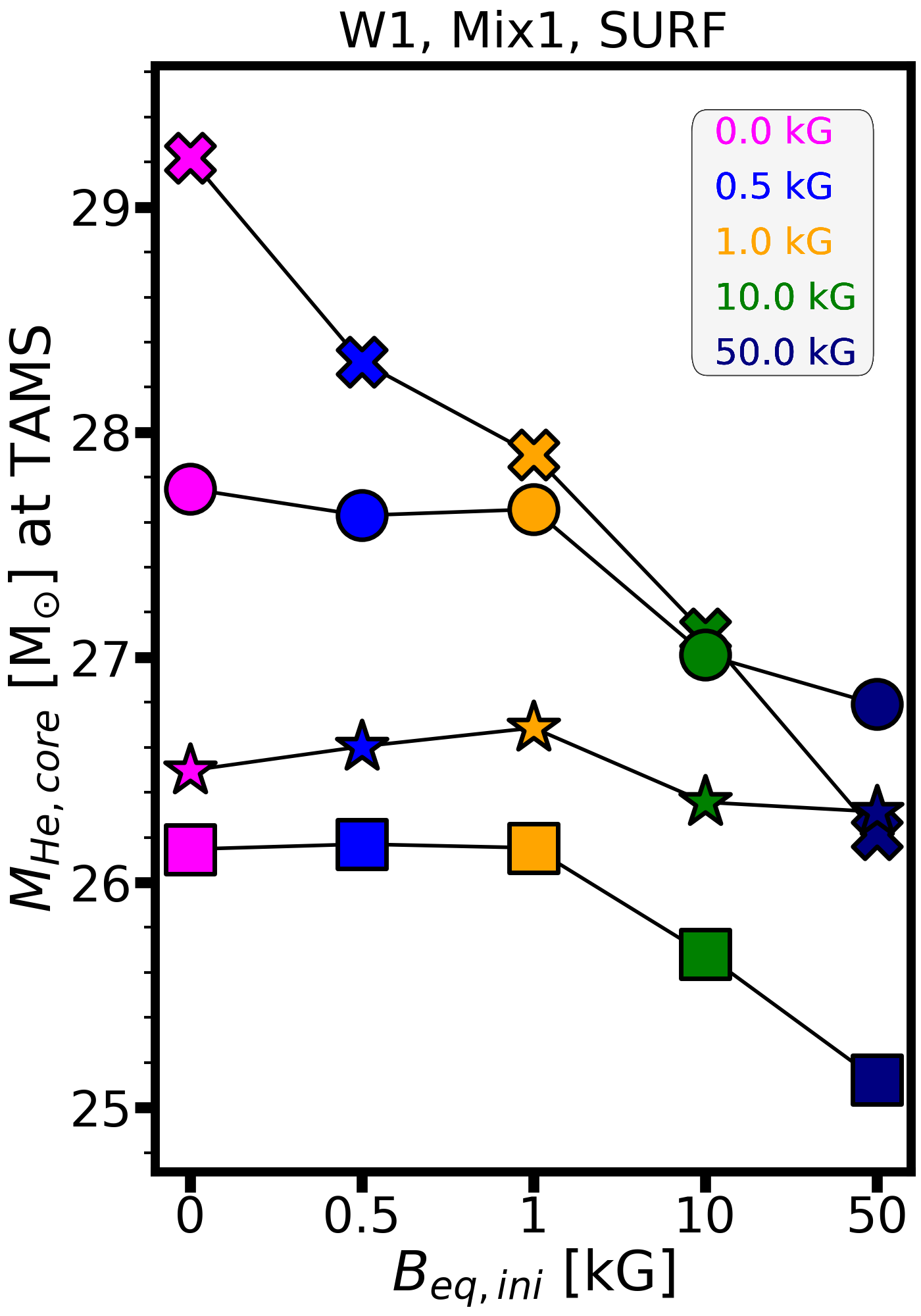}\includegraphics[width=0.25\textwidth]{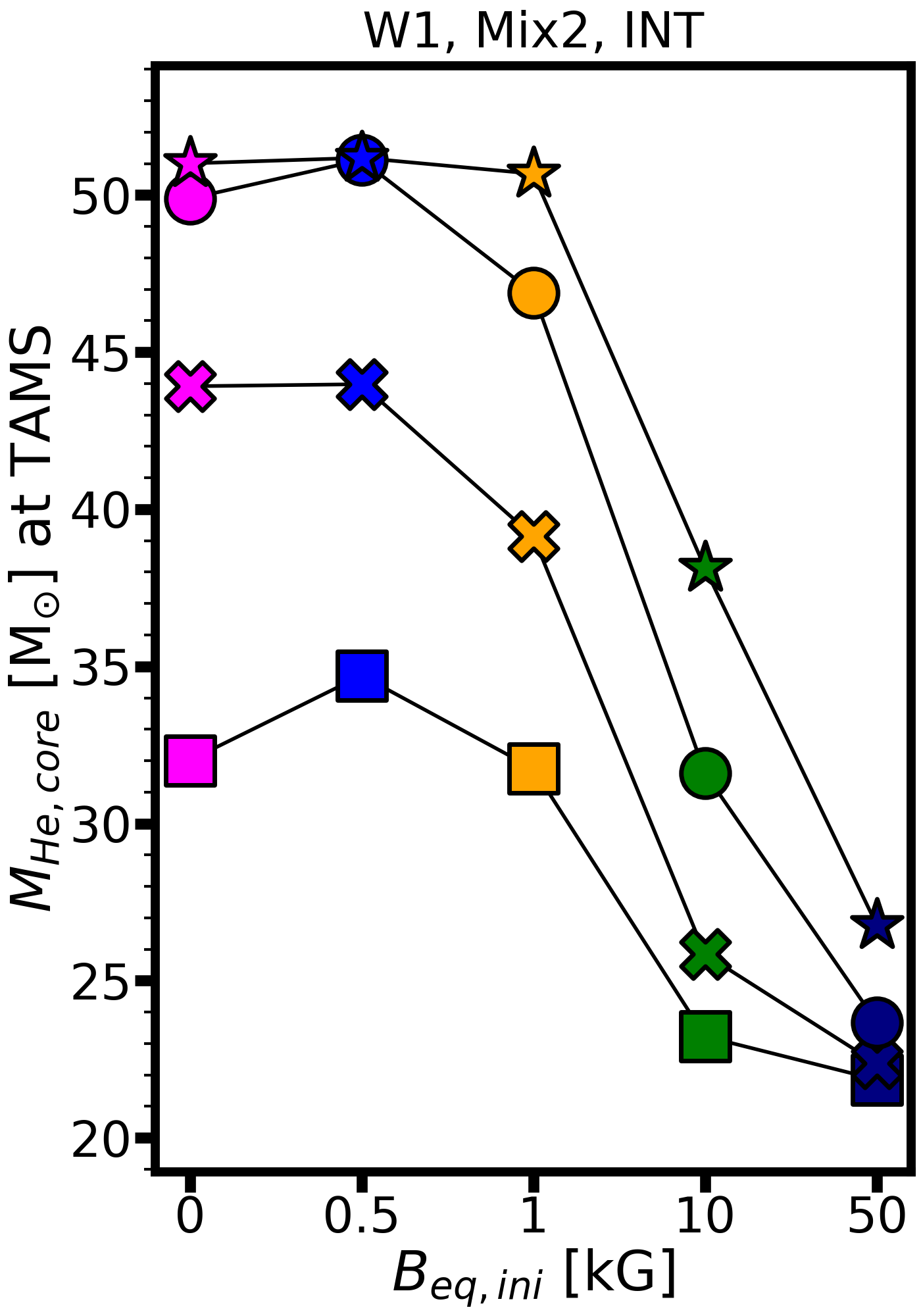}\includegraphics[width=0.25\textwidth]{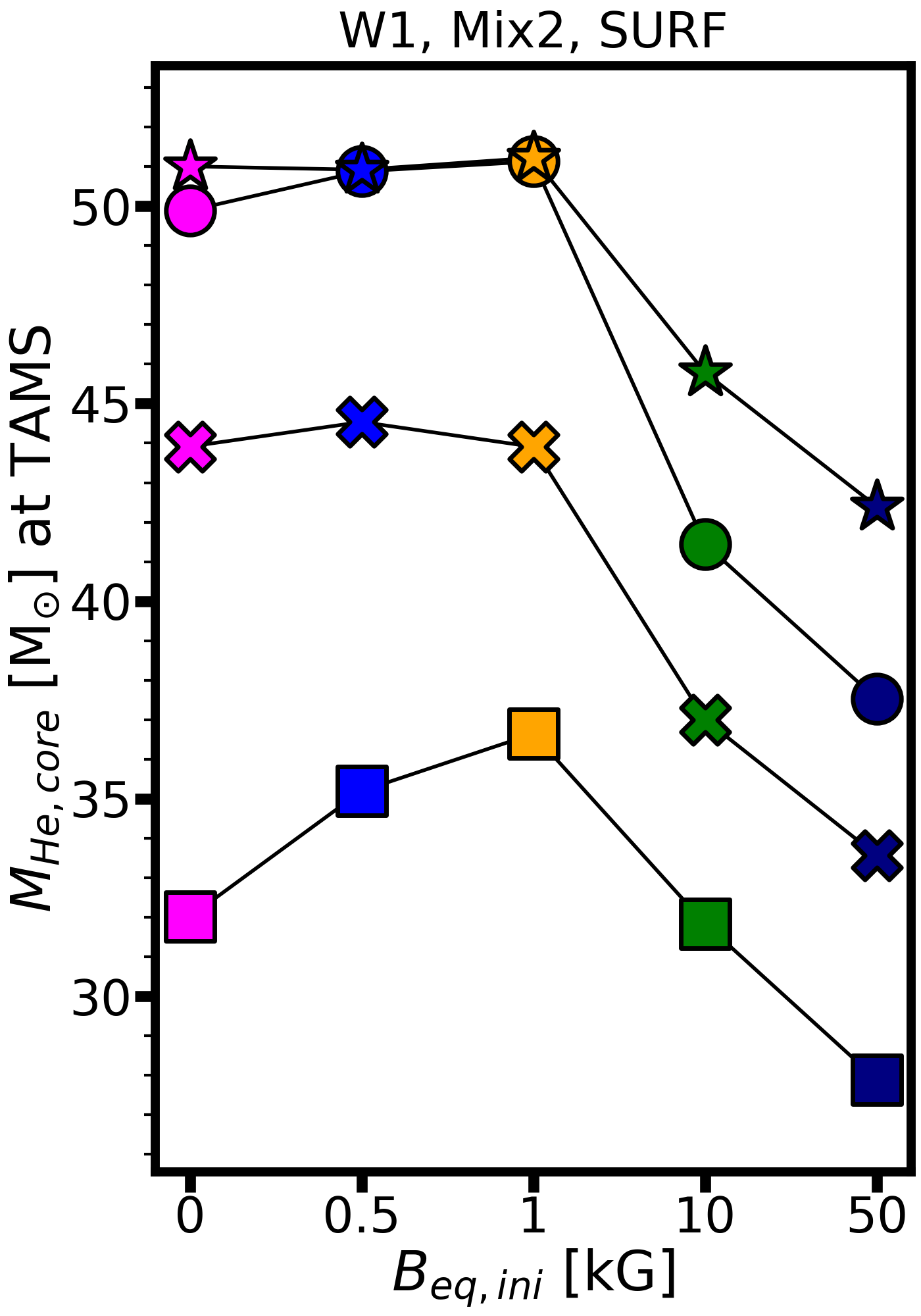}
\includegraphics[width=0.25\textwidth]{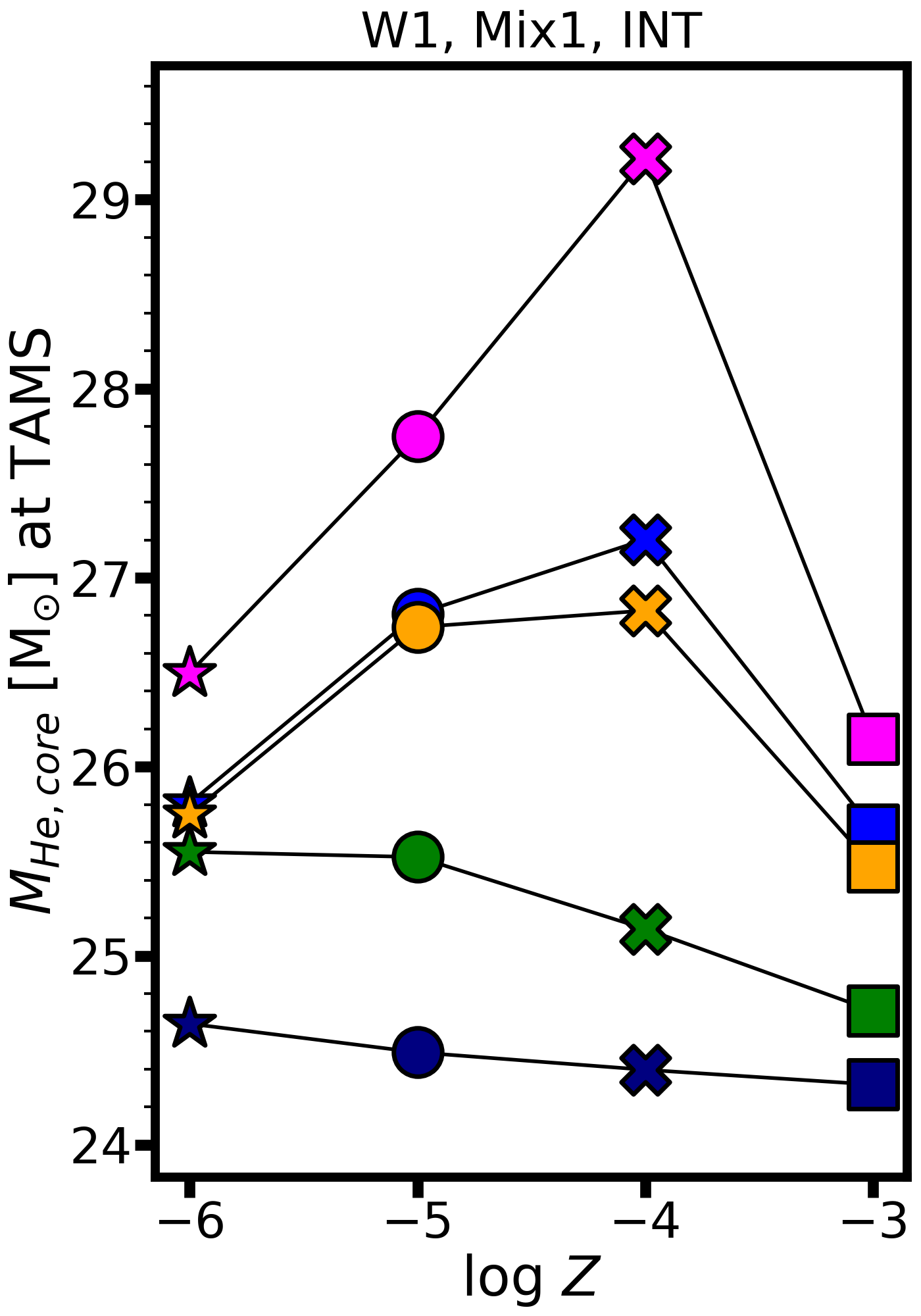}\includegraphics[width=0.25\textwidth]{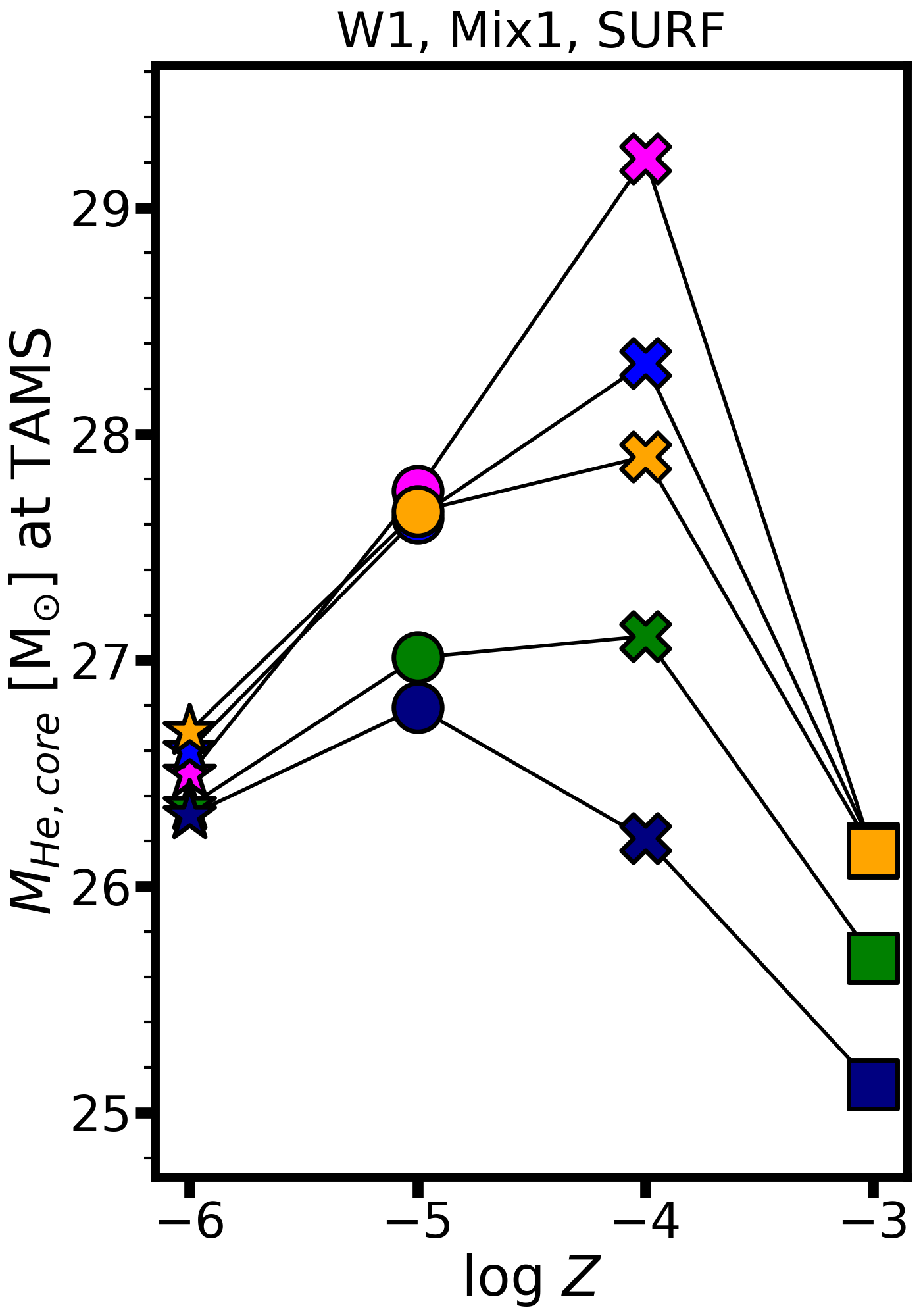}\includegraphics[width=0.25\textwidth]{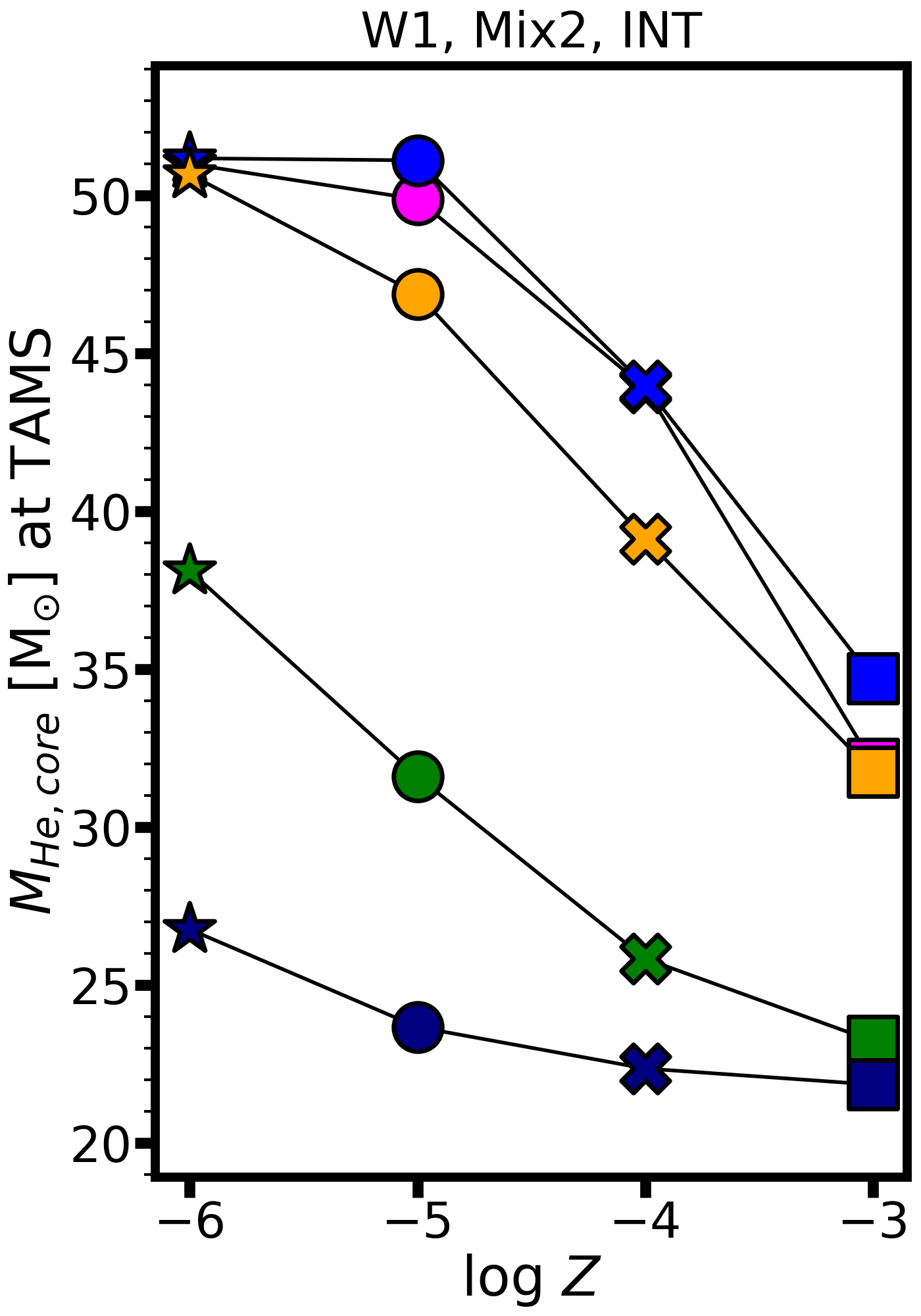}\includegraphics[width=0.25\textwidth]{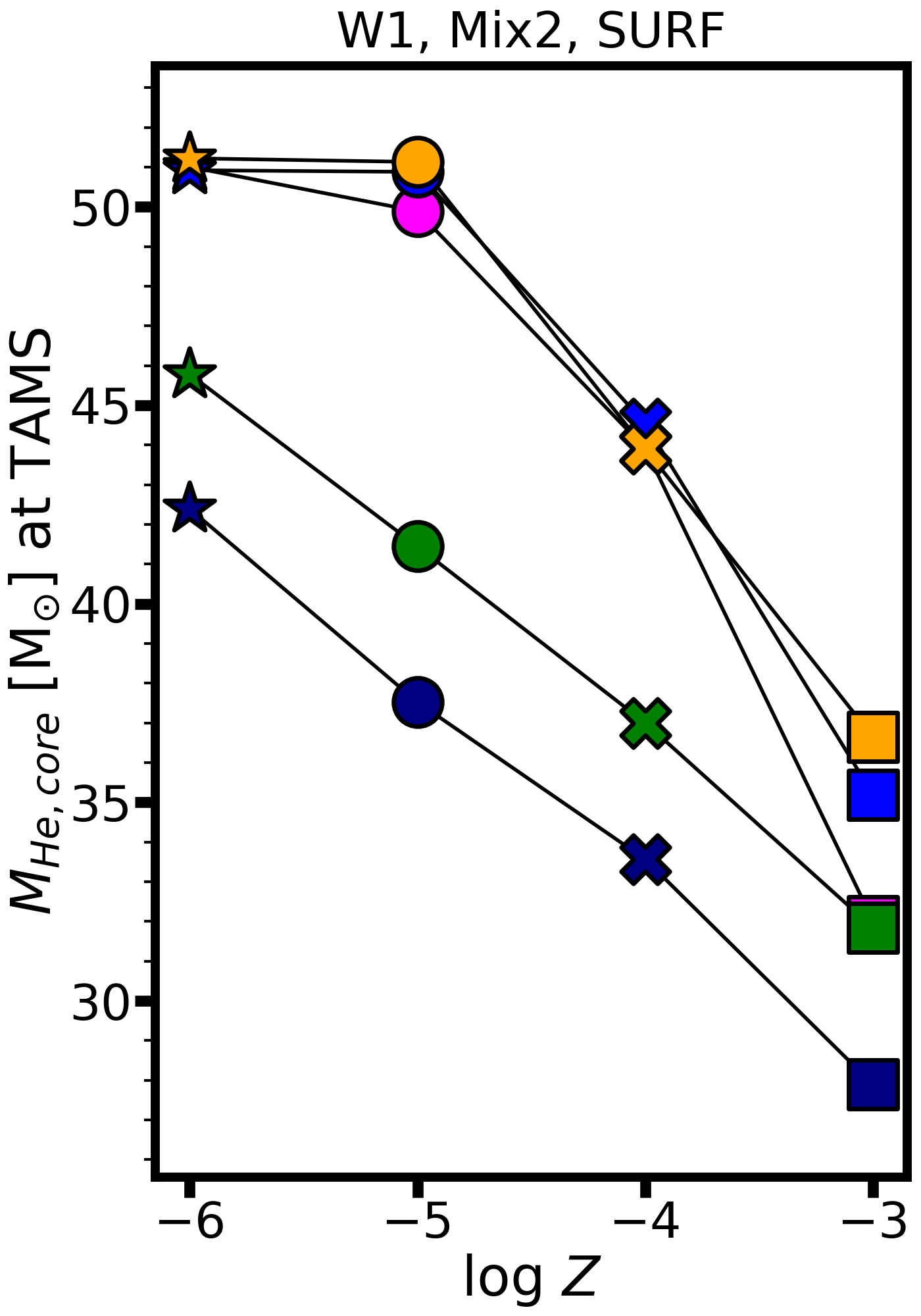}
\caption{Helium core mass of initially 60 M$_\odot$ models (with $\sim$43 M$_\odot$ ZAMS core mass) at the end of the main sequence as a function of the initial magnetic field strength (top) and metallicity (lower). The panel title indicates the wind, chemical mixing, and magnetic braking schemes. Note the different vertical axis scales.}
\label{fig:coremass}
\end{figure*}

%
%
\subsection{Implications for low-metallicity supernovae}\label{sec:d2}

The properties of the stellar core are important indicators of the star's subsequent evolution. We evaluate the convective core masses in our models at the TAMS to establish its parameter-space dependence. At this stage the composition of the core is essentially only helium.

Figure \ref{fig:coremass} shows our results combined across various initial magnetic field strengths, the metallicity ranges, the angular momentum transport and magnetic braking schemes, and the two chemical mixing schemes. 
In initially 60~M$_\odot$ models, the convective core mass at the ZAMS is $\approx$43~M$_\odot$.
Stars following a classical evolution in the HRD have a convective core that decreases during the main sequence. Overshooting and rotational mixing are able to extend the core size. Since overshooting is assumed with a fixed value in our models (c.f. \PaperIV), the main difference is the strength of rotational mixing in the radiative envelope. To a large extent, this is controlled by the Mix1-Mix2 chemical mixing schemes, and the rotational evolution of the star, constrained by magnetic braking. In the limit of QCHE, the core mass can increase during most of the main sequence evolution.

We find the following main trends. 
Models using the INT braking scheme tend to have smaller core masses compared to models using the SURF braking scheme. This is directly the consequence of the INT models braking the rotation of the entire star. Hence they rotate slower in the layers near the core, which makes the core-boundary mixing less efficient. 
Models utilising the Mix1 scheme are less efficient in chemical mixing. Therefore the predicted core masses in the Mix1 scheme are fairly similar. The Mix2 models tend to have much larger helium cores than Mix1 models when the magnetic field is weak and the metallicity is high. Indeed, in general, weaker magnetic fields and lower metallicity also lead to higher core masses.
Interestingly, in the Mix1 chemical mixing scheme (both for INT and SURF magnetic braking scenarios), the highest core mass is reached at $Z=10^{-4}$ (cross symbols). This is unlike the Mix2 case, where the core mass increases as a function of lowering metallicity.

The most striking result of this parameter study is the quantitative difference between the models. For example, for a given initial magnetic field strength or initial metallicity, the helium core masses can differ by up to about 30~M$_\odot$ for an initially 60~M$_\odot$ stellar model when considering various modelling assumptions. 

The core mass typically changes moderately during the core helium burning phase. For example, recent models of \cite{laplace2021} indicate that binary effects might slightly increase the helium core mass, while in typical single-stars a slight decrease is expected.
Although we do not follow the post-main sequence evolution, the TAMS helium core mass may give some hints towards the expected CO-core masses. In general, the explodability of these models as typical core-collapse events is difficult to predict and will require future investigations until the pre-supernova stages.
The onset of pair-instability supernova (PISN), which requires high CO-core masses\footnote{During oxygen core burning, pair-instability and pulsational pair-instability supernovae (PPISN) may occur if the central density is low enough and the central temperature is high enough to initiate electron-positron pair production. This can be approximated by a minimum mass of the carbon-oxygen core of about 40~M$_\odot$ for PPISN and 60~M$_\odot$ for PISN \citep{wheeler2015}, although this limit depends on several factors, for example, rotation (see, e.g., \citealt{chatzopoulos2012,chatzopoulos2015}).}, may be favoured in low-metallicity environments \citep[e.g.,][]{langer2012}.
We find that strongly magnetised models tend to have smaller He core masses (and thus likely smaller CO core masses). Therefore, in addition to low-metallicity environments, a weak magnetic field would also favour reaching the PISN regime in the parameter space considered in this work.

%
%
\section{Future work}\label{sec:future}

Extending high-metallicity stellar evolution model calculations to low-metallicity environments is not straightforward. We have demonstrated that changes in the wind, rotational, and magnetic properties largely affect the model predictions. Thus far the physics of stellar winds and magnetospheres are mostly constrained in the local Universe. If line-driving becomes inefficient to launch the wind, unusual conditions may arise. A low wind density implies a high magnetic confinement for a given strength of a magnetic field. To some extent, the star then becomes a conservative system in terms of mass and angular momentum. For this reason, magnetic and non-magnetic model predictions become identical under the assumption of the W2 wind scheme (e.g., the lowest metallicity models in Figure~\ref{fig:critrotW2}). If the decrease in wind strength and magnetic braking are gradual, then we expect systematic shifts in the physical quantities, such as surface abundances, ejected masses, and rotational velocities. In the future, we plan to apply sensible considerations and extend this work to zero-metallicity Population III stars. 

In order to help constrain primary nitrogen enrichment and study the pre-supernova stages with implication to low-metallicity supernovae, gamma-ray bursts, etc., computations until advanced burning stages are required. Amongst several challenging factors, the stability and evolution of fossil magnetic fields remain elusive beyond the main sequence. A point of interest in this context is the study of convective expulsion \citep{gough1966,spruit1979,schussler2001}. Namely, if strong convection develops in a stellar layer, the magnetic field lines cannot reconnect, leading to the removal of magnetic flux from a given region. Recently, such stability criterion was studied in one-dimensional stellar evolution codes \citep{macdonald2019,jermyn2020}. Since extended convective layers develop in the stellar envelope during the post-main sequence, it is a important to consider this when calculating models until advanced burning stages.

%
%
\section{Conclusions}\label{sec:concl}

In this work we extend our formalism to calculate surface fossil magnetic field effects in massive star evolutionary models at metallicities lower than that of the SMC. 
We explore a relevant parameter space to map out key uncertainties in evolutionary model predictions. Our main conclusions are: 
\begin{itemize}
    \item The separation between classical and quasi-chemically homogeneous evolution, as well as the evolution of surface abundances strongly depend on metallicity, the chemical mixing scheme, and the magnetic field strength. Stronger mixing, lower metallicity, and weaker magnetic fields favour the development of QCHE. 
    \item Generally, the surface magnetic field strength weakens by an order of magnitude on the main sequence, following the increase of the stellar radius. However, during QCHE, the stellar radius can remain almost unchanged, leading to a surface magnetic field strength that is constant in time. 
    \item In diffusive angular momentum schemes, the speed of rotation is the driver of chemical mixing in radiative regions. Similar to high-metallicity environments, the rotational evolution is drastically impacted by magnetic braking. However, following our assumptions, this process becomes weaker as a function of metallicity. Instead, the QCHE channel might lead to the onset of Wolf-Rayet type winds, which are able to brake the surface rotation by hundreds of km\,s$^{-1}$ on the main sequence.
    \item We systematically calculate the critical de-coupling mass-loss rates in our models. This shows that already at metallicities of $Z=10^{-4}$ and $Z=10^{-5}$ the models with initially 20 and 60~M$_\odot$, respectively, may not be able to launch a wind. If we were to assume that there is no mass loss in such a case (W2 scheme), then the models would tend towards critical rotation. 
    \item The ejected mass of chemical species shows a significant dependence on magnetic field strength. Generally, the stronger the magnetic field, the lower the amount of ejected mass. A strong magnetic field slows the rotation and thus suppresses the surface nitrogen enrichment even if the star initially rotates fast or mixes material efficiently.
    \item The core masses by the end of the main sequence can vastly differ, depending on the chemical and rotational evolution. For an initial mass of 60~M$_\odot$, we find helium core masses in the range of 22 - 52~M$_\odot$. We expect that in low-metallicity environments pair-instability supernova are favoured from initially non- or weakly magnetised progenitors.
\end{itemize}

Our study is an exploration of physical ingredients in massive star evolutionary models that are unconstrained at low metallicities. As such, we demonstrate that the interpretation based on stellar models is uncertain and requires careful consideration of the underlying assumptions. For this reason, the interpretation of nitrogen excess of distant galaxies could strongly depend on specific stellar progenitor models. Our results for the ejected mass of nitrogen from main sequence single-star models show a large scatter of several orders of magnitude, highlighting the need to better understand physical processes in massive stars at low metallicities.

\section*{Acknowledgements}
We thank the anonymous referee for providing us with constructive comments that led to improving the manuscript. We thank the \textsc{mesa} developers for making their code publicly available. 
%
Z.K. acknowledges support from JSPS Kakenhi Grant-in-Aid for Scientific Research (23K19071).
G.C. and H.N. are supported by the NINS International Research Exchange Support Program. H.N. is also supported by Grant-in-Aid for Scientific Research (23K03468).
A.u.-D. acknowledges NASA ATP grant number 80NSSC22K0628 and support by NASA through Chandra Award number TM4-25001A issued by the Chandra X-ray Observatory 27 Center, which is operated by the Smithsonian Astrophysical Observatory for and on behalf of NASA under contract NAS8-03060.
Numerical computations were carried out on the PC cluster at the Center for Computational Astrophysics, National Astronomical Observatory of Japan.

\section*{Data Availability}

A full reproduction package is available on Zenodo at \textbf{link will be provided after the permanent MNRAS identifier is allocated.} 


\bibliographystyle{mnras}
\bibliography{ref}


\appendix

\section{\textsc{mesa} microphysics}\label{sec:micro}
%
%

%
%
The MESA EOS is a blend of the OPAL \citep{Rogers2002}, SCVH
\citep{Saumon1995}, FreeEOS \citep{Irwin2004}, HELM \citep{Timmes2000},
PC \citep{Potekhin2010}, and Skye \citep{jermyn2021b} EOSes.

Radiative opacities are primarily from OPAL \citep{Iglesias1993,
Iglesias1996}, with low-temperature data from \citet{Ferguson2005}
and the high-temperature, Compton-scattering dominated regime by
\citet{Poutanen2017}.  Electron conduction opacities are from
\citet{Cassisi2007}.

Nuclear reaction rates are from JINA REACLIB \citep{Cyburt2010}, NACRE \citep{Angulo1999} and
additional tabulated weak reaction rates; \citet{Fuller1985, Oda1994,
Langanke2000}.  Screening is included via the prescription of \citet{Chugunov2007}.
Thermal neutrino loss rates are from \citet{Itoh1996}.

\bsp	
\label{lastpage}
\end{document}